\documentclass[twocolumn,amssymb,amsmath,
nofootinbib,tightenlines,showpacs,floatfix,
superscriptaddress,aps,prb]{revtex4-2}
\usepackage{amsfonts}
\usepackage{txfonts}
\usepackage{upgreek}
\usepackage{amsbsy}
\usepackage{bm}
\usepackage[colorlinks=true,linkcolor=blue,filecolor=blue,menucolor=yellow,urlcolor=blue,citecolor=blue,anchorcolor=blue]{hyperref}
\usepackage{graphicx}
\usepackage{times} 
\usepackage{dcolumn}

\newcommand{\bsigma}{\mbox{\boldmath$\sigma$}}

\newcommand{\fsf}{F$_1$SF$_2$}
\newcommand{\sff}{SF$_1$F$_2$}

\newcommand{\fff}{F$_1$F$_2$F$_3$}

\newcommand{\fnf}{F$_1$N F$_3$}
\newcommand{\sfffs}{SF$_1$F$_2$F$_3$S}

\begin{document}

\title{Supercurrent Diode Effect, Spin Torques, and Robust Zero-Energy Peak in
 Planar Half-Metallic Trilayers } 
\author{Klaus Halterman} 
\email{klaus.b.halterman.civ@us.navy.mil}
\affiliation{Michelson Lab, Physics Division, Naval Air Warfare Center, China Lake, California 93555}
\author{Mohammad Alidoust} 
\affiliation{Department of Physics, Norwegian University of Science
  and Technology, N-7491 Trondheim, Norway}
\author{Ross Smith} 
\affiliation{Army Research Lab, Aberdeen Proving Ground, MD 21005}
\author{Spencer Starr} 
\affiliation{General Dynamics Information Technology, Falls Church, VA 22042}

\date{\today} 
\begin{abstract}
We consider a Josephson junction with
 \fff~ferromagnetic trilayers in the ballistic regime, where the
magnetization in each
ferromagnet  ${\rm F}_i~(i=1,2,3)$,
can have  arbitrary orientations and 
magnetization strengths.
The trilayers are 
 sandwiched between two $s$-wave superconductors with a macroscopic phase
difference $\Delta \varphi$.
  With our generalised theoretical and numerical techniques,
  we are able to
study the planar spatial profiles and $\Delta \varphi$-dependencies of
the charge supercurrents, spin supercurrents, spin torques, and density of states
for complex systems that are finite in two dimensions.
A broad range of magnetization strengths of  the central  $\rm F_2$ layer are  considered, from an
unpolarized normal metal (N)  to a half-metallic phase, supporting only one spin species.
Our results reveal that when 
  the magnetization configuration in \fff~has three orthogonal components, a
   supercurrent can flow  at $\Delta \varphi=0$,
    and a strong second harmonic in the  current-phase relation appears.
    Remarkably, upon increasing the magnetization strength in the central ferromagnet layer up to the half-metallic limit,
     the self-biased current and second harmonic component 
      become dramatically enhanced, and the critical supercurrent reaches its maximum value.
      The 
      higher harmonics in the current-phase relations can be controlled
      by the relative magnetization orientations, with negligible current damping compared to the
      corresponding  \fnf~counterparts.    
   Additionally,  for a broad range of exchange field strengths in the central  ferromagnet ${\rm F}_2$,
     the ground state of the system can be tuned to an arbitrary phase difference $\varphi_0$, e.g.,
     by rotating the magnetization in the outer ferromagnet $\rm F_3$.
    For  intermediate exchange field strengths in 
     ${\rm F}_2$,  a  $\varphi_0$ state can arise that creates
 a  superconducting  diode effect,
 whereby  $\Delta\varphi$
 can be tuned to
create a one-way 
 dissipationless current flow.
     The spatial maps of the spin currents and effective magnetic moments reveal a 
      long-ranged spin torque  in the half-metallic phase. 
      Moreover, the density of states studies  demonstrate
      the  emergence of zero energy peaks
       for the 
      mutually orthogonal magnetization configurations, which is strongest in the half-metallic phase. Our results suggest that this simple trilayer Josephson junction 
      can be an excellent candidate for producing
     experimentally accessible signatures for
      long-ranged self-biased supercurrents, and supercurrent 
     diode-effects.
  
\end{abstract}
\maketitle
\section{Introduction} \label{intro}

The competition of ferromagnetic 
and superconducting orders in hybrid junction systems has been the focus of extensive research over the past decades. When 
constructing devices that consist of
superconductor (S) and
ferromagnet (F) elements,
the \fsf~and \sff~spin valves are
some of  the simplest configurations 
with externally controlled  
system properties.
Such 
platforms  have been extensively studied experimentally and theoretically  \cite{L.R.Tagirov,fsf1,fsf2,fsf3,fsf41,fsf42,fsf5,fsf6,fsf7,sff1,sff2,jap_2015,andeev,Ryazanov2,Zdravkov,Antropov,Khaydukov,zutic}.
In addition to the geometrical properties like 
the layer thicknesses, 
the relative magnetization misalignment in these systems plays a
pivotal role in determining the
 end functionality of these systems.
  To effectively control the magnetization misalignment, one needs to judiciously choose materials with 
  the proper
   magnetization strengths so that an external magnetic field can rotate the magnetization in one of the F regions while the other one remains essentially intact \cite{fsf5,Ryazanov2,fsf6}.

The presence of magnetization inhomogeneity and spin-orbit coupling can result in long-ranged spin-triplet superconducting correlations \cite{fsf2}. 
These long-ranged correlations are predicted to 
leave their imprint on various measurable quantities, including  the critical supercurrent \cite{Bergeret2005:RMP,jap_2015,Halterman2015:PRB,buzzed,Buzdin2005,sfs4,Keizer2006} and zero-energy density of states \cite{zep,zep2,zep3,zep4,half,half2}. 
An unambiguous observation of the triplet correlations in the DOS  still has
yet  to be found. However, the 
critical supercurrent has revealed compelling evidence of triplet correlations
  by employing a relatively thick half-metallic (H) layer of $\rm CrO_2$ (0.3-1 $\upmu$m) in a {\rm NbTiN}-$\rm CrO_2$-{\rm NbTiN} Josephson configuration \cite{Keizer2006}. 
  Additionally, a magnetization-orientation dependent supercurrent response on the order of $40~\upmu $A 
  was observed,  hinting at a controlled long-ranged spin supercurrent.
  When the ferromagnet is 
   in a half-metallic phase, 
    the very large magnetization strength (on the order of Fermi energy) 
    permits only one spin to exist \cite{J.Coey1}.
     In  recent years, heterostructures consisting of H layers have attracted much attention and caused further advances in this research field \cite{bernard1,singh,C.Visani1,Z.Sefrioui1,K.Dybko1,Y. Kalcheim1}. 
     The H layer has shown that both theoretically and experimentally,
     it can provide an
     enhancement to the critical temperature,
      and induce a strong
       nonuniform response to the
       magnetization misalignment in multilayer structures \cite{bernard1,singh,Mironov,half,half2}. 
       In an experiment involving a 
{\rm MoGe}-{\rm Ni}-{\rm Cu}-{\rm CrO}$_2$ spin valve,
 the superconducting critical temperature $T_c$ showed variations
  on the order of $\Delta T_c\!\sim\!800$mK with variations in
   the magnetization misalignment angle, in excellent agreement with theoretical results \cite{half,half2}. 
A $\rm La_{0.6}Ca_{0.4}MnO_3$ half-metal counterpart, 
consisting of 
a {\rm LCMO}-{\rm Au}-{\rm Py}-{\rm Cu}-{\rm Nb} stack,
showed $\Delta T_c \!\sim\! 150$mK, 
which is a slight
 improvement over  $\Delta T_c\!\sim\! 50$mK
 found
 in
 systems that use weaker ferromagnetic counterparts \cite{sff2}.
 The former experiment \cite{singh} employed a
  large out-of-plane external magnetic field of $H\sim 2$T whereas the latter experiment \cite{bernard1} used a relatively 
low in-plane magnetic field of $H \!\sim\! 3.3$mT that improves device reliability. 

The Josephson effect is traditionally understood to be the supercurrent that is generated when there is a difference between the
macroscopic phases of two S banks that are separated by an intrinsically non-superconducting region. 
Once a supercurrent is established,
its flow depends on 
the geometrical parameters, phase differences, and material properties of the system.
 In a conventional  ferromagnetic Josephson junction,
  the ground state energy can switch between 
  specific superconducting phase differences, typically 
  $\Delta\varphi=0$ and $\pi$, due to the dominant first harmonic 
in the current-phase relation: $\sin \Delta \varphi$ \cite{A.G.Golubov,Ryazanov2}. 
Additionally, close to any
 $0$-$\pi$ crossover, higher harmonics, i.e., $\sin 2\Delta\varphi$, $\sin 3\Delta \varphi \ldots$,  can appear. The higher harmonics can cause a continuous transition from $0$ state to the $\pi$ state \cite{Alidoust:PRB2021,goldobin}.
A finite supercurrent at zero phase difference, $\Delta\varphi=0$, can also arise, and the
Josephson ground-state can  be characterized by a
superconducting phase difference $\varphi_0$.
 There are two $\varphi_0$ ground states 
 located symmetrically around $\Delta\varphi=0$ associated with this continuous transition, and can have 
 no spontaneous current.\cite{Alidoust:PRB2021,goldobin} Note that a
$\varphi_0$ junction
 introduces excellent opportunities to introduce quantum computer bits other than $0$ and $\pi$. 
Another mechanism to induce a $\varphi_0$ ground state
is a proper combination of magnetization direction and spin-orbit coupling.\cite{A.I.Buzdin2,A.A.Reynoso,D.B.Szombati,zu1,zu2,zu3,Assouline2019:NC,Alidoust2018:PRB1,Alidoust2020:PRB2,K.Kulikov,Alidoust2018:PRB2,Alidoust2018:PRB3,Alidoust2020:PRB1,Alidoust:PRB2021,J.F.Liu1,I.Margaris,M.A.Silaev1,N.Mohanta,H.Wu,C.Baumgartner,J.J.He,N.F.Q.Yuan} In this case, the $\varphi_0$ is associated with a self-biased spontaneous supercurrent and located on one side of $\Delta\varphi=0$, depending on the direction of magnetization and spin-orbit coupling.\cite{Alidoust:PRB2021} 
This mechanism is found to appear in a wide range of materials platform in the presence and absence of nonmagnetic impurities, such as topological insulators \cite{zu1,zu2,zu3,Assouline2019:NC}, Weyl semimetals \cite{Alidoust2018:PRB1,Alidoust2020:PRB2,K.Kulikov}, black phosphorus \cite{Alidoust2018:PRB2,Alidoust2018:PRB3}, linear Rashba-Dressehalus spin-orbit coupled platforms for revealing a persistent spin helix \cite{Alidoust2020:PRB1}, and cubic Rashba-Dressehalus systems \cite{Alidoust:PRB2021}. 
Another configuration that hosts a self-biased supercurrent and has received far 
less attention is a spin-polarized Josephson junction with a simple arrangement of three ferromagnets 
having 
their magnetization orientations each orthogonal to one another \cite{J.F.Liu1,I.Margaris,M.A.Silaev1}.

\begin{figure}[t]
\centering
\includegraphics[width=1\columnwidth]{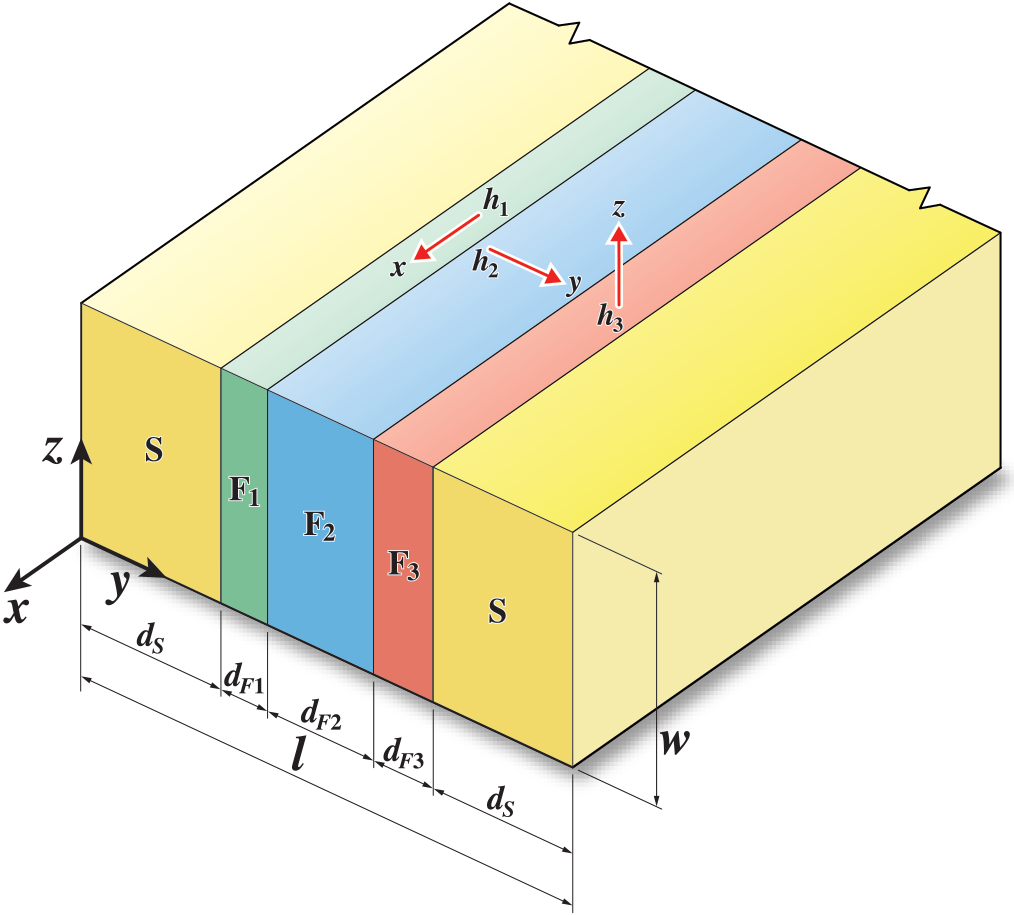}
\caption{(Color online). Schematic of the \fff~ferromagnetic trilayer structure sandwiched between two $s$-wave superconductors 
($\rm S$) with thickness $d_S$. The system is finite in the $y$-$z$ plane with 
width $w$ and length $l$, while  infinite along $x$.  
The ferromagnets $\rm F_i \,(i=1,2,3)$ are of thicknesses $d_{Fi}$
and have arbitrary exchange-field strengths $h_i$.
The typical magnetic configuration is shown here, where ${\bm h}_1$, ${\bm h}_2$,
and ${\bm h}_3$ 
are oriented along the $x$, $y$, and $z$ directions, respectively (the `$xyz$' configuration),
although
the  exchange-field vectors can have
arbitrary  orientations determined by
${\bm h}_{i} = h_{i}(\cos\theta_i, \sin\theta_i\sin\phi_i, \sin\theta_i\cos\phi_i)$.
} 
\label{diagram}
\end{figure}

The supercurrent $J(\Delta\varphi)$ in spin-polarized Josephson junctions can be controlled by a variety of mechanisms,
including
through  magnetization rotations, and incorporating different types of
magnets with  mismatched exchange-field strengths.
For trilayer ferromagnetic Josephson junctions, including those that contain half-metallic layers,
and where only two of the exchange field vectors are
orthogonal, the supercurrent direction can
be altered by changing the relative magnetization orientation \cite{Halterman2015:PRB,C.-T.Wu2018}.
If the ground state of the Josephson junction is at $\varphi_0=0$,
 the supercurrent generally obeys
$|J(+\Delta\varphi)| = |J(-\Delta\varphi )|$.
 In some special cases,
a superconducting diode effect can arise, whereby
$|J(+\Delta\varphi + \varphi_0)| \neq |J(-\Delta\varphi + \varphi_0)|$.
When this occurs, 
 changing
  the direction of the superconducting phase gradient, i.e.,
   $\Delta\varphi \rightarrow -\Delta\varphi$, 
  the amplitude for the supercurrent is no longer invariant.
   We find that remarkably, for the simple trilayer ferromagnet \sfffs~structure shown in Fig.~\ref{diagram},
where the exchange field vectors in the  ferromagnet layers are orthogonal to one another,
   a diode effect emerges for the supercurrent 
at intermediate exchange field strengths of the central ferromagnet. 

The main aim of this paper is therefore to present
an extensive  investigation of the influence of
the many relevant physical  parameters on the spin and charge
 transport  in  \sfffs~ferromagnetic Josephson
 configurations. 
 We will use for these purposes a tailor made
numerical method that allows for the exact
solutions of the relevant microscopic equations for 
Josephson strictures that are finite in two-dimensions.
 To systematically explore a broad range of systems
  and determine experimentally relevant parameter sets 
  resulting in a tunable 
  $\varphi_0$ state and  supercurrent flow, 
  the magnetization  in one of the ferromagnet layers
  will have its magnitude 
   continuously increased up to the half-metallic phase, and several key
   orientations will be investigated. 
   To further understand  the  quantum size-effects inherent to structures constrained in two-dimensions,
   a variety of structure lengths and widths will be studied (see Fig.~\ref{diagram}).     
Moreover, the often extreme ranges in energy scales 
for these types of systems require
a microscopic theory  without the approximations inherent to quasiclassical approaches. 
To properly simulate these  experimentally relevant systems, we have therefore
generalized the numerical Bogoliubov de Gennes (BdG) approach to planar geometries where the system is confined in two dimensions and infinite along the third dimension \cite{Alidoust2020:PRB0,H.Meng,half,half2,bernard1}.
The microscopic method used here accounts for the significant band
curvature near the Fermi energy arising from the strong spin-splitting effects of the half-metallic layers. 
This numerical approach has also found excellent agreement 
with results 
that rely on asymptotic approximations, such as the
Andreev approximation \cite{Alidoust2020:PRB0}, {\it ab initio} calculations \cite{H.Ness}, and supports the half-metallic phase where the  magnetization strength  is as large as the Fermi level \cite{half,half2,bernard1}. 
This latter phase is inaccessible to approximate methods, such as the quasiclassical approach, which considers the Fermi level to be the largest energy in the 
system \cite{Z.Shomali,Zh.Devizorova1,Zh.Devizorova2,Mironov,Bergeret2005:RMP,Buzdin2005}. 

We shall
explore the whole of the parameter range that accounts for the
effects that finite sample size, relative magnetizations (magnitude and orientation), 
and macroscopic phase differences
have on the singlet and triplet pair amplitudes, magnetic moments, and  local density of states (DOS).
We will identify the ground state 
 with  phase difference $\varphi_0$ 
 that is tunable
by
     changes  to the magnetization orientation of the outer ferromagnet.
    For  intermediate exchange field strengths in 
     the central $F$ layer, we find the emergence of a  $\varphi_0$ state with a
     one-way supercurrent flow, i.e., 
 a  superconducting  diode effect.
 Further increases of 
 the magnetization strength in the $\rm F_2$ layer enhances 
the
supercurrent so that the critical supercurrent reaches a
maximum at the half-metallic phase. This magnetization increase induces 
$0$-$\pi$ transitions and controls the induction of higher harmonics in the current-phase-relation. 
When the
magnetization in the F layers has mutually orthogonal components, a spontaneous supercurrent 
flows through the system, which is strongest when the $\rm F_2$ layer hosts a half-metal. The spatial maps of the
spin currents and magnetic moments reveals
 a  long-ranged magnetic moment and equal-spin triplet correlations, propagating from the $\rm F_1$ layer into the right superconducting region. The spatial profiles of 
 the 
 DOS reveals that 
 when the trilayer junction region has
 mutually orthogonal magnetizations, 
  robust zero energy peaks appear in the
 middle of the outer ferromagnets in the  trilayers,
with the largest peaks appearing at the interfaces between the central and outer ferromagnets.
 Similar to the critical supercurrent, the zero energy peaks are largest when the $\rm F_2$ layer is a half-metal.

The paper is organized as follows. In Sec.~\ref{method}, 
we summarize the generalized BdG approach and present expressions for the charge current, spin current, superconducting correlations, and magnetic moment. In Sec.~\ref{results}, the numerical results and findings shall be presented. Finally, in Sec.~\ref{conclusions} we provide a brief summary of the main results.

\section{Methods} \label{method}
We first present the Hamiltonian, BdG formalism, the associated field operators, and Fourier expansion method in Sec.~\ref{hamil}. In Sec. \ref{tripcorr}, the spin-singlet and spin-triplet correlations are given in terms of BdG transformations. The definitions and derivations of magnetic moment, DOS, charge current, and spin current and spin torque are presented in Secs. \ref{magnetization}, \ref{dossec}, \ref{chargetrans}, and \ref{spintrans}, respectively.
\subsection{Hamiltonian}\label{hamil}
We now briefly summarize the BdG approach used to describe our
spatially inhomogeneous nanostructure.
We consider a Josephson configuration that is finite in the $y$-$z$ plane, and infinite in the $x$ direction 
[see Fig.~\ref{diagram}].
 The effective Hamiltonian ${\cal H}_{\rm eff}$
 that describes this system is given by
\begin{eqnarray}
\label{ham}
{\cal H}_{\rm eff}&=&\int d^3r \left\{ \sum_s
\psi_s^{\dagger}\left(\bm{r}\right)H_0
\psi_s\left(\bm{r}\right)\right.\nonumber \\
&+&\left.\frac{1}{2}\left[\sum_{s\:s'}\left(i\sigma_y\right)_{ss'}
\Delta\left(\bm{r}\right)\psi_s^{\dagger}
\left(\bm{r}\right)\psi_{s'}^{\dagger}
\left(\bm{r}\right)+H.c.\right]\right.\nonumber \\
&-&\left.\sum_{s\:s'}\psi_s^{\dagger}
\left(\bm{r}\right)\left(\bm{h}\cdot\bm{\sigma}
\right)_{ss'}\psi_{s'}\left(\bm{r}\right)\right\},
\end{eqnarray}
where $s$ and $s'$ are spin indices, and $\Delta({\bm r})$ is the pair potential, given by,
\begin{align}
\label{del}
\Delta({\bm r}) = &\frac{V_s ({\bm r})}{2}\left[\left\langle
\psi_{\uparrow}({\bm r}) \psi_{\downarrow} ({\bm r})\right\rangle-
\left\langle \psi_{\downarrow}({\bm r}) \psi_{\uparrow} ({\bm r})\right\rangle\right],
\end{align}
Here we have assumed an
  $s$-wave on-site potential
 with attractive interaction:
$V ({\bm r}-{\bm r'})={V_s ({\bm r})}\delta ({\bm r}-{\bm r'})$,
with  $V_s({\bm r})$  the interaction strength, which is nonzero only
for energies less than a characteristic ``Debye'' energy, $\omega_D$,
and nonzero only within  the superconductor regions. 
The single-particle 
Hamiltonian $H_0$ for our finite-sized system in the two directions $y$ and $z$ [see Fig.~\ref{diagram}] is written:
\begin{align}
H_0=-\frac{1}{2m}\left(\frac{\partial^2}{\partial z^2} +\frac{\partial^2}{\partial y^2}\right)+\frac{1}{2m}k_x^2 + U(y,z)-\mu,
\end{align}
 in which $\mu$ is the Fermi energy, and $U(y,z)$ is the spin-independent scattering potential. 
The exchange  field $\bm{h}\equiv (h_x,h_y,h_z)$
describes the magnetic exchange interaction,
and $\bm{\sigma}\equiv (\sigma_x,\sigma_y,\sigma_z)$ are Pauli matrices.
To diagonalize ${\cal H}_{\rm eff}$, 
the field operators $\psi_s$ and $\psi_s^\dagger$ are expanded by means of the Bogoliubov transformation:
\begin{equation}
\begin{aligned} \label{trans}
\psi_\uparrow ({\bm r}) & = \sum_n \left( u_{n\uparrow}({\bm r}) \gamma_n - 
v^*_{n\uparrow}({\bm r}) \gamma_n^\dagger\right), \\
\psi_\downarrow ({\bm r}) & = \sum_n \left( u_{n\downarrow}({\bm r}) \gamma_n +
v^*_{n\downarrow}({\bm r}) \gamma_n^\dagger\right),
\end{aligned}
\end{equation}
where $u_{n s}$, $v_{n s}$ represent the quasiparticle amplitudes,
and $\gamma_n$, $\gamma_n^\dagger$ are the Bogoliubov creation and annihilation operators, respectively.
The transformations in Eq.~(\ref{trans}) are required to diagonalize ${\cal H}_{\rm eff}$
such that,
\begin{equation}
\begin{aligned}\label{diag}
\left[{\cal H}_{\rm eff} ,\gamma_n\right] &= - \epsilon_n \gamma_n, \quad
\left[{\cal H}_{\rm eff} ,\gamma^\dagger_n\right] &= \epsilon_n \gamma^\dagger_n,
\end{aligned}
\end{equation}
which leads to 
the spin-dependent
Bogoliubov-de Gennes (BdG) equations as,
\begin{widetext}
\begin{align} 
&\begin{pmatrix}
H_0 -h_z(y,z)&-h_x(y,z)+ih_y(y,z)&0&\Delta(y,z) \\[5 pt]
-h_x(y,z)-ih_y(y,z)&H_0 +h_z(y,z)&\Delta(y,z)&0 \\[5 pt]
0&\Delta^*(y,z)&-(H_0 -h_z(y,z))&-h_x(y,z)-ih_y(y,z) \\[5 pt]
\Delta^*(y,z)&0&-h_x(y,z)+ih_y(y,z)&-(H_0+h_z(y,z)) \\
\end{pmatrix}
\left( \begin{array}{c}
u_{n,\uparrow}(y,z)\\[5 pt]
u_{n,\downarrow}(y,z)\\[5 pt]
v_{n,\uparrow}(y,z)\\[5 pt]
v_{n,\downarrow}(y,z)
\end{array}\right)
=
\epsilon_n
\left( \begin{array}{c}
u_{n,\uparrow}(y,z)\\[5 pt]
u_{n,\downarrow}(y,z)\\[5 pt]
v_{n,\uparrow}(y,z)\\[5 pt]
v_{n,\downarrow}(y,z)
\end{array}\right)
\label{Hk},
\end{align}
\end{widetext}
The pair potential $\Delta(y,z)$ in Eq.~(\ref{Hk}) 
 is assigned  an  initial value,
 taken to be the bulk gap, $\Delta_0$, in  
${S_1}$
and  $\Delta_0 e^{i\varphi_2}$ in $S_2$,
so that the macroscopic phase difference is $\Delta\varphi = \varphi_2$.
We will investigate zero-phase (``anomalous'') spin and charge currents
for
 $\Delta\varphi=0^\circ$,
as well as controllable Josephson $\varphi_0$ states that occur when the ground state of
the system occurs when  $\Delta\varphi=\varphi_0$ (where 
 $\varphi_0$ does not necessarily equal  0 or $\pi$).
 The ferromagnetic exchange field vector has its components expressed as:
\begin{align}
 {\bm h}_{i} = h_{i}(\cos\theta_i, \sin\theta_i\sin\phi_i, \sin\theta_i\cos\phi_i),
 \label{hex}
\end{align}
where $i=1,2,3$  denotes the ferromagnetic region shown in Fig.~\ref{diagram}.
With these inputs, Eq.~(\ref{Hk}) is then numerically
diagonalized to give
the quasiparticle eigenenergies $\epsilon_n$ and 
 the quasiparticle eigenfunctions ($u_{n \sigma}(y,z)$, $v_{n \sigma}(y,z)$).
 throughout the entire junction \cite{khold}.
 Since the structure in Fig.~\ref{diagram} is finite in the $y$ and $z$ directions,
we solve the  BdG equations
by expanding  the quasiparticle wavefunctions in a two-dimensional Fourier series basis,
\begin{equation}
\begin{aligned}
 \label{expanse}
u_{n, \sigma}(y,z)&=\dfrac{2}{\sqrt{w l}} \sum^{N_l}_p \sum^{N_w}_q u^{p,q}_{n, \sigma} \sin \Big(\frac{p \pi y}{l}\Big)
\sin \Big(\frac{q \pi z}{w}\Big), \\
v_{n, \sigma}(y,z)&=\dfrac{2}{\sqrt{w l}}  \sum^{N_l}_p \sum^{N_w}_q v^{p,q}_{n,\sigma} \sin \Big(\frac{p \pi y}{l}\Big)
\sin \Big(\frac{q \pi z}{w}\Big),
\end{aligned}
\end{equation}
where $\sigma=\uparrow,\downarrow$. Further details on this method can be found in Appendix~\ref{appA}

\subsection{Triplet correlations and pair amplitude}
\label{tripcorr}
As discussed in Introduction, when strong ferromagnetic 
layers, such as  half-metals, are part of a Josephson junction,
 the  spin-triplet Cooper pairs can play an important role
in both the thermodynamic  and transport properties of the system \cite{half,C.-T.Wu2018}.
We begin by writing 
the spin triplet pairing correlations in the usual  way \cite{Halterman2007}, 
\begin{subequations}
\label{trippy}
\begin{align}
{f_0}({\bm r},t) = &\frac{1}{2}\left[\left\langle
\psi_{\uparrow}({\bm r},t) \psi_{\downarrow} ({\bm r},0)\right\rangle+
\left\langle \psi_{\downarrow}({\bm r},t) \psi_{\uparrow} ({\bm r},0)\right\rangle\right],\\
\label{f1}
{f_1}({\bm r},t) = &\frac{1}{2}\left[\left\langle
\psi_{\uparrow}({\bm r},t) \psi_{\uparrow} ({\bm r},0)\right\rangle
-\left\langle \psi_{\downarrow}({\bm r},t) \psi_{\downarrow} ({\bm r},0)\right\rangle\right], \\
\label{f2}
{f_2}({\bm r},t) = &\frac{1}{2}\left[\left\langle
\psi_{\uparrow}({\bm r},t) \psi_{\uparrow} ({\bm r},0)\right\rangle
+\left\langle \psi_{\downarrow}({\bm r},t) \psi_{\downarrow} ({\bm r},0)\right\rangle\right],
\end{align}
\end{subequations}
where the subscript $0$ corresponds to $m_s=0$, and 
the subscripts $1$ and $2$ refer to the
$m_s=\pm1$ projections on the spin quantization axis \cite{half2,trip2}. 
It was shown
in the previous works that using this approach to find both the opposite-spin and equal-spin triplet pairs,
satisfy the Pauli exclusion principle, and that the spin triplet pairs
vanish at $t=0$. If the exchange fields between the $\rm F$ layers 
are not aligned, 
the total spin operator of the pairs does not commute with the effective Hamiltonian, and
the spin-polarized components $f_1$ and $f_2$ acquire nonzero values. 

As mentioned earlier, 
the pair potential 
gives valuable information regarding the superconducting correlations within the superconductors only,
 as it vanishes outside of those regions where $V_s({\bm r})=0$. 
To reveal the fullest details
of the spin singlet correlations throughout the entire system, which includes 
 proximity effects between layers, 
 we evaluate the pair
amplitude from Eq.~(\ref{del}),
$f_3 = {\Delta({\bm r})}/{V_s ({\bm r})}$.
Inserting the Bogoliubov expansions (Eq.~(\ref{trans})) into Eq.~(\ref{del})
gives the 
  pair amplitude $f_3(y,z)$
  in terms of the quasiparticle amplitudes:
\begin{align} 
\label{swave}
f_3({y,z}) {=} \frac{1}{4}\sum_n 
\big[u_{n\uparrow}({y,z}) v^*_{n\downarrow}({y,z})  
{+}u_{n\downarrow}({y,z})  v^*_{n\uparrow}({y,z})  \big]\tanh
\Big(\frac{\epsilon_n}{2 T} \Big),
\end{align}
where the sum is cut off for states with energies 
that exceed $\omega_D$.
Here
the identity $1-2f(\varepsilon) = \tanh (\varepsilon/2T)$ 
 has been used  ($f$  is the Fermi function),
along with the expectation values: $\langle \gamma^\dagger_n \gamma_m \rangle=\delta_{nm} f(\epsilon_n)$,
$\langle \gamma_m \gamma_n^\dagger \rangle=\delta_{nm} (1-f(\epsilon_n))$,
and $\langle \gamma_n \gamma_m  \rangle=0$.

To express the spin triplet correlations [Eqs.~(\ref{trippy})] in terms of the quasiparticle energies and amplitudes, 
we first 
write  the 
 Heisenberg equations
of motion for the Bogoliubov creation and annihilation operators:
$i {\partial \gamma_n}/{\partial t} = [\gamma_n,{\cal H}_{\rm eff} ]$, 
$i {\partial \gamma^\dagger_n}/{\partial t} = [\gamma^\dagger_n,{\cal H}_{\rm eff}  ]$.
Using the conditions in Eq.~(\ref{diag}),
we have the solutions, $\gamma_n(t) = \gamma_n e^{-i \epsilon_n t}$ and
$\gamma^\dagger_n(t) = \gamma^\dagger_n e^{i \epsilon_n t}$.
Inserting these solutions into the 
generalized Bogoliubov transformations [Eq.~(\ref{trans})],
it becomes  possible to 
write the field operators in Eqs.~(\ref{trippy}) 
\begin{subequations}
\label{fall}
\begin{align}
f_{0}(y,z,t) &=  \frac{1}{2}\sum_{n}
\left[u_{n \uparrow}(y,z) v^{\ast}_{n\downarrow}(y,z)
-u_{n \downarrow}(y,z) v^{\ast}_{n\uparrow}(y,z)
\right] \zeta_n(t), \label{f0} \\
f_{1}(y,z,t) & =-\frac{1}{2} \sum_{n}
\left[
u_{n \uparrow}(y,z) v^{\ast}_{n\uparrow}(y,z)
+u_{n \downarrow}(y,z) v^{\ast}_{n\downarrow}(y,z)
\right]\zeta_n(t), \label{f1} \\
f_{2}(y,z,t) & =-\frac{1}{2} \sum_{n}
\left[
u_{n \uparrow}(y,z) v^{\ast}_{n\uparrow}(y,z)
-u_{n \downarrow}(y,z) v^{\ast}_{n\downarrow}(y,z)
\right]\zeta_n(t), \label{f2}
\end{align}
\end{subequations}
where 
the sums are over {\it all} energy values, and 
$\zeta_n(t) \equiv \cos(\epsilon_n t)-i\sin(\epsilon_nt)\tanh(\epsilon_n/2 T)$.
It can be more insightful to find the triplet correlations projected along the local 
spin axis \cite{K.Halterman_ss2016}, as dictated by the exchange field direction, instead of along the 
conventional $z$ direction. 
Performing the requisite spin rotations, we find:
\begin{align}
f_{0}' &= f_0 \cos\phi \sin\theta - f_1 \cos\theta- i f_2\sin\theta\sin\phi, \\
f_{1}' & = f_0 \cos \theta \cos\phi+f_1 \sin\theta - i f_2 \cos\theta \sin\phi, \\
f_{2}' & = -i f_0 \sin \phi + f_2 \cos\phi.
\end{align}
Thus, e.g., considering  the principle axes, we have for
along $x$ ($\theta=0^\circ$, $\phi=0^\circ$):
$f_0' = -f_1 $, $f_1'=f_0$, and $f_2' = f_2$. For along $y$ 
($\theta=90^\circ$, $\phi = 90^\circ$):
$f_0' = -i f_2 $, $f_1'=f_1$, and $f_2' = -i f_0$, and finally, along 
the  quantization $z$ axis
($\theta=90^\circ$, $\phi=0^\circ$),
 all spin triplet components are unchanged: $f_0' = f_0$, $f_1' = f_1$,
 $f_2'=f_2$.
 The spin singlet pair amplitude $f_3$ is always invariant under spin rotations.

\subsection{Magnetic moment}\label{magnetization}
 In our structures spin transport is influenced by the 
leakage of magnetic moment out of the $\rm F$ layers and into
the superconductors. This can be characterized
by the local magnetic moment ${\bm m}({\bm r})$, 
\begin{align} \label{mag}
{\bm m}({\bm r})  =-\mu_B\, \langle {\bm \eta}({\bm r})   \rangle,
\end{align}
where ${\bm \eta}({\bm r}) $ is the spin density operator,
\begin{align} \label{spinop}
{\bm \eta}({\bm r})  = \psi^\dagger({\bm r})  {\bsigma} \psi({\bm r}),
\end{align}
and $\mu_B$  the Bohr magneton. 
Again, by using the generalized Bogoliubov transformation, 
each component of ${\bm m}$ can be written 
in terms of the quasiparticle and quasihole wavefunctions:
\begin{subequations}
\label{mmcomp}
\begin{align}
m_x(y,z) & =- 2\mu_B \sum_n {\rm Re}\Big[ u_{n\uparrow}(y,z) u_{n\downarrow}^\ast(y,z) f(\epsilon_n)\nonumber \\
&-v_{n\uparrow}(y,z) v_{n\downarrow}^\ast(y,z)(1-f(\epsilon_n)) \Big],  
\end{align}
\begin{align}
m_y(y,z) & = 2\mu_B \sum_n {\rm Im}\Big[u_{n\uparrow}(y,z) u_{n\downarrow}^\ast(y,z) f(\epsilon_n) \nonumber \\
&+v_{n\uparrow}(y,z) v_{n\downarrow}^\ast(y,z)(1-f(\epsilon_n)) \Big], 
\end{align}
\begin{align}
m_z(y,z)=& - \mu_B \sum_n \Bigg[ \left(\left|u_{n\uparrow}(y,z)\right|^2- \left|u_{n\downarrow}(y,z)\right|^2\right) f(\epsilon_n)  \nonumber \\
&+\left(\left|v_{n\uparrow}(y,z)\right|^2- \left|v_{n\downarrow}(y,z)\right|^2\right)(1-f(\epsilon_n)) \Bigg].
\end{align}
\end{subequations}

\subsection{Density of states}
\label{dossec}
The proximity-induced electronic density of states (DOS)
can reveal signatures of the energy gap and localized Andreev
bound states. One important experimental  quantity 
 involves tunneling spectroscopy experiments
which can probe the local single particle spectra encompassing the proximity-induced DOS. The 
total DOS,  $N(y,z,\varepsilon)$ is the sum
$N_{\uparrow} (y,z,\varepsilon)+N_\downarrow(y,z,\varepsilon)$, involving the
spin-resolved local DOS $N_\sigma(y,z,\varepsilon)$:
\begin{align} \label{dos}
N_\sigma(y,z,\varepsilon) \!=\! - \! \sum_n 
\Biggl[ |u_{n\sigma} (y,z)|^2 f' (\varepsilon \!- \! \epsilon_n) 
+ |v_{n\sigma}(y,z)|^2 f'(\varepsilon \!+ \! \epsilon_n) \Biggr ],
\end{align}
where $f'(\varepsilon) = \partial f/\partial \varepsilon$ is
the derivative of the Fermi function,
$f(\varepsilon)=(e^{\varepsilon/T} + 1)^{-1}$.
It is also  convenient to use the following
relation: $f'(\varepsilon) = -1/(4 T \cosh^2 (\varepsilon/(2T))$.
The sum above is over the quasiparticle amplitudes 
and  eigenenergies $\epsilon_n$.
Thus, the DOS  calculated in Eq.~(\ref{dos}) can
provide  both a spatial mapping and energy-resolved
characterization of 
the number  of quasiparticle states in Josephson junctions.

\subsection{Charge transport}\label{chargetrans}
 
For the system shown in Fig.~\ref{diagram},
we
compute the supercurrent
 in the $y$ and $z$ directions
 by starting with
  the Heisenberg equation for the charge density $\rho({\bm r})$:
${{\partial}_t\left\langle\rho({\bm r})\right\rangle 
=i\left\langle\left[{\cal H}_{\rm eff},\rho({\bm r})\right]\right\rangle}$,
where,
\begin{align}
\rho({\bm r}) =&  2\sum_n 
\Biggl[ \left(|u_{n\uparrow} ({\bm r})|^2+|u_{n\downarrow} ({\bm r})|^2\right) f (\epsilon_n) \nonumber \\
& +\left ( |v_{n\uparrow}({\bm r})|^2+|v_{n\downarrow}({\bm r})|^2\right) \left(1- f(\epsilon_n) \right)\Biggr ].
\end{align}
The continuity equation for the charge supercurrent density ${\bm J}$
in the ferromagnetic junction region is written
\begin{equation}
\frac{\partial}{\partial t}\left\langle\rho(\bm r)\right\rangle
+\nabla\cdot{\bm J}=0,
\label{current}
\end{equation}
where each component  $J_i$ ($i=y,z$)
 is given by,
\begin{widetext}
 \begin{align}
&J_i=-\frac{e}{m}\!\sum_{n} {\rm Im }\Biggl\{ \left[u_{n\uparrow}(y,z) \partial_i 
u^{*}_{n\uparrow}(y,z)+ u_{n\downarrow}(y,z) \partial_i 
u^{*}_{n\downarrow}(y,z)  \right]f(\epsilon_n)
+
\left[v_{n\uparrow}(y,z) \partial_i
v^{*}_{n\uparrow}(y,z)+ v_{n\downarrow}(y,z) \partial_i 
v^{*}_{n\downarrow}(y,z)  \right]\left(1 - f(\epsilon_n)\right) \Biggr\}, 
\label{jj}
\end{align}
\end{widetext}
For the trilayer in the steady state, i.e., $\partial_t \left\langle\rho(\bm r)\right\rangle =0$,
 we have numerically verified  $\nabla\cdot{\bm J}=0$ when calculating 
current-phase-relations.

\subsection{Spin transport}
\label{spintrans}

We now extend the  above considerations to spin
transport.
As in the case of the charge density, the Heisenberg picture is 
utilized  to determine the time evolution of the spin density, 
${\bm \eta}({\bm r},t)$,
$\frac{\partial}{\partial t} \langle {\bm \eta}({\bm r},t) \rangle = i \langle 
[{\cal H}_{\rm eff},{\bm \eta}({\bm r},t)] \rangle$,
where ${\bm \eta}$ is given in Eq.~(\ref{spinop}).
The associated continuity equation now reads,
\begin{align} \label{scon}
\frac{\partial}{\partial t} \langle {\bm \eta}({\bm r},t) \rangle + {\bm \nabla}\cdot {\bm {\mathcal S}}&= 
{\bm \tau}, 
\end{align}
where ${\bm {\mathcal S}}$ is the spin current tensor.
This fundamental conservation law  balances the spin current gradients and the spin torque ${\bm \tau}$.
The expression for the 
spin-current, ${\bm {\mathcal S}}$, is found from taking the commutator 
above  Eq.~(\ref{scon}) and inserting  Eq.~(\ref{ham}):
\begin{align}
{\bm {\mathcal S}} = -\frac{i}{2m} \Bigl \langle \psi^\dagger({\bm r}){\bm \nabla} {\bm \sigma}\psi({\bm r})
- {\bm \nabla}  \psi^\dagger({\bm r}) {\bm \sigma} \psi({\bm r})
\Bigr \rangle,
\end{align}
where the tensor components ${\bm {\mathcal S}}_{\alpha\beta}$
represents spin current with spin $\beta$ flowing along the $\alpha$ direction in real-space.
We can now expand each spin component of the spin current in terms of
the quasiparticle amplitudes to obtain:
\begin{widetext}
\begin{subequations}
\begin{align}
&{\cal{S}}_{yx} = -\frac{i}{2m}\sum_n \Biggl\lbrace f_n\Bigl[u_{n\uparrow}^* \frac{\partial u_{n \downarrow}}{\partial y}+
u_{n\downarrow}^* \frac{\partial u_{n \uparrow}}{\partial y}-
u_{n\downarrow} \frac{\partial u^*_{n \uparrow}}{\partial y}-
u_{n\uparrow}\frac{\partial u^*_{n \downarrow}}{\partial y} \Bigr ] 
-(1-f_n)
\Bigl[v_{n\uparrow} \frac{\partial v^*_{n \downarrow}}{\partial y}+
v_{n\downarrow} \frac{\partial v^*_{n \uparrow}}{\partial y}-
v^*_{n\uparrow} \frac{\partial v_{n \downarrow}}{\partial y}-
v^*_{n\downarrow} \frac{\partial v_{n \uparrow}}{\partial y} \Bigr ] \Biggr\rbrace, \\
&{\bm S}_{yy} = 
-\frac{1}{2m}\sum_n \Biggl\lbrace f_n\Bigl[u_{n\uparrow}^* \frac{\partial u_{n \downarrow}}{\partial y}-
u_{n\downarrow}^* \frac{\partial u_{n \uparrow}}{\partial y}-
u_{n\downarrow} \frac{\partial u^*_{n \uparrow}}{\partial y}+
u_{n\uparrow}\frac{\partial u^*_{n \downarrow}}{\partial y} \Bigr ] 
-(1-f_n)
\Bigl[v_{n\uparrow} \frac{\partial v^*_{n \downarrow}}{\partial y}-
v_{n\downarrow} \frac{\partial v^*_{n \uparrow}}{\partial y}+
v^*_{n\uparrow} \frac{\partial v_{n \downarrow}}{\partial y}-
v^*_{n\downarrow} \frac{\partial v_{n \uparrow}}{\partial y} \Bigr ]\Biggr\rbrace, \\
&{\bm S}_{yz} =
-\frac{i}{2m}\sum_n \Biggl\lbrace f_n\Bigl[u_{n\uparrow}^* \frac{\partial u_{n \uparrow}}{\partial y}-
u_{n\uparrow} \frac{\partial u^*_{n \uparrow}}{\partial y}-
u^*_{n\downarrow} \frac{\partial u_{n \downarrow}}{\partial y}+
u_{n\downarrow}\frac{\partial u^*_{n \downarrow}}{\partial y} \Bigr ] 
-(1-f_n)
\Bigl[-v_{n\uparrow} \frac{\partial v^*_{n \uparrow}}{\partial y}+
v^*_{n\uparrow} \frac{\partial v_{n \uparrow}}{\partial y}+
v_{n\downarrow} \frac{\partial v^*_{n \downarrow}}{\partial y}-
v^*_{n\downarrow} \frac{\partial v_{n \downarrow}}{\partial y} \Bigr ]\Biggr\rbrace, 
\end{align}
\begin{align}
&{\cal{S}}_{zx} = -\frac{i}{2m}\sum_n \Biggl\lbrace f_n\Bigl[u_{n\uparrow}^* \frac{\partial u_{n \downarrow}}{\partial z}+
u_{n\downarrow}^* \frac{\partial u_{n \uparrow}}{\partial z}-
u_{n\downarrow} \frac{\partial u^*_{n \uparrow}}{\partial z}-
u_{n\uparrow}\frac{\partial u^*_{n \downarrow}}{\partial z} \Bigr ] 
-(1-f_n)
\Bigl[v_{n\uparrow} \frac{\partial v^*_{n \downarrow}}{\partial z}+
v_{n\downarrow} \frac{\partial v^*_{n \uparrow}}{\partial z}-
v^*_{n\uparrow} \frac{\partial v_{n \downarrow}}{\partial z}-
v^*_{n\downarrow} \frac{\partial v_{n \uparrow}}{\partial z} \Bigr ] \Biggr\rbrace, \\
&{\bm S}_{zy} = 
-\frac{1}{2m}\sum_n \Biggl\lbrace f_n\Bigl[u_{n\uparrow}^* \frac{\partial u_{n \downarrow}}{\partial z}-
u_{n\downarrow}^* \frac{\partial u_{n \uparrow}}{\partial z}-
u_{n\downarrow} \frac{\partial u^*_{n \uparrow}}{\partial z}+
u_{n\uparrow}\frac{\partial u^*_{n \downarrow}}{\partial z} \Bigr ] 
-(1-f_n)
\Bigl[v_{n\uparrow} \frac{\partial v^*_{n \downarrow}}{\partial z}-
v_{n\downarrow} \frac{\partial v^*_{n \uparrow}}{\partial z}+
v^*_{n\uparrow} \frac{\partial v_{n \downarrow}}{\partial z}-
v^*_{n\downarrow} \frac{\partial v_{n \uparrow}}{\partial z} \Bigr ]\Biggr\rbrace, \\
&{\bm S}_{zz}=
-\frac{i}{2m}\sum_n \Biggl\lbrace f_n\Bigl[u_{n\uparrow}^* \frac{\partial u_{n \uparrow}}{\partial z}-
u_{n\uparrow} \frac{\partial u^*_{n \uparrow}}{\partial z}-
u^*_{n\downarrow} \frac{\partial u_{n \downarrow}}{\partial z}+
u_{n\downarrow}\frac{\partial u^*_{n \downarrow}}{\partial z} \Bigr ] 
-(1-f_n)
\Bigl[-v_{n\uparrow} \frac{\partial v^*_{n \uparrow}}{\partial z}+
v^*_{n\uparrow} \frac{\partial v_{n \uparrow}}{\partial z}+
v_{n\downarrow} \frac{\partial v^*_{n \downarrow}}{\partial z}-
v^*_{n\downarrow} \frac{\partial v_{n \downarrow}}{\partial z} \Bigr ]\Biggr\rbrace,
\end{align}
\end{subequations}
\end{widetext}
The introduction of spatial inhomogeneity in magnetization texture or inversion-breaking spin-orbit coupling result in a net spin current imbalance that is finite even in the absence of a Josephson current. \cite{Halterman2015:PRB,alidoust1,alidoust2}

To compute the spin torque, 
another approach involves
using 
the continuity equation in the steady state
to determine the spin torque by
evaluating the
derivatives of the spin current as a function of position.
In the steady state, the continuity equation reads, 
${\bm \nabla}\cdot {\bm {\mathcal S}}= 
{\bm \tau}$.
Since the structure is finite in the $y$ and $z$ directions, we have
the following components of the torque vector:
\begin{align} \label{torky}
\tau_x = \frac{\partial {\cal S}_{yx}}{\partial y}+\frac{\partial {\cal S}_{z  x}}{\partial z}, \,
\tau_y = \frac{\partial {\cal S}_{y y}}{\partial y}+\frac{\partial {\cal S}_{z  y}}{\partial z}, \,
\tau_z = \frac{\partial {\cal S}_{y z}}{\partial y}+\frac{\partial {\cal S}_{z  z}}{\partial z}.
\end{align}
When the junctions are in 
static equilibrium and there
 is no charge current, 
the spin-current 
does not necessarily vanish 
because any inhomogeneous
magnetization leads to a non-zero spin torque 
that generates  a net spin current.

\section{Results and discussions} \label{results}

\begin{figure*}[t!] 
\centering
{
\includegraphics[width=0.32\textwidth,scale=0.01]{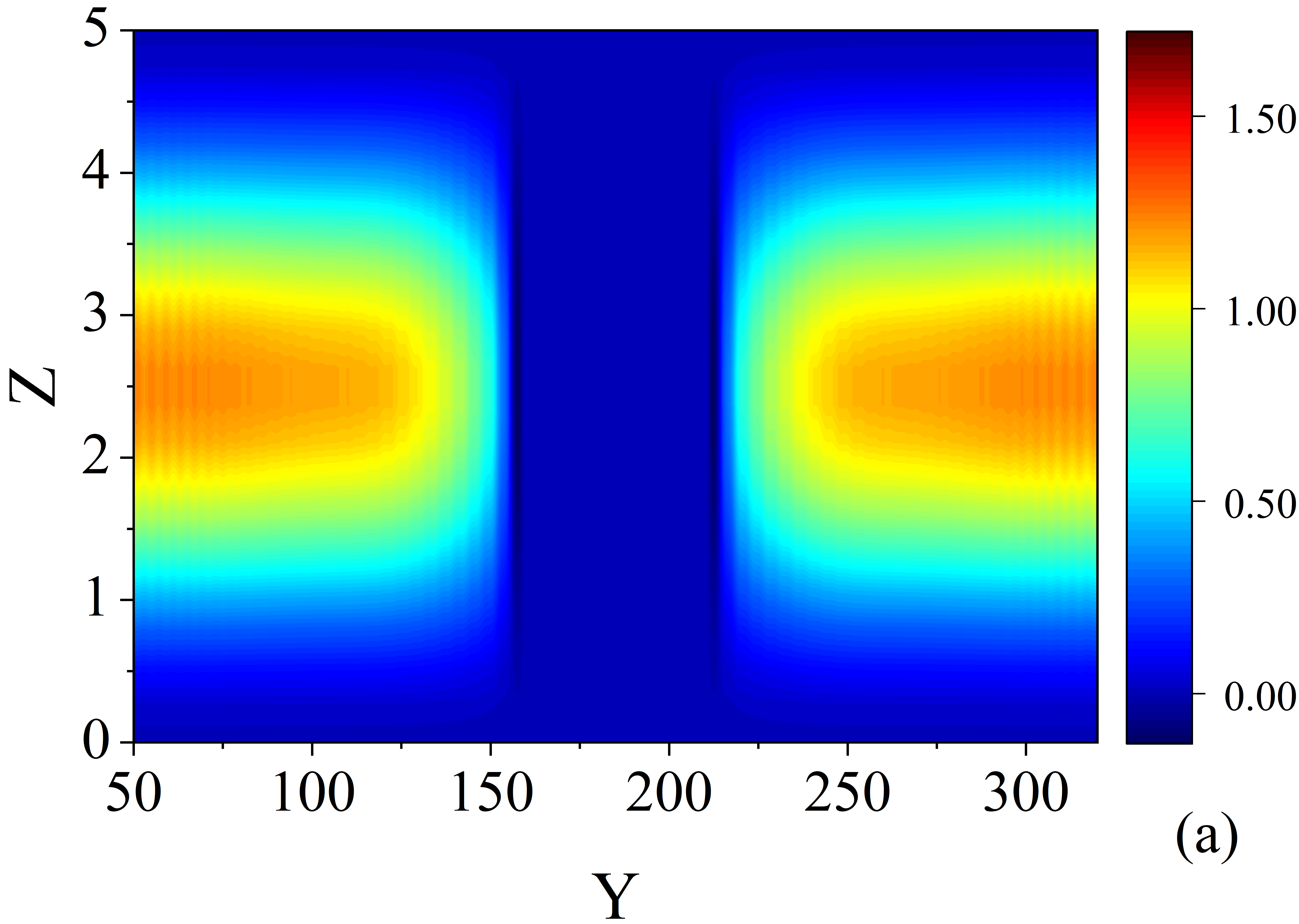} 
\includegraphics[width=0.32\textwidth,scale=0.01]{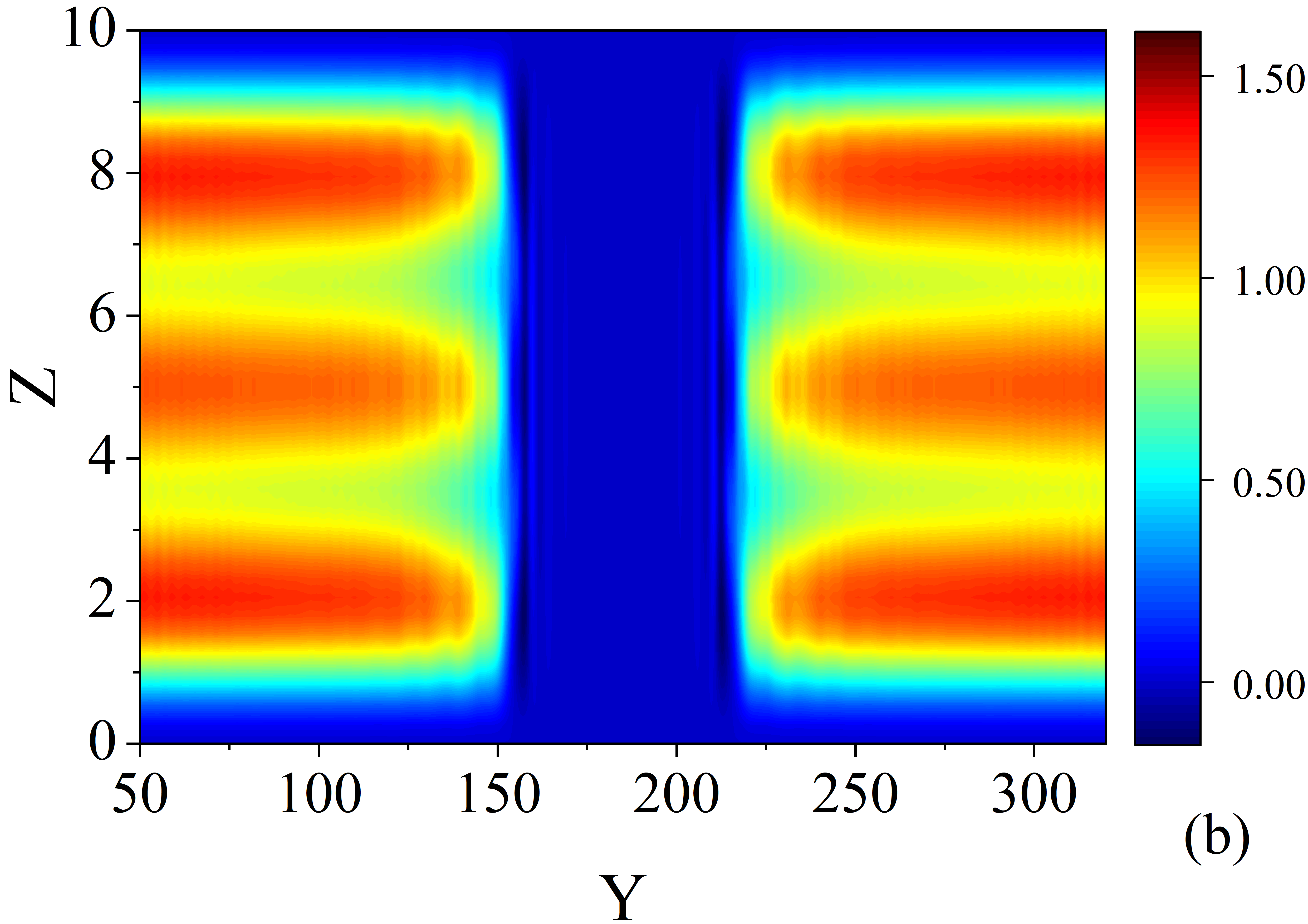} 
\includegraphics[width=0.32\textwidth,scale=0.01]{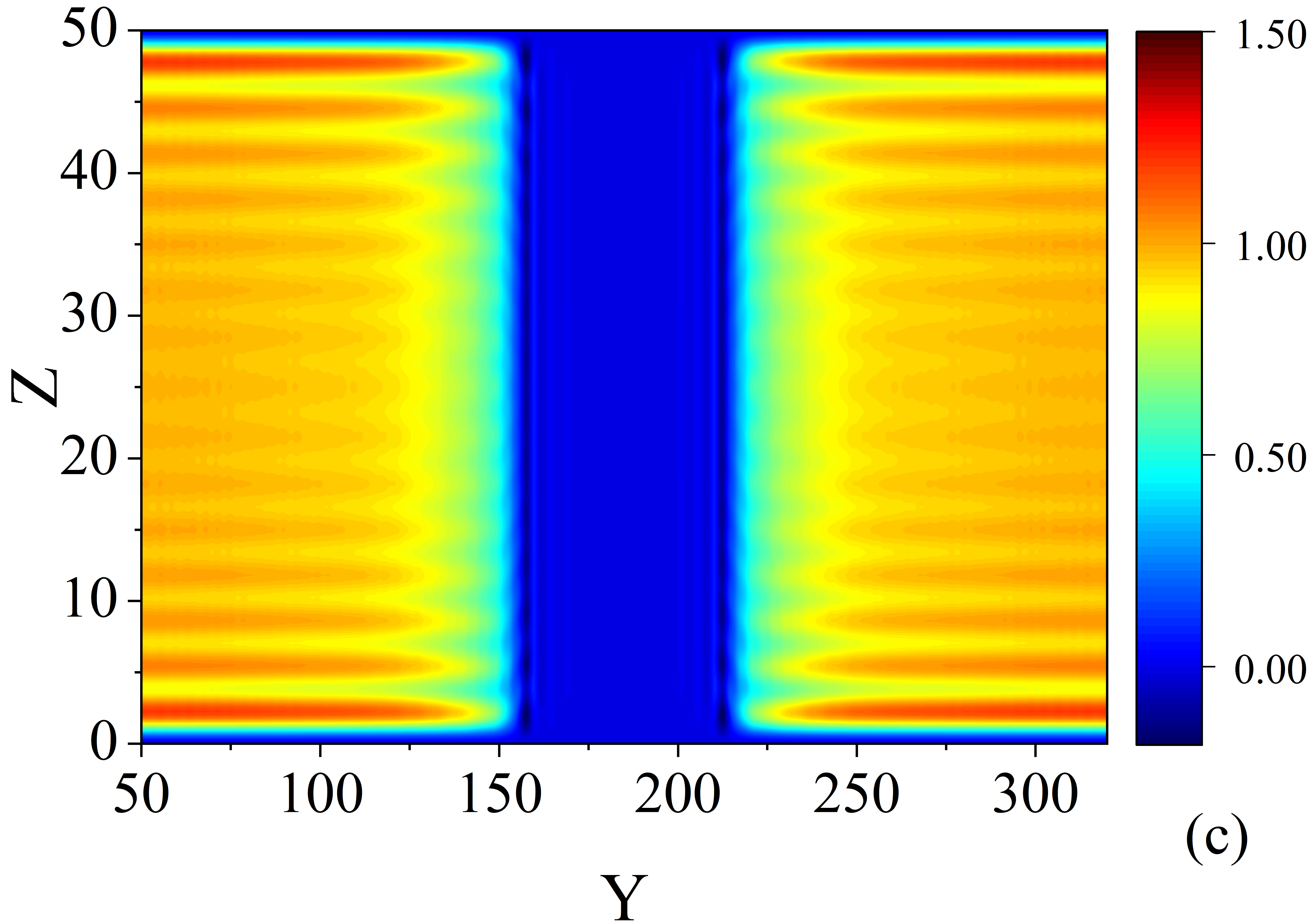} 
\includegraphics[width=0.32\textwidth,scale=0.01]{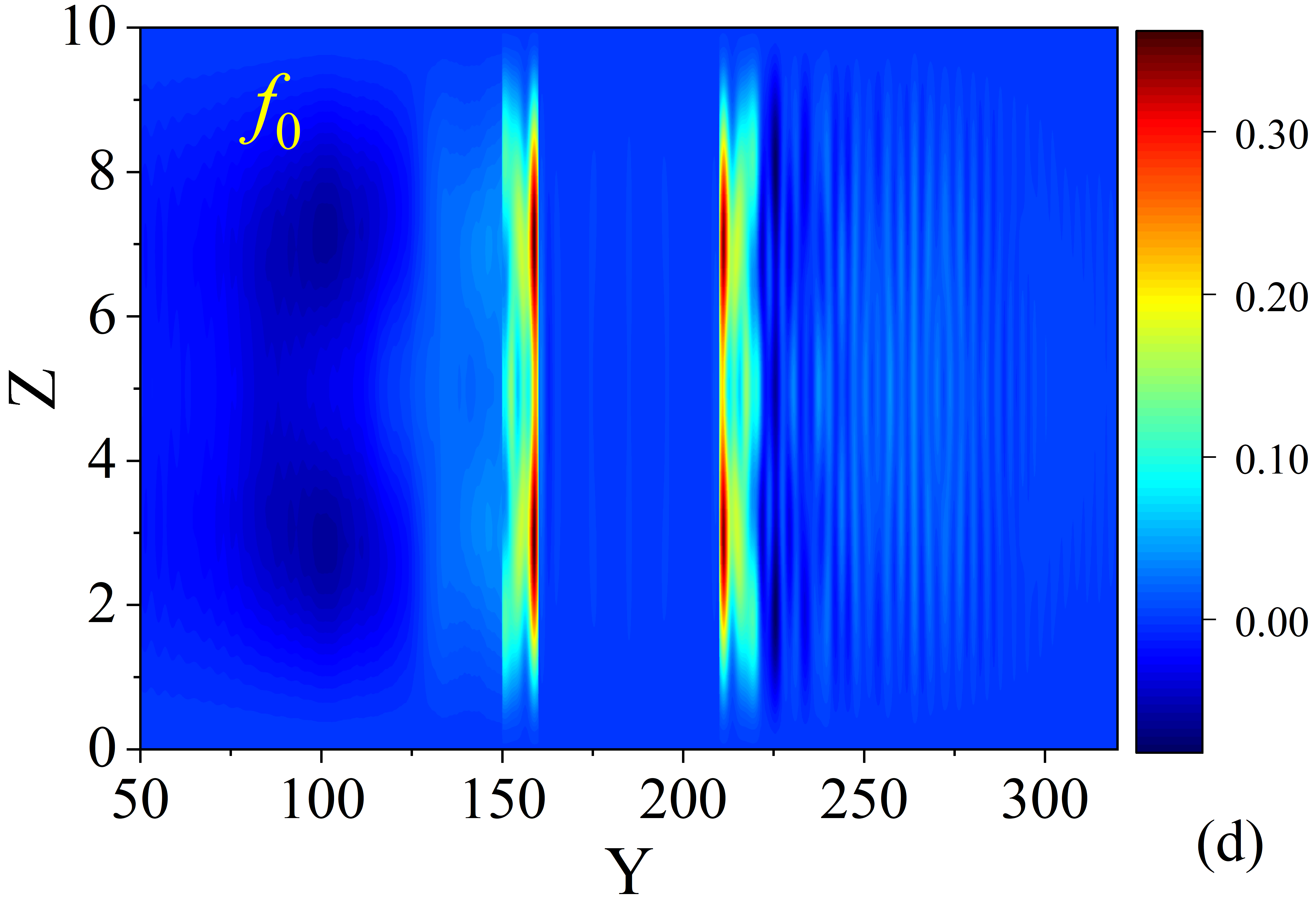} 
\includegraphics[width=0.32\textwidth,scale=0.01]{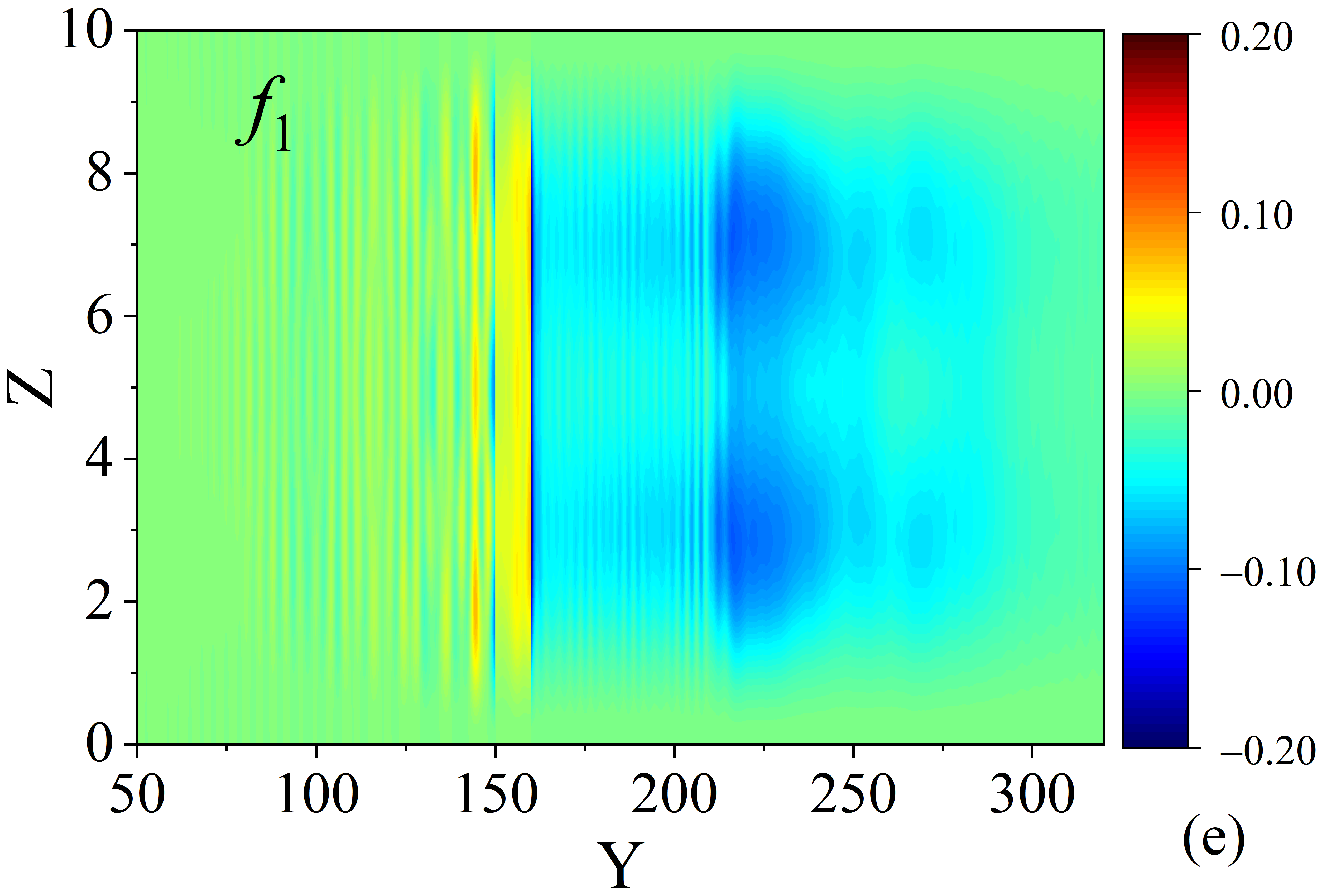} 
\includegraphics[width=0.32\textwidth,scale=0.01]{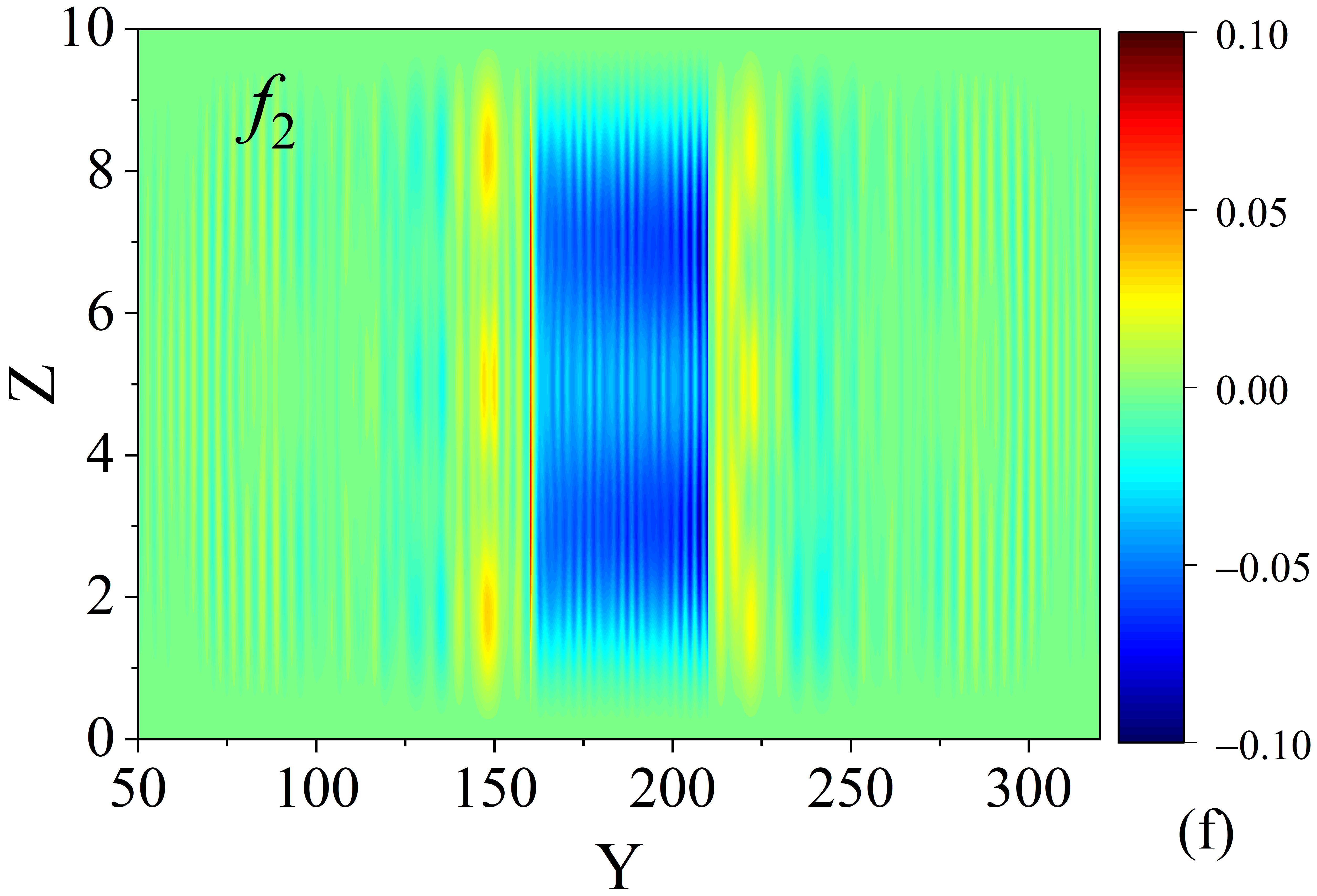} 
 }
\caption
{
Pair amplitude for a half-metallic Josephson junction
with an initial   phase difference of $\Delta\varphi =0$. 
The exchange field magnitudes are set to $h_1=h_3=0.1 E_F$,
and $h_2=E_F$. 
The exchange fields
in each ferromagnet are orthogonal to one another, i.e., 
 directed along the $x$, $y$, and $z$ directions for $\rm F_1$, $\rm F_2$, and $\rm F_3$, respectively. 
The geometrical parameters are as follows:
$D_{F1} = D_{F3} = 10$, and $D_{F2}=50$.  
Three different widths $W$ are considered, $W=5$ [(a)], $W=10$, [(b)],
and $W=50$ [(c)].
Spin triplet amplitudes $f_0$, $f_1$ and $f_2$ for the Josephson junction in panel (b).
with an initial phase difference of $\Delta\varphi =0$, and
 relatively narrow width, $W=10$.
 }
\label{2Dpa}
\end{figure*}    

The results of our systematic investigations are presented
below in terms of convenient dimensionless quantities. Our
choices are as follows: all length scales, including the 
spatial coordinates $Y = k_F y$, $Z=k_F z$,  and widths $D_{Fi} \equiv k_F d_{Fi} (i = 1,2,3)$ are
normalized by the Fermi wave-vector, $k_F$. 
For the superconducting correlation length $\xi$ we choose the value $k_F \xi_0 = 50$,
and the computational region occupied by the S electrodes
corresponds to a width of $k_Fd_S=150$ 
(see Appendices \ref{appA}, \ref{appB}, and \ref{appC} for numerical details). 
The outer ferromagnet layers must be thin to effectively generate the 
spin-triplet correlations \cite{Halterman2015:PRB,zep,K.Halterman_ss2016}.
 We found $D_{F1}=D_{F3}=10$ to be effective lengths for generating a large population of spin-triplet correlations. 
All temperatures are measured in units of $T_{c0}$,
the transition temperature of bulk S material, and we consider
the low temperature regime, $T/T_{c0} = 0.05$, throughout the paper. 
Energy scales
are normalized by the Fermi energy, $E_F$, including the Stoner
energy $h_i$   and the energy cutoff, $\omega_D$, used in 
calculating the pair amplitude, Eq.~(\ref{swave}). 
The latter is set at 0.1, with the main results relatively insensitive to
 this cutoff choice. The strength of the magnetic exchange fields, $h_{1,3}$ is taken to be
the same in both magnets: we set its dimensionless value to a
representative $h_{1,3}/E_F = 0.1$. The exchange field in the central ferromagnet layer
will typically correspond to 
a half-metal ($h_2=E_F$) but other exchange field strengths will be studied as well.
The orientation angles of the
magnetic exchange field in each of the ferromagnet regions can vary, depending
on the quantity being studied, however most cases consider the $xyz$ configuration,
whereby the exchange fields in $\rm F_1$, $\rm F_2$, and $\rm F_3$ are 
aligned on the $x$, $y$, and $z$ directions, respectively.
All results for the DOS in 
this work are presented normalized to the superconductor’s normal state DOS at the Fermi level.
The magnetization is normalized by 
$\mu_B n_e$, and
the normalization $\tau_0$
 for the spin torque follows from the normalizations for ${\bm h}_i$ and ${\bm m}$ in Eq.~(\ref{tau1}).
For the transport quantities,
we normalize  the charge currents by $J_0$, where
$J_0 = e n_e v_F$,  $v_F = k_F/m$ is the Fermi velocity, obtained through dividing Fermi wave vector $k_F$ by the mass of quasiparticles $m$, and $n_e$ is the electron
density, written as $n_e = k_F^3/(3\pi^2)$.
We focus on the supercurrent along the $y$ direction $J_y$, 
evaluated  at $Z=W/2$, the center of the junction. 
Nearly identical  results are found
if $J_y$ is spatially integrated along the width $W$.

In describing inhomogeneous superconductivity in multilayer \sfffs~Josephson structures,
it is insightful to examine 
 the spatial properties of the singlet correlations.
 Thus we present in Fig.~\ref{2Dpa}, the real parts of the pair amplitudes $f_3$
 normalized by their bulk value for three different geometrical configurations.
Three different widths of the junction are shown:
Figures~\ref{2Dpa}(a) and \ref{2Dpa}(b) correspond to relatively narrow junctions with 
 $W=5$, and  $W=10$, respectively.
 Figure~\ref{2Dpa}(c) is for a junction width of
$W=50=k_F\xi_0$.
 The pair amplitudes are determined  by
 summing the  quasiparticle
 amplitudes and energies  
[Eq.~(\ref{swave})]  calculated    
from Eq.~(\ref{Hk}).
The central ferromagnet is half metallic ($h_2/E_F=1$) with a dimensionless width of $D_{F2}=50$.
The  exchange field vectors are oriented along the $x$, $y$, and $z$ directions in $\rm F_1$, $\rm F_2$, and $\rm F_3$, respectively  (the $xyz$ configuration).
Along the $y$ direction, 
the outer superconductors occupy the regions $0< Y < 150$ and $220 < Y<  370$.
Figures~\ref{2Dpa}(a)-\ref{2Dpa}(c), the pair amplitude vanishes in the half-metal region ($160 < Y < 210$),
where only one spin channel is permitted.
For the narrow channel Fig.~\ref{2Dpa}(a), superconductivity reaches its maximal value 
along the midline of the width side ($Z=2.5$) and nearly vanishes in the vicinity of the outer boundaries ($Z=0,5$). Boundary and size-effects from quasiparticle reflections at the outer walls are mainly responsible for the observed contours.
 The number of peaks in the pair amplitude profiles  is seen to increase for the wider junctions
 shown in Figs.~\ref{2Dpa}(b) and \ref{2Dpa}(c), as 
  electron trajectories involve larger paths,
and  interference effects become diminished.

The spatial behavior of the spin triplet correlations $f_0$, $f_1$, and $f_2$ are shown in
the bottom row of  Fig.~\ref{2Dpa}.
The system parameters are the same as those used for 
the $W=10$ case shown in Fig.~\ref{2Dpa}(b). Beginning with 
the opposite-spin triplets $f_0$,  Fig.~\ref{2Dpa}(d) demonstrates 
that the regions occupied by the thin outer ferromagnets ($150<Y<160$ and $210<Y\leq 220$)
contain  a considerable enhancement of the $f_0$ triplet component that decays substantially in the other regions.
The equal-spin triplet correlations $f_1$ and $f_2$ in Figs.~\ref{2Dpa}(e,f)
exhibit behavior that is strikingly different, where they permeate the entire
half-metal ($160\leq Y \leq210$) where spin-polarized pair correlations can survive. Moreover,
the $f_1$ triplets are seen to propagate into the superconductor ($Y > 220$) at the given time.
Note that 
 long-range  transport of triplet pairs can occur in Josephson structures containing only two magnets,
 i.e., \sfffs~junctions,
 as will be seen below, trilayer  ferromagnet junctions are required
  to generate a spontaneous supercurrent with a zero phase difference $\Delta\varphi=0$.

\begin{figure}[t!]
\centering
\includegraphics[width=0.495\textwidth]{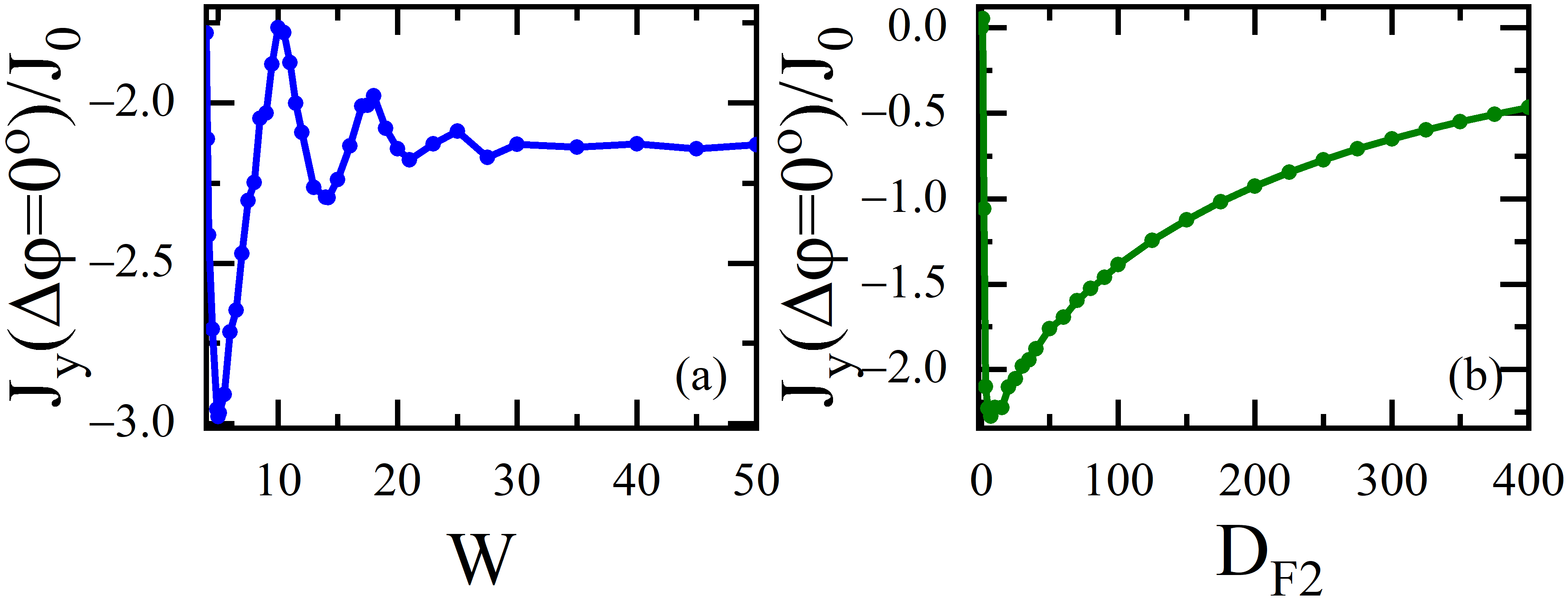}
\caption{
(a) Normalized  anomalous  current $J_y(\Delta\varphi=0)$
versus the transverse width $W$. The lengths of the ferromagnet segments are as 
follows: $D_{F1} =D_{F3} =10$, and $D_{F2}=50$.
In (b) the width $W=10$ is fixed, while the length of the central ferromagnet, $D_{F2}$,
changes. The surrounding ferromagnets $\rm F_1$ and $\rm F_3$ have the same lengths as in (a).
The junction for each scenario has a macroscopic phase difference of zero, $\Delta\varphi=0$, between the  outer
superconducting electrodes, and the exchange fields in the  ferromagnets
are all orthogonal to one another (i.e., the $xyz$ configuration)
} 
\label{W_DF2}
\end{figure}
As observed in Fig.~\ref{2Dpa}, differing junction widths lead to considerable differences in the behavior of
the singlet correlations. To examine how geometrical changes affect Cooper pair transport, 
we present in Figs.~\ref{W_DF2}(a,b) the normalized supercurrent
in the $y$ direction  as a function of the normalized width $W$ and
central ferromagnet layer thickness $D_{F2}$, respectively.
Here we fix $\Delta\varphi =0$,  and the middle ferromagnet is half-metallic, i.e., $h_2/E_F=1$.
It was found theoretically and experimentally that equal-spin triplet
pairs can result in a more robust Josephson supercurrent that has a
weak sensitivity to ferromagnet layer thicknesses due to their long ranged nature \cite{Keizer2006,C.-T.Wu2018}.
If one of the ferromagnets in the junction is
half-metallic, the equal-spin triplet correlations are expected
to play an even greater role in the behavior of the  supercurrent. 
In Fig.~\ref{W_DF2}(a) the zero-phase current is calculated 
for a wide range of
junction widths  $W$, while keeping the
lengths  fixed at $D_{F2}=50$ and $D_{F1}=D_{F3}=10$.
It is evident that 
damped oscillatory 
 quantum size-effects are discernible for $W\lesssim 30$. 
For larger $W$, the size-effects vanish, and the zero-phase current 
levels off at its bulk value.
Related to this,
oscillations over the Fermi length scale 
were observed in the critical temperatures of 
thin films, attributed to quantum size effects \cite{orr,newtc}.
Next, the normalized width is fixed to $W=10$, and the  length
of the central half-metal  is varied.
The spontaneous supercurrent vanishes when $D_{F2}=0$ since that leaves only two ferromagnets with orthogonal 
exchange fields. Increasing $D_{F2}$ leads to a rapid increase in the magnitude of $J_y(\Delta\varphi=0)$ that peaks at $D_{F2}\approx 10$. As $D_{F2}$ is increased further, the supercurrent tends towards zero slowly.
This slow decay is indicative of a spin-polarized current flowing through the junction, as
a current comprised of spin singlet pairs would decay completely within the half-metal region, supporting the strongest magnetization strength.

\begin{figure}[t!] 
\centering
{
\includegraphics[width=0.47\textwidth,scale=0.01]{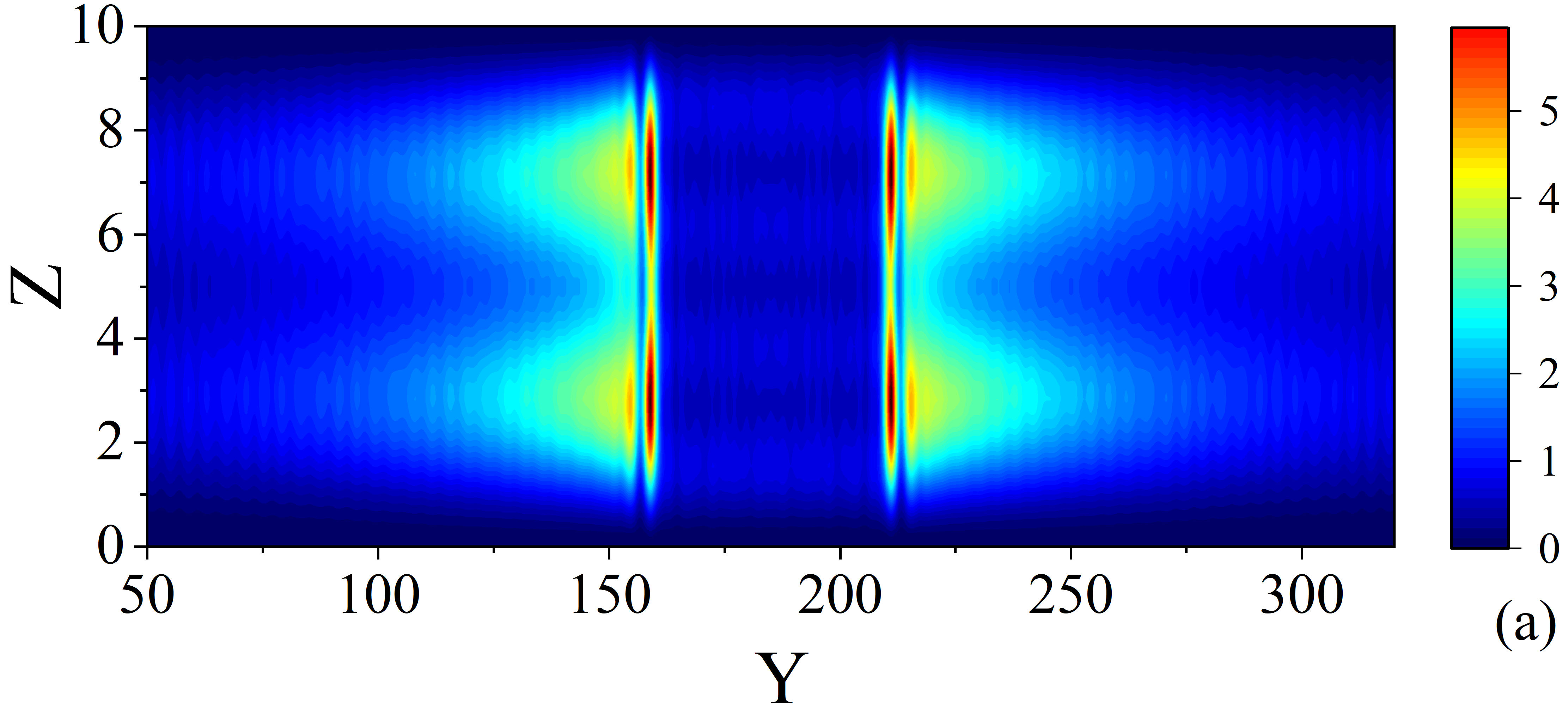} 
\includegraphics[width=0.47\textwidth,scale=0.01]{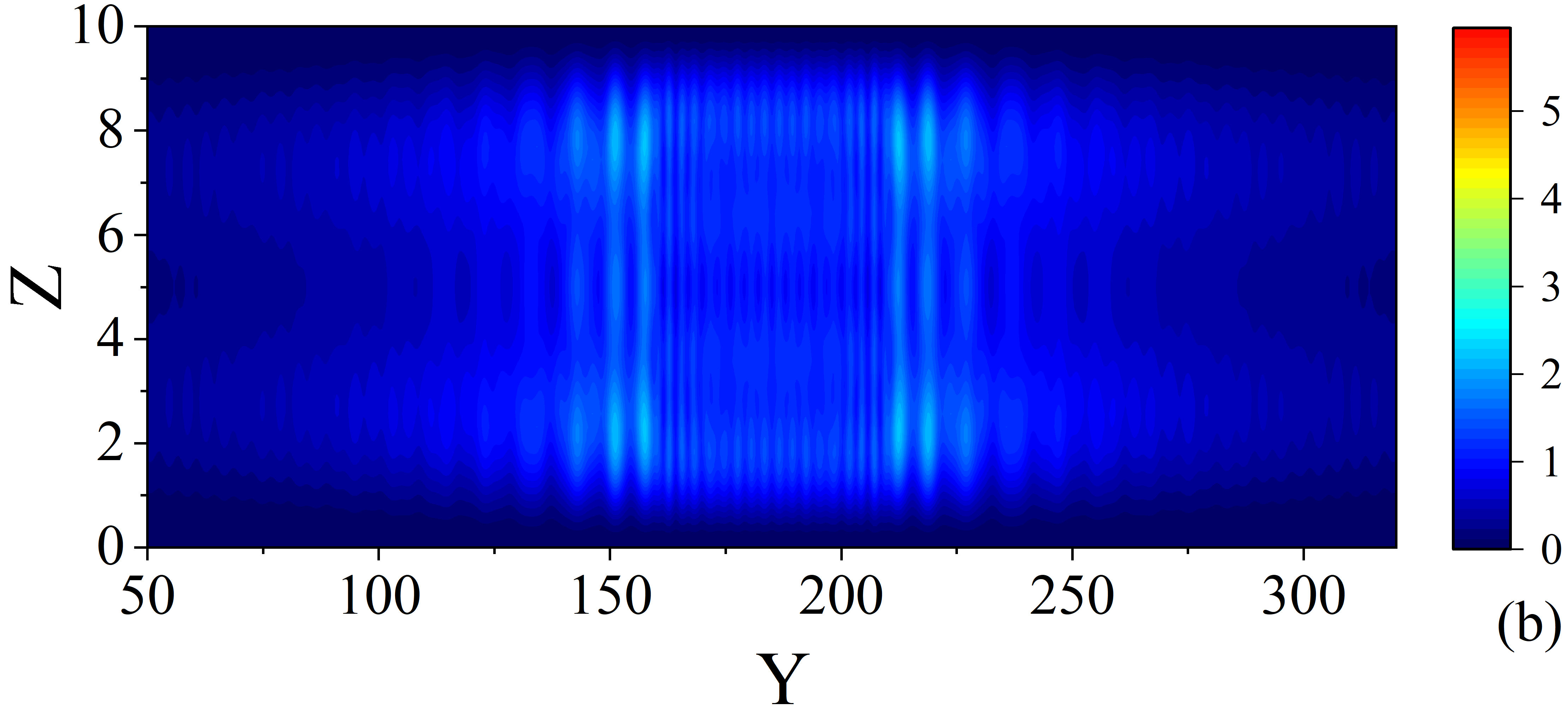} 
\includegraphics[width=0.47\textwidth,scale=0.01]{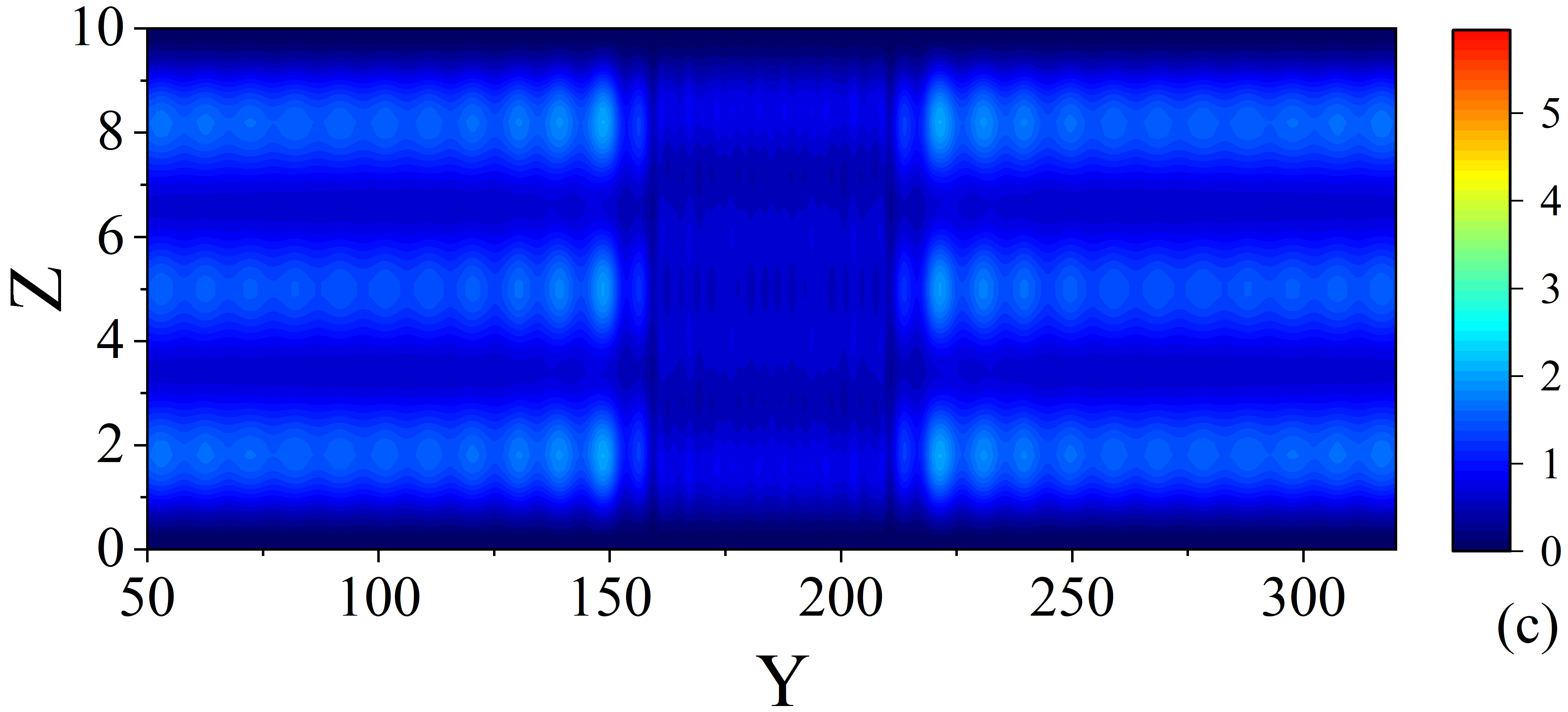} 
 }
\caption
{
Local density of states
 at three different normalized energies:
 (a) $\epsilon/\Delta_0 = 0$, (b) $\epsilon/\Delta_0= 0.5$, and (c) $\epsilon/\Delta_0=1$.
An initial phase difference of $\Delta\varphi =0$ is considered. The exchange field of the central $\rm F$ layer is set to 
 $h_2=E_F$, and the outer two magnets 
have $h_1=h_3=0.1 E_F$.
The exchange field orientations in each ferromagnet are orthogonal to one another, 
corresponding to the $x$, $y$, and $z$ directions for $\rm F_1$, $\rm F_2$, and $\rm F_3$, respectively. 
A narrow junction width is considered, with $W=10$.
 }
\label{2DdosW10}
\end{figure}   

\begin{figure*}[] 
\centering
{
\includegraphics[width=0.32\textwidth,scale=0.01]{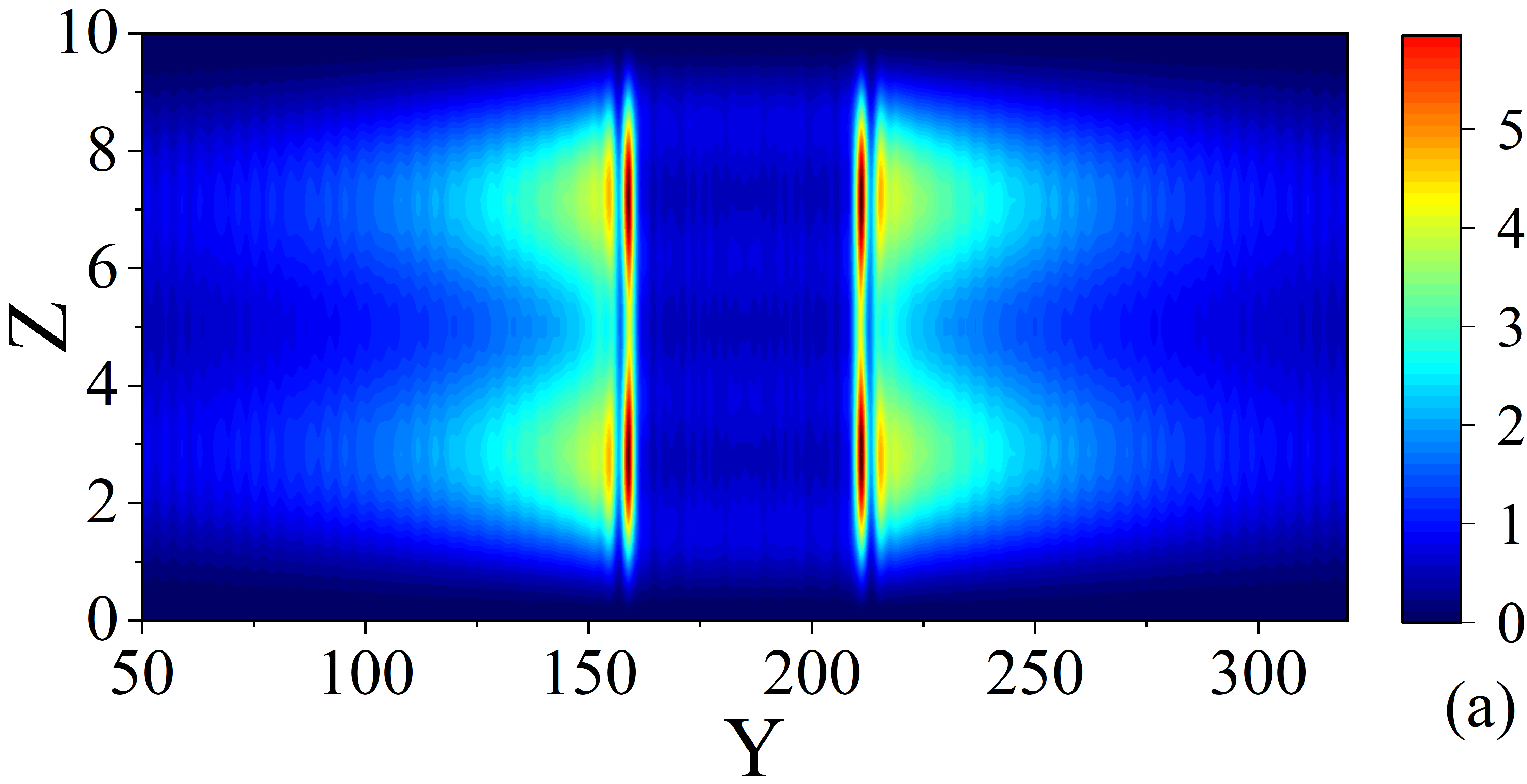} 
\includegraphics[width=0.32\textwidth,scale=0.01]{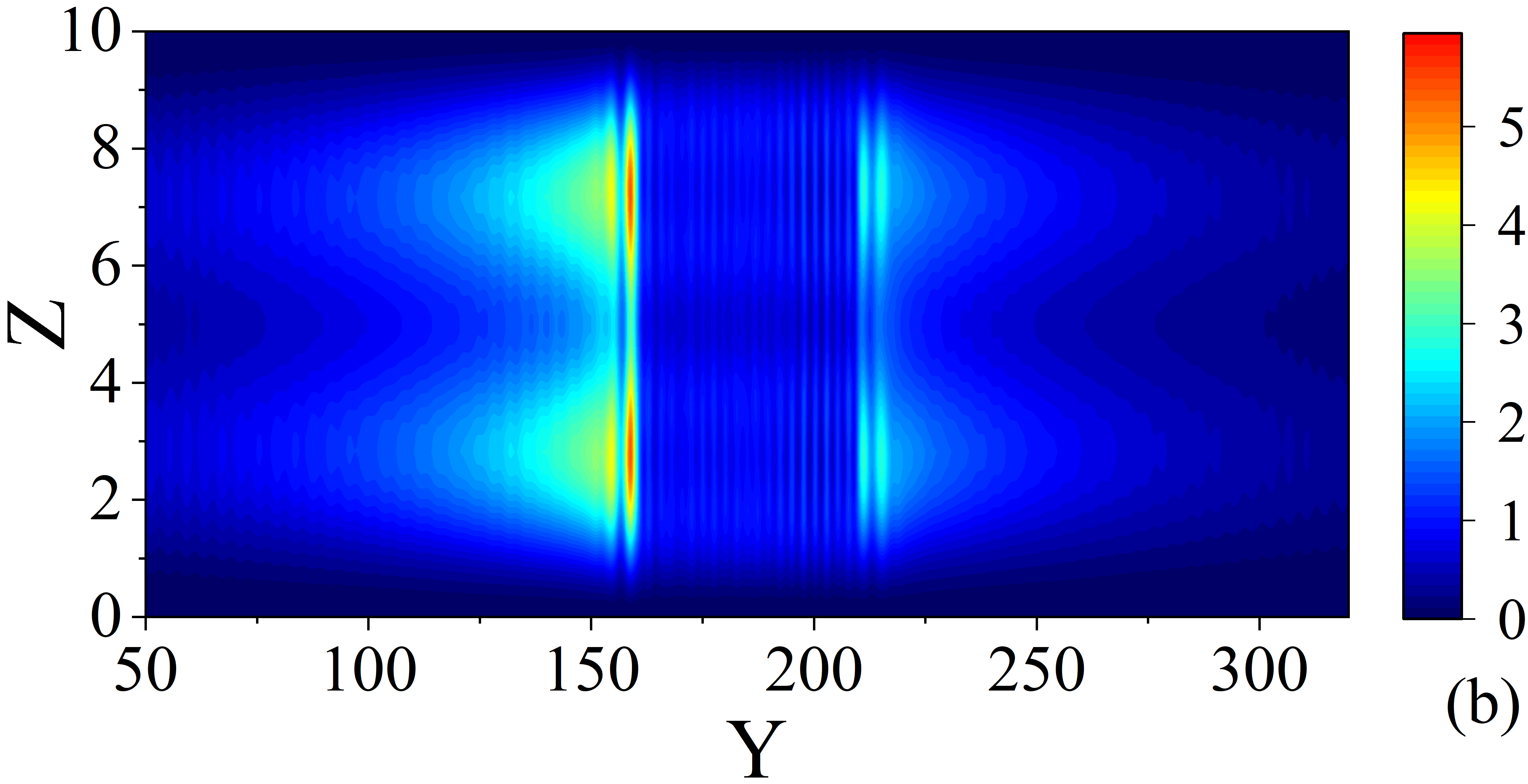} 
\includegraphics[width=0.32\textwidth,scale=0.01]{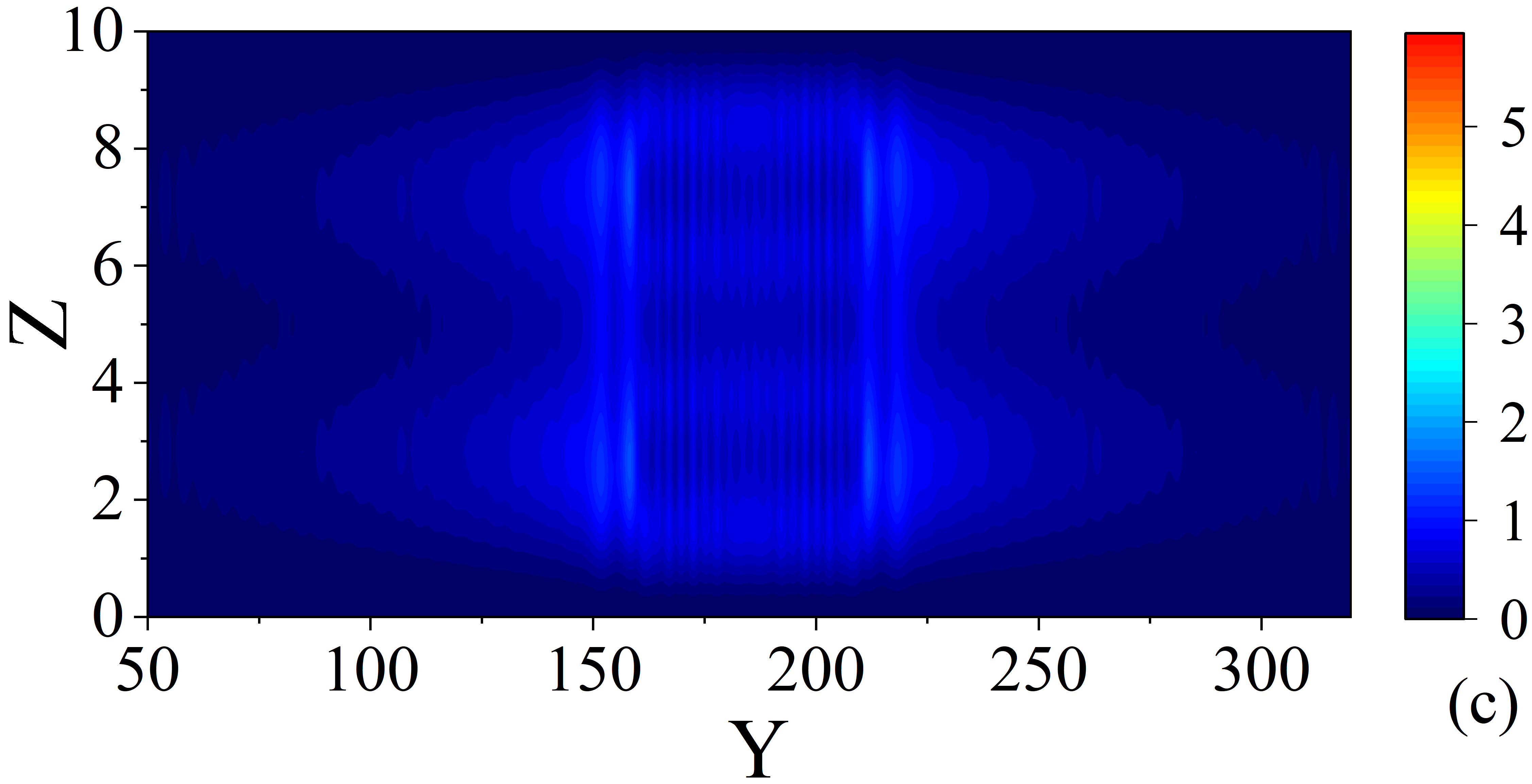} 
}
{
\includegraphics[width=0.32\textwidth,scale=0.01]{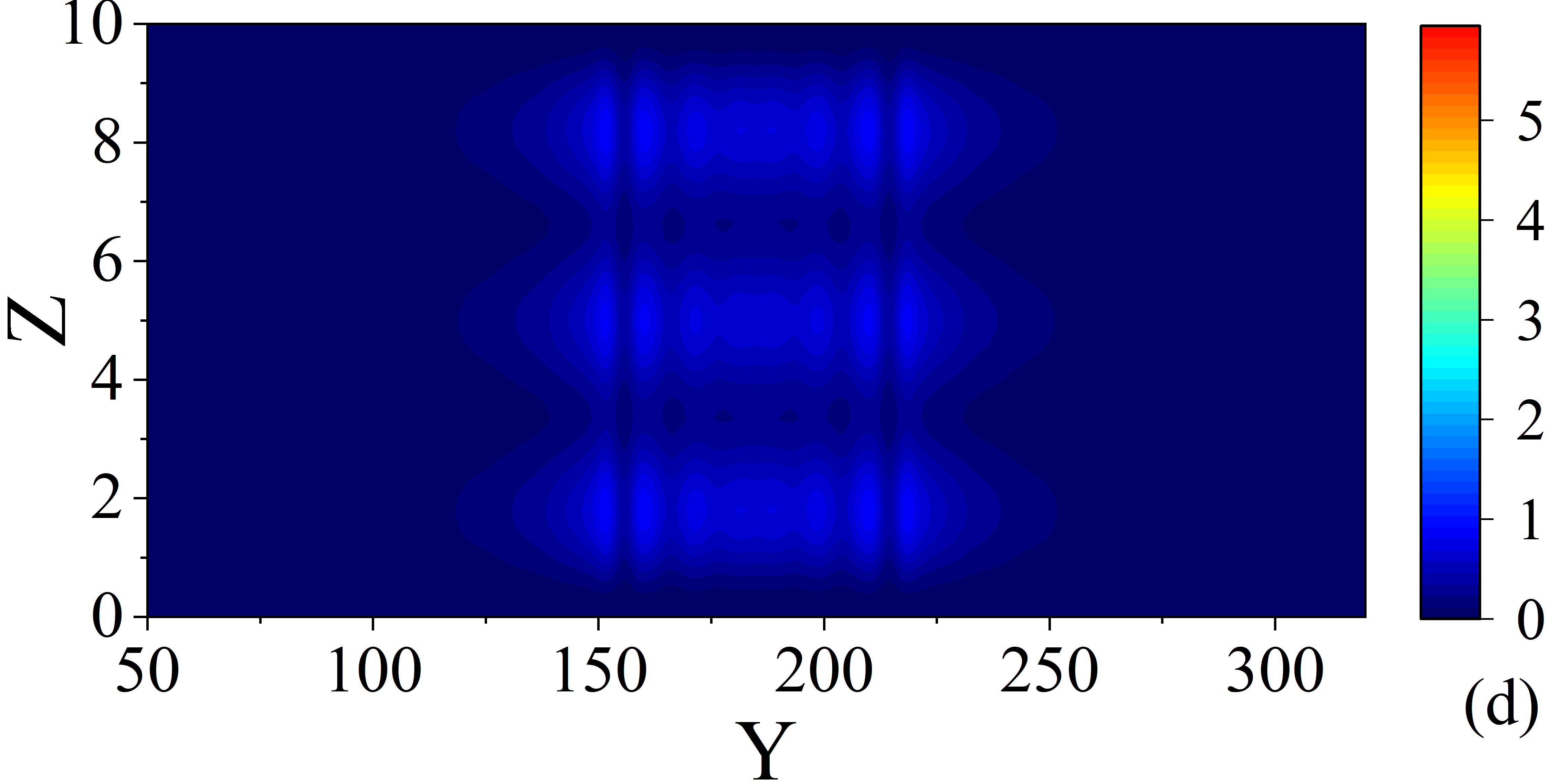} 
\includegraphics[width=0.32\textwidth,scale=0.01]{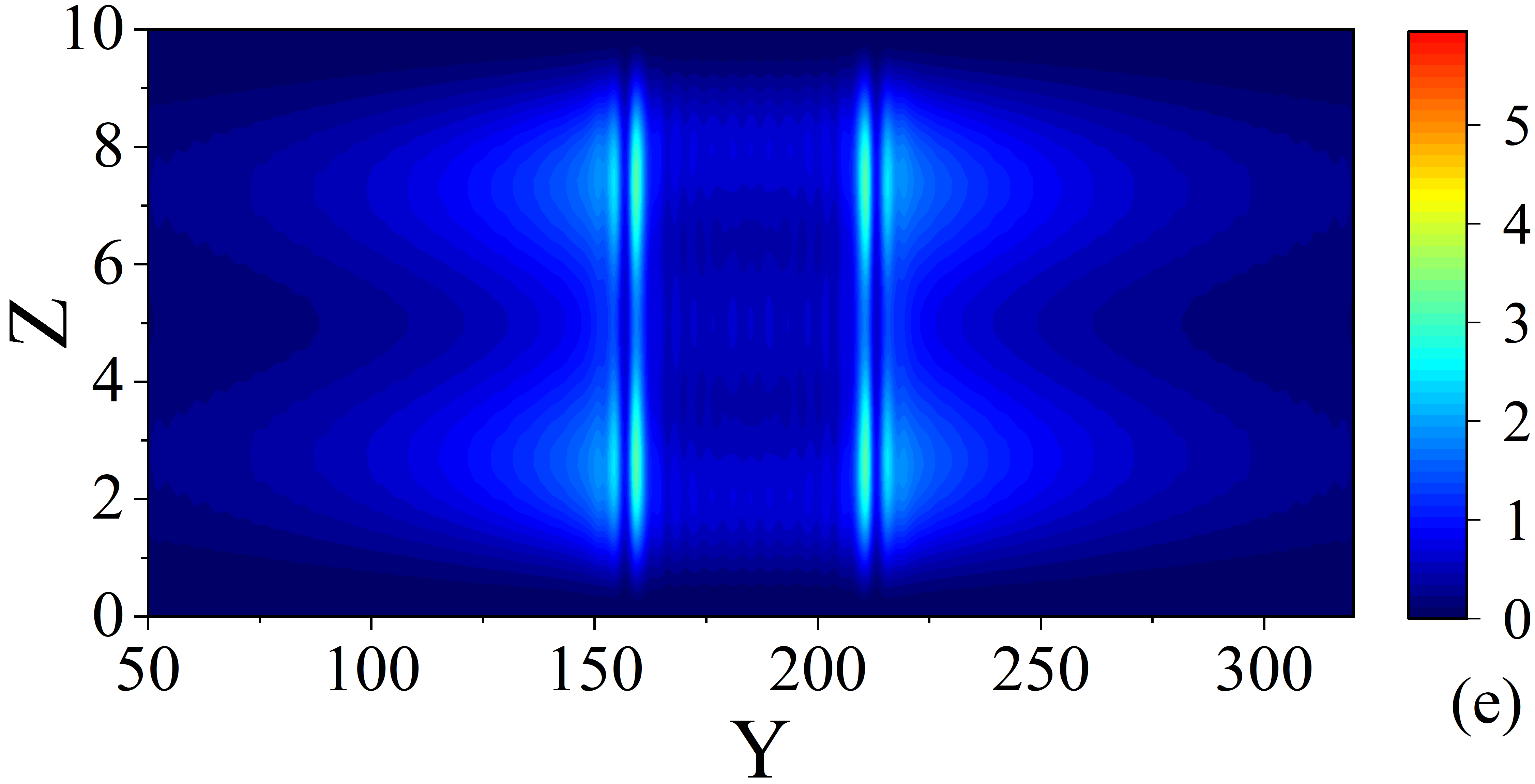} 
\includegraphics[width=0.32\textwidth,scale=0.01]{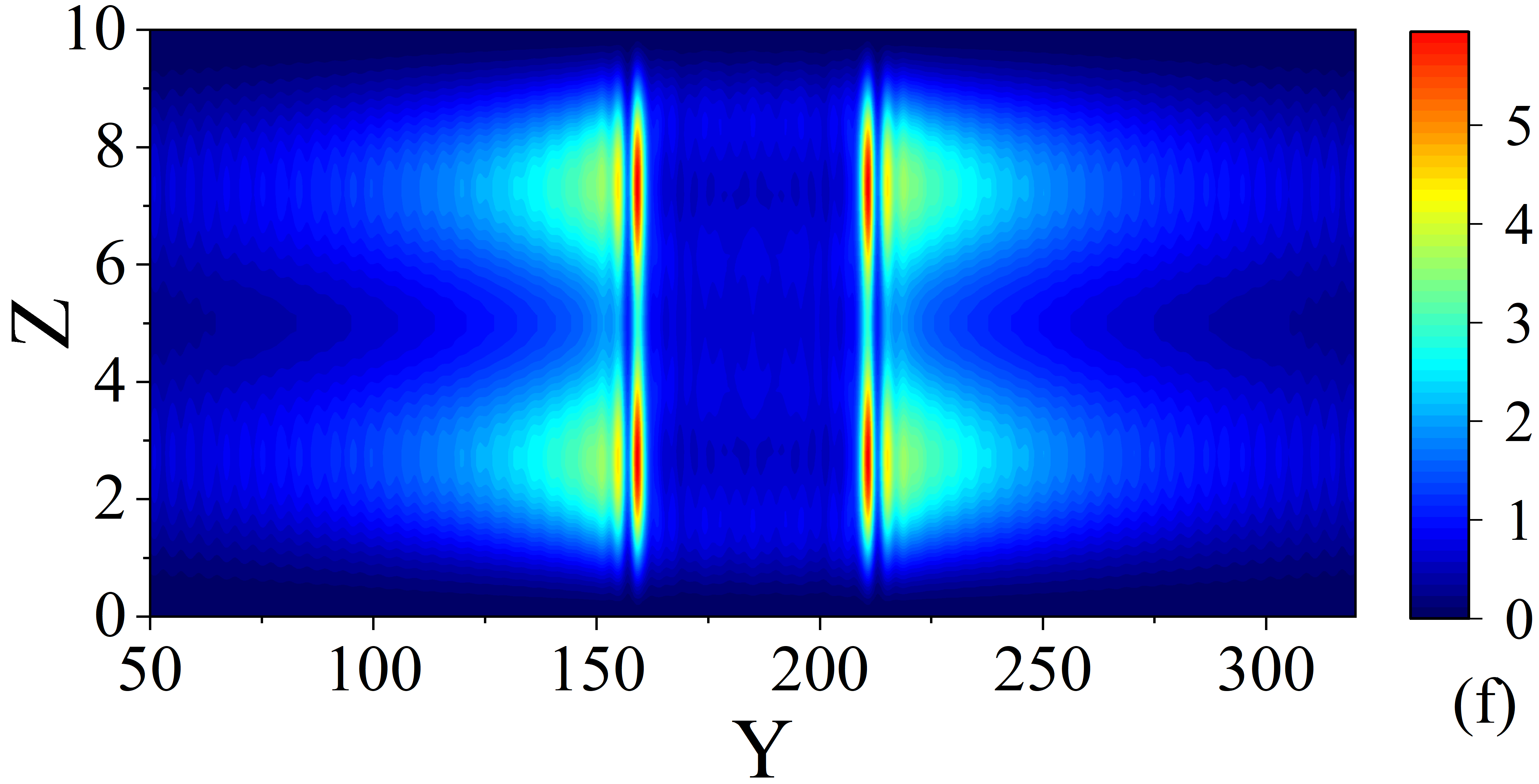} 
 }
\caption
{
Local density of states at zero energy.
Top row:
 Three different relative orientations of the exchange field in the outer ferromagnet $\rm F_3$ are considered:
  In (a),(b)
  ${\bm h}_1$ and ${\bm h}_2$ are directed along the $x$ and $y$ directions, respectively,
 while in (a) ${\bm h}_3$ is along $z$ ($\theta_3=90^\circ$,$\phi_3=0^\circ$), and (b) ${\bm h}_3$ is along $y$
($\theta_3=90^\circ$,$\phi_3=90^\circ$).
In (c), the exchange fields in all three magnets are aligned along the $y$ direction.
An  initial phase difference of $\Delta\varphi =0$ is assumed. The magnitude of
the exchange field of the central $\rm F$ layer is set to 
 $h_2=E_F$, and the outer two magnets 
have  $h_1=h_3=0.1 E_F$.
A narrow junction  width is considered, with $W=10$.
Bottom row: Three different  exchange fields in the central ferromagnet $\rm F_2$ are considered:
(d) $h=0$, (e) $h=0.5$, and (f) $h=0.7$.
A
zero phase difference is considered, i.e.,
 $\Delta\varphi =0$. The magnitude of
the exchange field in the outer two magnets 
have  $h_1=h_3=0.1 E_F$.
A narrow junction width is considered, with $W=10$.
 }
\label{dosW10_phi3vary}
\end{figure*}    
 
The electronic density of states (DOS) provides another avenue for detecting 
signatures of single-particle localized Andreev states. 
The study of single-particle excitations in these systems can
reveal indirect signatures of the proximity induced spin singlet
and spin triplet pair correlations.
Experimentally, this involves tunneling spectroscopy experiments
which can probe the local proximity-induced DOS.
From a theoretical perspective, it is revealing 
to study the DOS spatially throughout the entire junction at certain key energies.
Thus in Figs.~\ref{2DdosW10}(a-c) the local DOS, normalized by the normal state DOS, $N_0$, 
is shown at three normalized 
energies. For $\epsilon/\Delta_0=0$ in Fig.~\ref{2DdosW10}(a) there is a considerable enhancement 
in the DOS in the vicinity of the interfaces separating the superconductors from 
the ferromagnetic junction region. In particular the DOS peaks within the weaker outer ferromagnets
$\rm F_1$ and $\rm F_3$, and decays abruptly within the half-metallic region. There is a much slower decay within the superconductor regions. This profile is indicative of the generation
of equal-spin triplet correlations.
In Fig.~\ref{2DdosW10}(c) for $\epsilon/\Delta_0=1$, there are BCS-like peaks that exhibit quantum interference patterns
due to the finite size of the junction. These peaks remain regular throughout the outer superconducting banks,
and the DOS is much smaller in the ferromagnetic regions, where it is weak but nonzero as there is no 
energy gap for the parameters used here.

Having established the existence of strong zero energy peaks for the half-metallic junction
with each of the exchange fields of the three magnets orthogonal to one another,
we now examine the local DOS at zero energy 
for other orientation angles and magnitudes  of the exchange field.
In the top row of Fig.~\ref{dosW10_phi3vary}, the normalized exchange energy is set at $h_2/E_F=1$,
while the angle $\phi_3$ is $\phi_3=0^\circ$ in Fig.~\ref{dosW10_phi3vary}(a) and $\phi_3=90^\circ$ in Fig.~\ref{dosW10_phi3vary}(b).
In Fig.~\ref{dosW10_phi3vary}(c), each ferromagnet has its exchange field aligned along the $y$ direction.
Clearly the most dominant zero energy peaks occur in Fig.~\ref{dosW10_phi3vary}(a)
within the thin outer ferromagnets ($150<Y<160$ and $210<Y<220$) where
the magnets are all orthogonal. In Fig.~\ref{dosW10_phi3vary}(b), the rightmost two magnets $\rm F_1$ and $\rm F_2$ (see Fig.~\ref{diagram})
have their exchange field along the $y$ direction. This reduces the magnetic inhomogeneity, leading to an
asymmetric profile for the quasiparticle excitations that is dominant on the left side.
Finally in Fig.~\ref{dosW10_phi3vary}(c), when there exists 
the possibility to describe the system by a single spin-quantization axis,
the spin-polarized triplet correlations vanish~\cite{trip2}, and there is
a strongly diminished DOS at zero energy.
For the bottom row of Fig.~\ref{dosW10_phi3vary},
the ferromagnets are all fixed in the $xyz$ configuration, while the central
exchange field strength is varied according to $h_2/E_F=0,0.5,0.7$ in Figs.~\ref{dosW10_phi3vary}(e-f), respectively. 
For the unpolarized normal
 metal case in Fig.~\ref{dosW10_phi3vary}(d), there is no
zero energy peaks within the junction, but instead a series of weak, repeating bands 
that arise from proximity effects and a superposition of quasiparticle states confined by the outer superconductor 
banks.
Increasing the exchange field to $h_2/E_F=0.5$ in Fig.~\ref{dosW10_phi3vary}(e), there is a slight development of zero energy states
along the boundaries that becomes amplified for $h_2/E_F=0.7$ [Fig.~\ref{dosW10_phi3vary}(f)].

\begin{figure}[t] 
\centering
{
\includegraphics[width=0.48\textwidth]{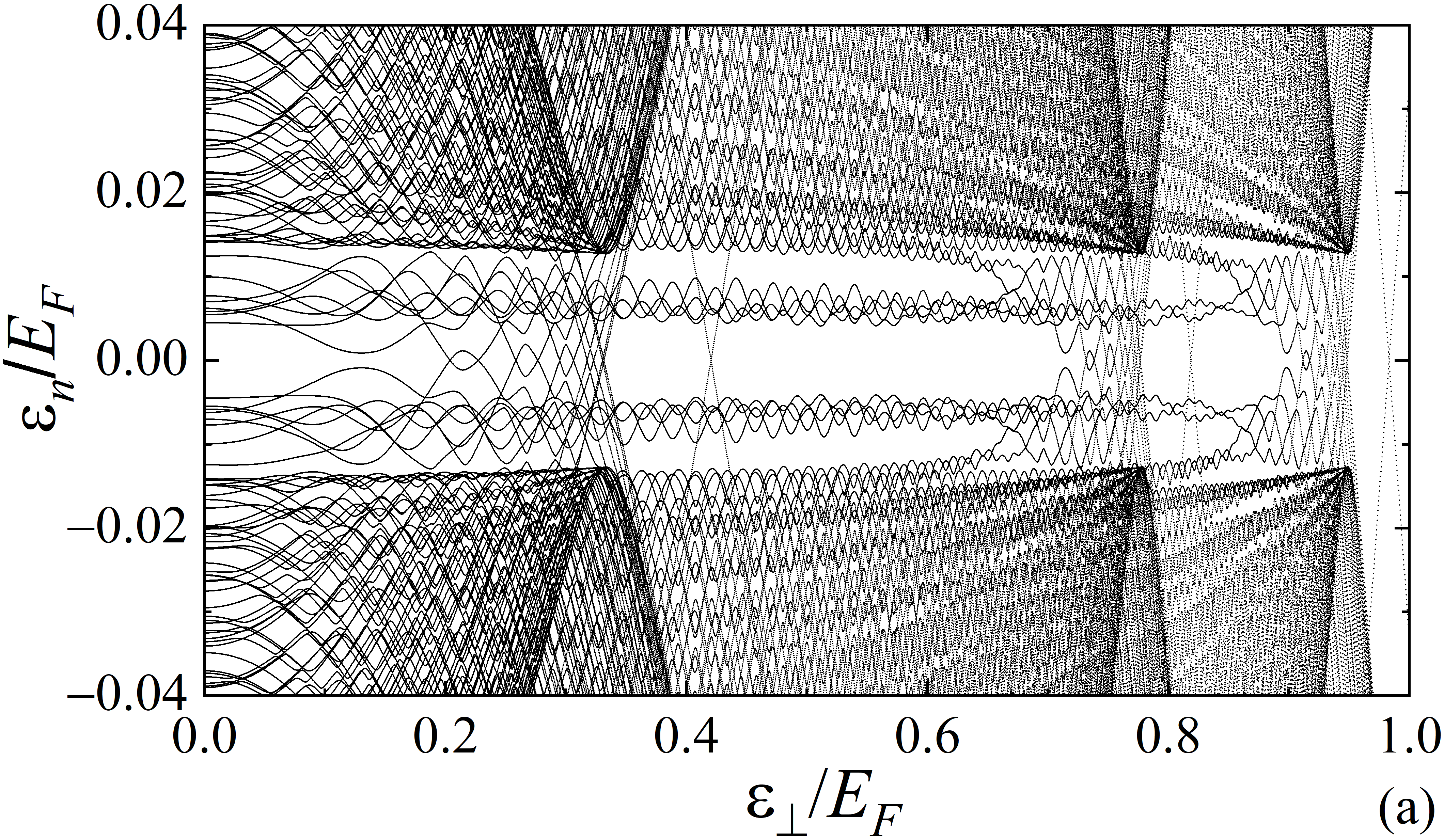} 
\includegraphics[width=0.48\textwidth]{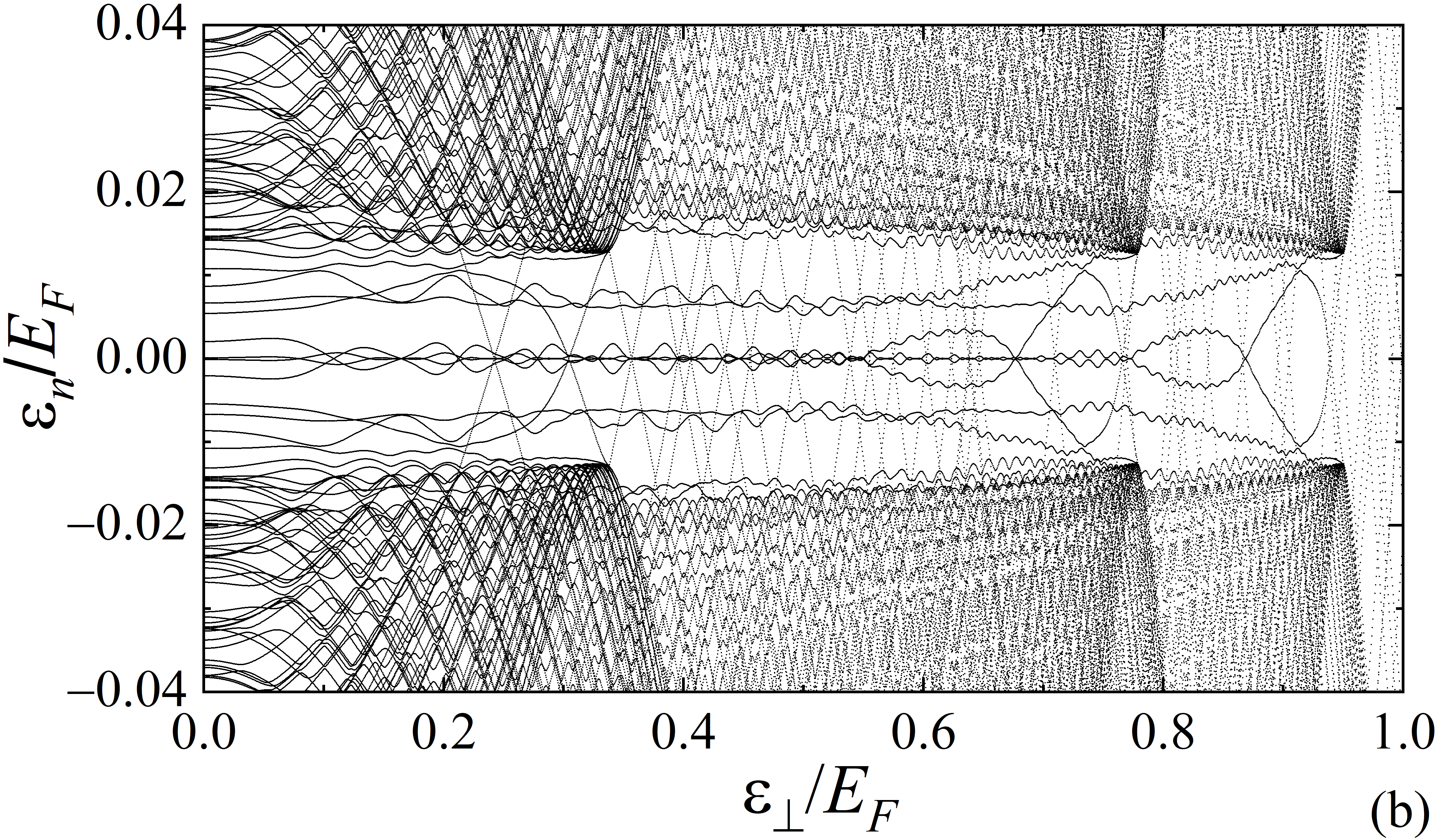} 
 }
\caption
{
Energy spectra as a function of the transverse single particle energy 
$\varepsilon_\perp\equiv k_x^2/(2m)$.
Two type of materials are considered
for the central $\rm F$ region:(a)
 a nonmangetic normal metal with $h/E_F=0$, and (b) a fully spin-polarized
 ferromagnet, $h/E_F=1$.
The  macroscopic phase difference is fixed at $\Delta\varphi =0$. The magnitude of
the exchange field in the outer two magnets 
have $h_{1,3}/E_F=0.1$.
A narrow junction width is considered, with $W=10$.
 }
\label{spectra_hvary}
\end{figure}    
Turning now to the electronic spectra, 
we present in Fig.~\ref{spectra_hvary} the quasiparticle energies $\varepsilon_n$, normalized by $E_F$, 
versus the normalized transverse energy $\varepsilon_\perp/E_F$.
We consider two extreme values of the exchange field in the central junction region:
 Fig.~\ref{spectra_hvary}(a) corresponds to a normal metal with $h_2/E_F=0$, 
and Fig.~\ref{spectra_hvary}(b) contains the spectra for the half-metal case $h_2/E_F=1$.
For a homogeneous bulk superconductor,
a continuum of states occupy energies that fall outside of the bulk  gap $\Delta_0$.
 For our system that includes multiple junction layers 
 and quantum size-effects
  there will be substantial modifications to the  scattering states 
  and an emergence of discrete bound states in 
the  excitation spectrum that approximately fall within 
the scaled gap $\Delta_0/E_F=2/(\pi k_F \xi_0)$,  corresponding to $\Delta_0/E_F \approx 0.0127$.
Indeed as Fig.~\ref{spectra_hvary}(a) shows, the eigenenergies $\varepsilon_n$ exhibit an intricate
oscillatory behavior within the bulk gap region, although with limited quasiparticle 
excitations with energies near zero.
This is in stark contrast to the half-metallic case in Fig.~\ref{spectra_hvary}(b), where 
there are a considerable number of states that reside at zero energy.
These results are consistent with the zero-energy peaks in the DOS profiles seen in Figs.~\ref{dosW10_phi3vary}(a,d).

\begin{figure}[t!]
\centering
\includegraphics[width=1\columnwidth]{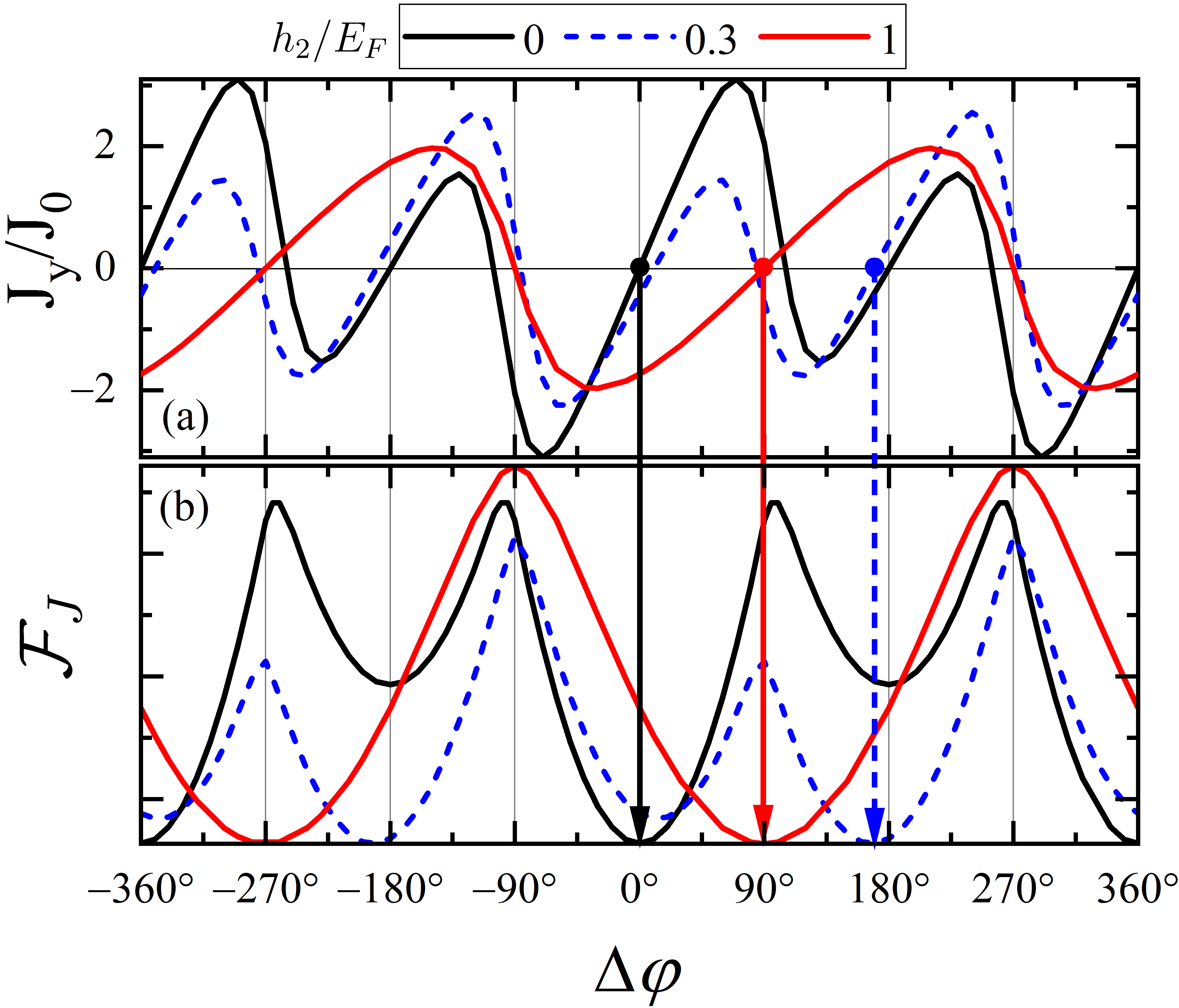}
\caption{
In (a) the current phase relations for different exchange fields of the central ferromagnet layer ${\rm F}_2$
are shown.
A range of  normalized exchange fields are 
considered (see legend), ranging from nonmagnetic ($h_2/E_F=0$) to
half-metallic ($h_2/E_F=1$). 
The exchange fields for the outer two magnets 
are  fixed at  $h_1=h_3=0.1 E_F$ and the $xyz$ configuration is considered.
A narrow junction width is considered, with $W=10$.
In (b) the free energy ${\cal F}_J$ is shown
in arbitrary units over the same $\Delta\varphi$ in (a). The minima of each curve are
shifted to the same level for clarity.
The arrows identify the phase differences that correspond to
the ground states of the systems in (a), with only the ground states for positive 
$\Delta \varphi$ shown.
} 
\label{Jy_vs_phase}
\end{figure}

Based on the findings of the DOS and energy spectra, it is clear
that
the exchange field in the central ferromagnet plays a crucial role in the 
formation of zero-energy modes and the existence of a spontaneous supercurrent.
From 
the quasiparticle energy diagrams discussed above [Fig.~\ref{spectra_hvary}],
 both the bound state and continuum state
spectra shown
 contribute to the total supercurrent flow.
We show in Fig.~\ref{Jy_vs_phase}(a) how the supercurrent flow is modified by changes to the 
magnetic strength in the $\rm F_2$ region by presenting the current-phase relations
for three representative values of $h_2/E_F$ (as labeled).
It is evident that in going from $h_2/E_F=0$ to $h_2/E_F=1$, 
new harmonics emerge, and
the
$2\pi$ periodic CPR changes considerably,
 due in part to
 the proximity-induced long-ranged
 spin-polarized triplet correlations.
In Fig.~\ref{Jy_vs_phase}(a), the zero-phase state for $h_2/E_F=0$,
 exhibits no
zero-phase supercurrent, 
since in this case there are only two orthogonal magnetizations present.
Increasing the normalized exchange field up to 0.3 results in the emergence
and gradual increase in the supercurrent when
$\Delta\varphi=0$.
 In Fig.~\ref{Jy_vs_phase}(b), stronger magnets are considered, with scaled values ranging from $0.4-1$.
 The largest jump in the anomalous current occurs when $h_2/E_F$ goes 
 from $0.4$ to $0.6$, with subsequent increases giving smaller supercurrent increments.
 In particular, when $h_2/E_F$ increases to the half-metallic state from $h_2/E_F=0.8$,
 the anomalous current saturates, and further increases effect little change.
 The same trends are observed for the maximal supercurrent flow,
 not just the zero-phase current.
 Thus,
 the spontaneous supercurrent ramps up
 as the magnetization strength of middle ferromagnet
increases.

When analyzing the CPR of
finite-sized Josephson junctions, and 
 determining the ground state of the system, it is useful to study the
free energy, ${\cal F}$, which is comprised of two terms: ${\cal F}\equiv  {\cal F}_J + {\cal F}_\Delta$.
Here we define  ${\cal F}_\Delta=\int dy dz |\Delta(y,z)|^2/g$, and,
\begin{align}
{\cal F}_J = -2T\sum_n \ln \left[2 \cosh\left(\frac{\epsilon_n}{2T}\right)\right].
\end{align}
For the cases considered in this paper, where
$\partial {\cal F}_\Delta/\partial \Delta\varphi$ is small,
we found that
the contribution from ${\cal F}_J$ 
is sufficient to 
determine the ground state of the system.
In  Fig.~\ref{Jy_vs_phase}(b), we therefore
present ${\cal F}_J$ as a function of $\Delta\varphi$
for the same system parameters used in Fig.~\ref{Jy_vs_phase}(a).
The vertical arrows identify the ground states of each of the three CPRs,
which  always occur at zero current.
It is important to note  that the ground state phase difference $\varphi_0$
can vary from 0 to approximately $180^\circ$ by changing the strength of
the  ferromagnet in 
the central layer. In particular,  by increasing $h_2/E_F$, the ground state shifts to higher phase differences.
The next notable feature is that with the emergence of a $\varphi_0$ state,
the CPR in Fig.~\ref{Jy_vs_phase}(a) for $h_2/E_F=0.3$ is seen to exhibit a clear  diode effect, whereby the 
 supercurrent exhibits the following property:
$|\sin(+\Delta\varphi + \varphi_0)| \neq |\sin(-\Delta\varphi + \varphi_0)|$.
Thus, by appropriately  tuning the  phase difference between the superconductor banks,
the dissipationless current flow can be made to have a one-way flow.
This has practical applications as a nanoscale spintronics device with low energy consumption.

  \begin{figure}[t!]
\centering
\includegraphics[width=0.49\textwidth]{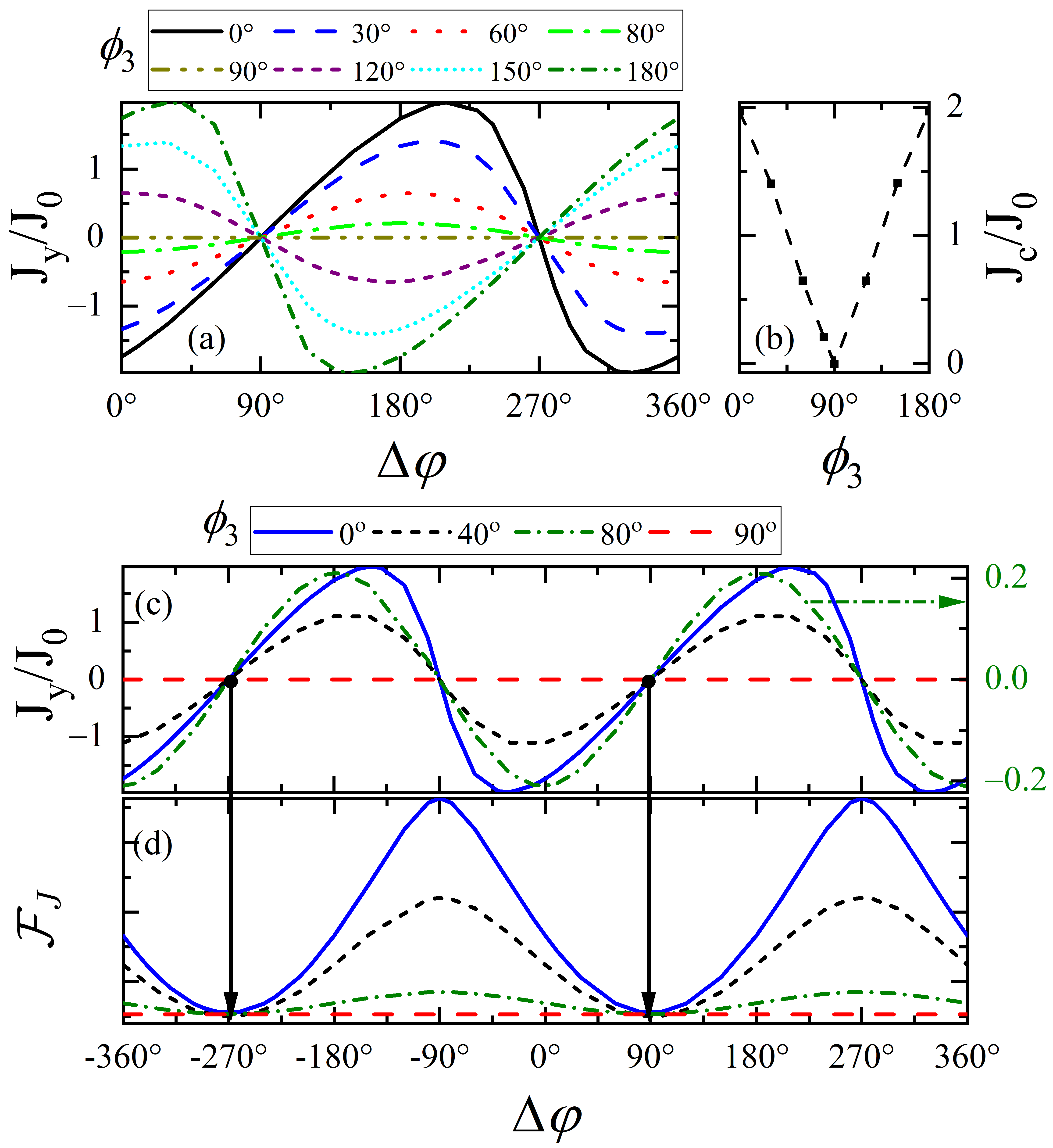}
\caption{
Top row:
(a) Current phase relations for several 
magnetization angles $\phi_3$ (shown in the   legend).
(b) The corresponding critical current as a function of $\phi_3$
for the system  in panel (a). 
In (c) an expanded view of the current phase relations is shown. 
A different scale
 for the $\phi_3=80^\circ$ case (labeled on the right vertical axis) is used
to illustrate how the CPR evolves to a simple sine curve. 
The bottom panel (d) shows the free energy ${\cal F}_J$, with the vertical
arrows identifying the ground state of the system.
In all cases a half-metallic junction is considered with $h_2/E_F =1$,
and the normalized exchange field in the outer two magnets 
is fixed at  $h_1/E_F=h_3/E_F=0.1$, with the  magnetizations along $x$ and $z$ directions, respectively.
Results shown are for a narrow  junction width of $W=10$.
} 
\label{j_vs_h}
\end{figure}

Next, the effects of rotating the magnetization on the supercurrent are investigated, and we
 address the effectiveness of exchange-field rotations in controlling charge flow,
 including on/off switching of the supercurrent.
 Controlling the magnetization rotation
can be achieved 
experimentally via, e.g., the application of external magnetic fields \cite{jara}. 
 The orientation of the exchange field vector ${\bm h}_3$ is described by the two angles
 $\theta_3$ and $\phi_3$ [Eq.~(\ref{hex})]. Here we fix $\theta_3$ so that $\theta_3=90^\circ$, 
 and 
variations in $\phi_3$ occur solely in the $y$-$z$ plane.
In Fig.~\ref{j_vs_h}(a), the CPR is shown for several angles $\phi_3$ in the outer magnet (see legend).
As $\phi_3$ varies from $\phi_3=0^\circ$
to $\phi_3=180^\circ$,
 the  vector ${\bm h}_3$ goes 
 from the $z$ direction to the  $-z$ direction.
The exchange field directions in the first and second ferromagnets remain fixed along
the $x$ and $y$ directions, respectively.
Here the $\rm F_2$ layer has a set exchange field of $h_2/E_F=1$.
As the rotation angle increases, starting from $\phi_3=0^\circ$, the 
oscillation amplitudes in the CPR diminish until vanishing completely at $\phi_3=90^\circ$,
at which point the exchange field in ${\rm F}_3$ is aligned with the central 
ferromagnet.
Further increases in $\phi_3 \geq 90^\circ$ results in a full reversal of the supercurrent
with the CPR profile changing sign and mirroring the results for $\Delta\varphi \leq 90^\circ$ 
about the center line $\Delta\varphi=180^\circ$.
Rotating  
 the exchange field angle, $\phi_3$,
 can be achieved through external means, and thus the resultant control of the 
 charge currents may be
beneficial in practical devices, including
nonvolatile memory elements.
The experimentally relevant critical current $J_c$ 
as a function of $\phi_3$ is
shown in Fig.~\ref{j_vs_h}(b),
where the data is extracted from the CPR in Fig.~\ref{j_vs_h}(a).
   The critical supercurrent is calculated
   by finding the maximum of the supercurrent over the
   entire phase difference interval, i.e., $J_c = \max[J_y(\Delta\varphi)]$.
The results show that
if the outer ferromagnet $\rm F_3$ has its exchange field vector initially 
pointing along $\hat{\bm y}$ ($\phi_3=90^\circ$), then
variations in $\phi_3$ about this point lead to an immediate linear increase 
in the critical current.
In Fig.~\ref{j_vs_h}(c), the CPR for a select $\phi_3$ are shown over an expanded range of phase differences
from $-360^\circ\leq \Delta\varphi \leq 360^\circ$. 
The scale on the right vertical axes corresponds to the $\phi_3=80^\circ$ case,
which is seen to exhibit  the conventional profile described by a sine function.
To compliment the current-phase relations, Fig.~\ref{j_vs_h}(d) contains the
free energies ${\cal F}_J$ for the curves in Fig.~\ref{j_vs_h}(c).
As was found in Fig.~\ref{Jy_vs_phase}, the ground state for $h_2/E_F=1$ corresponds to $\varphi_0=90^\circ,-270^\circ$.
Thus, while rotating the magnetization in the outer ferromagnet can result in the supercurrent
changing sign or vanishing altogether, the $\varphi_0$ ground state is robust and remains unchanged.

\begin{figure}[t!]
\centering
\includegraphics[width=0.49\textwidth]{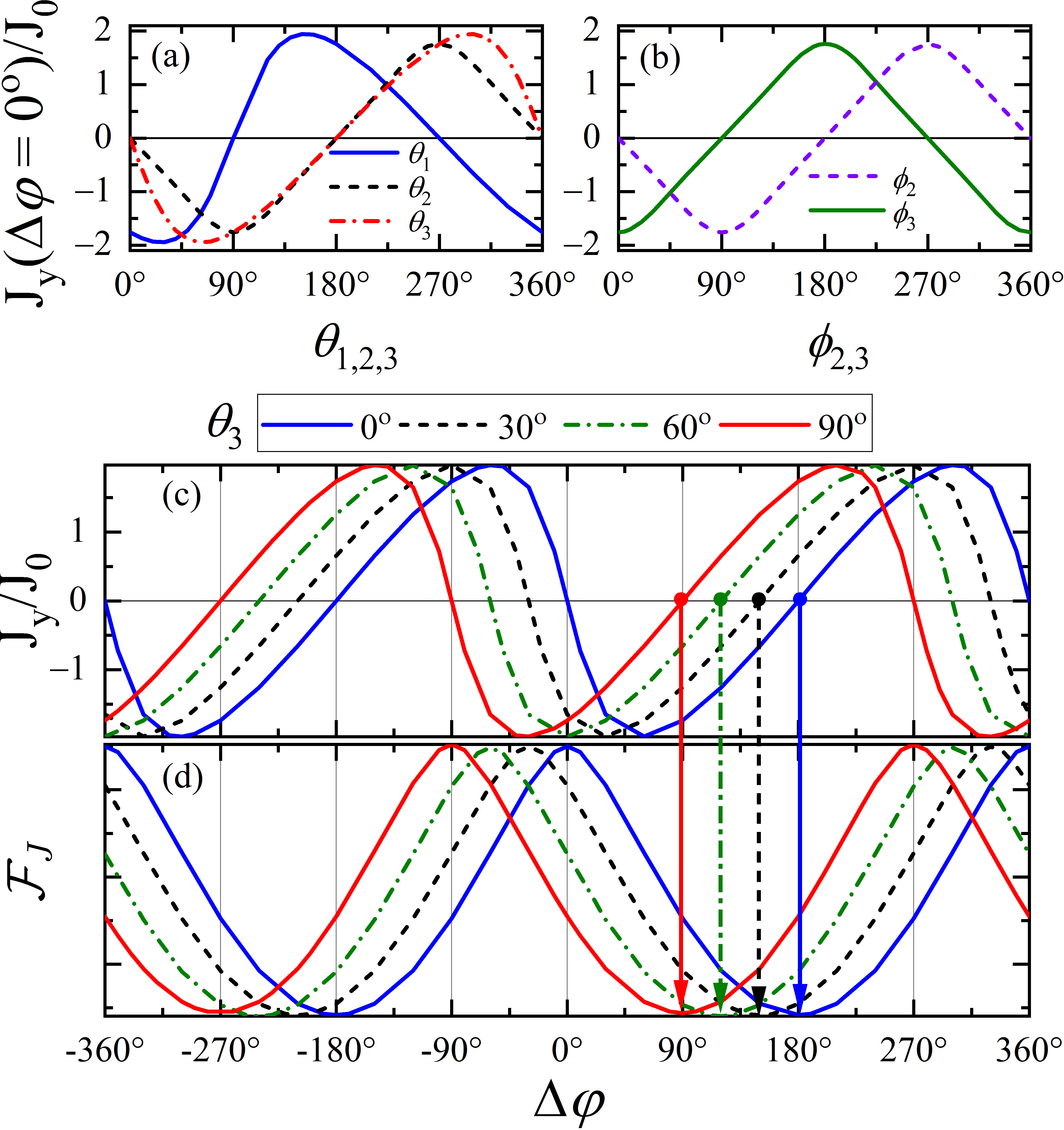}
\caption{
The top row (a) and (b) illustrate
the normalized anomalous supercurrent $J_y(\Delta\varphi=0^\circ)$ as a function of 
the angles $\theta_{1,2,3}$ and $\phi_{1,2}$ describing the exchange field vector orientations
in the respective ferromagnets ${\rm F}_{1,2,3}$.
In (c) the normalized Josephson current  is shown as a function of phase difference $\Delta \varphi$ for the selected angles $\theta_3=0^\circ, \theta_3=30^\circ, \theta_3=60^\circ, \theta_3=90^\circ$. The corresponding ground states are depicted by arrows pointing 
towards the free-energy minima in (d).
} 
\label{free_theta3}
\end{figure}

Next, we investigate how other orientations of the exchange field
vectors affect
the zero-phase  supercurrent response.
In Fig.~\ref{free_theta3}(a), the normalized supercurrent 
is shown as a function of the angles $\theta_{i}$ ($i=1,2,3$).
In most cases above, the exchange fields 
in $\rm F_1$, $\rm F_2$, and $\rm F_3$ were
in the $xyz$ configuration, respectively.
Now, when a given angle, e.g.,  $\theta_2$ is varied, 
$\rm F_1$ and $\rm F_3$ keep their exchange field vectors fixed along the $x$ and $z$ directions, respectively. 
When $\theta_1$ is varied, $\phi_1$ is set to $\phi_1=0^\circ$,
so $\theta_1$ sweeps occur in the $x$-$z$ plane.
Thus when rotating from $\theta_1=0^\circ$, to
$\theta_1=90^\circ$, 
 ${\bm h}_1$ goes from being aligned along the $x$ direction, to along the $z$ direction,
 respectively. This creates a situation with the two outer ferromagnets are both aligned along the $z$ direction,
 destroying the magnetic inhomogeneity needed for a finite anomalous current to flow,
 as observed in Fig.~\ref{free_theta3}(a).
As seen, the peak zero-phase current does not occur for the $xyz$ configuration as one might expect,
but rather when 
the exchange field vector in $\rm F_1$ is slightly skewed between the $x$ and $z$ directions.
When $\theta_2$ is now varying, the outer magnets are fixed according to  ${\bm h}_1=h_1 \hat{\bm x}$,
and ${\bm h}_3=h_3 \hat{\bm z}$. When $\theta_2$ is zero (or a multiple of $\pi$),
the leftmost two ferromagnets have colinear exchange fields, and thus the zero-phase current is zero in
those instances, as seen. As the exchange field rotates in the half-metal, it is observed that
$|J_y(\Delta\varphi=0)|$  is largest when $\theta_2 = \pi/2,3\pi/2$, i.e., precisely when the exchange field is 
strictly along $\pm y$.
Similar behavior is observed when $\theta_3$ varies, with the main difference being an offset
of the peak amplitude of the zero-phase current. Thus, for the outer magnets 
their optimal zero-phase supercurrent flow occurs when their exchange field vectors
point between the $x$ and $z$ directions.
Considering now magnetization rotations in the $y$-$z$ plane,
we present in Fig.~\ref{free_theta3}(b)
the zero-phase current as a function of
the angular variables $\phi_2$ and $\phi_3$. In each case, we fix
$\theta_{2,3} = 90^\circ$.
As $\phi_2$ or $\phi_3$ varies from 
 $0^\circ$ to $180^\circ$,
the corresponding exchange fields change their alignment from along $z$
to along $y$, respectively.
Therefore, we find that for the rotations of the
exchange field vector in the $y$-$z$ plane (see Fig.~\ref{diagram}),
plane, when $|J_y(\Delta\varphi)|$ is maximal for certain angles $\phi_2$,
it is zero when $\phi_3$ equals those angles.
More specifically, the curves have identical profiles that are offset from one another by $90^\circ$.

The CPR and free energy diagrams are illustrated in Fig.~\ref{free_theta3}(c) and 
Fig.~\ref{free_theta3}(d) respectively.  A range of orientations $\theta_3$ are considered,
as shown in the legend. The ${\rm F}_1$ and ${\rm F}_2$ magnets 
have their exchange fields aligned  along the $x$ and $y$ directions, respectively.
For each  $\theta_3$ shown,   $\phi_3=0^\circ$ is fixed, so that
$\theta_3$ variations result in ${\bm h}_3$ rotating in the $x$-$z$ plane.
For  $\theta_3=0^\circ$,   the exchange field vector in ${\rm F}_3$ is pointing along $x$,
while  $\theta_3=90^\circ$, has it
  along the $z$ direction.
The CPR in Fig.~\ref{free_theta3}(c) shows that it retains its profile as ${\bm h}_3$ rotates,
and simply undergoes an overall shift by an amount that is proportional to   $\theta_3$.
The free energy ${\cal F}_J$ has the same behavior
as Fig.~\ref{free_theta3}(d) shows, 
changing $\theta_3$ causes
the ground state phase difference $\varphi_0$
to 
shift similarly.
This tunablilty of the ground state by magnetization rotations can be achieved through an external magnetic
field, or a spin torque mechanism.

\begin{figure}[t!]
\centering
\includegraphics[width=1\columnwidth]{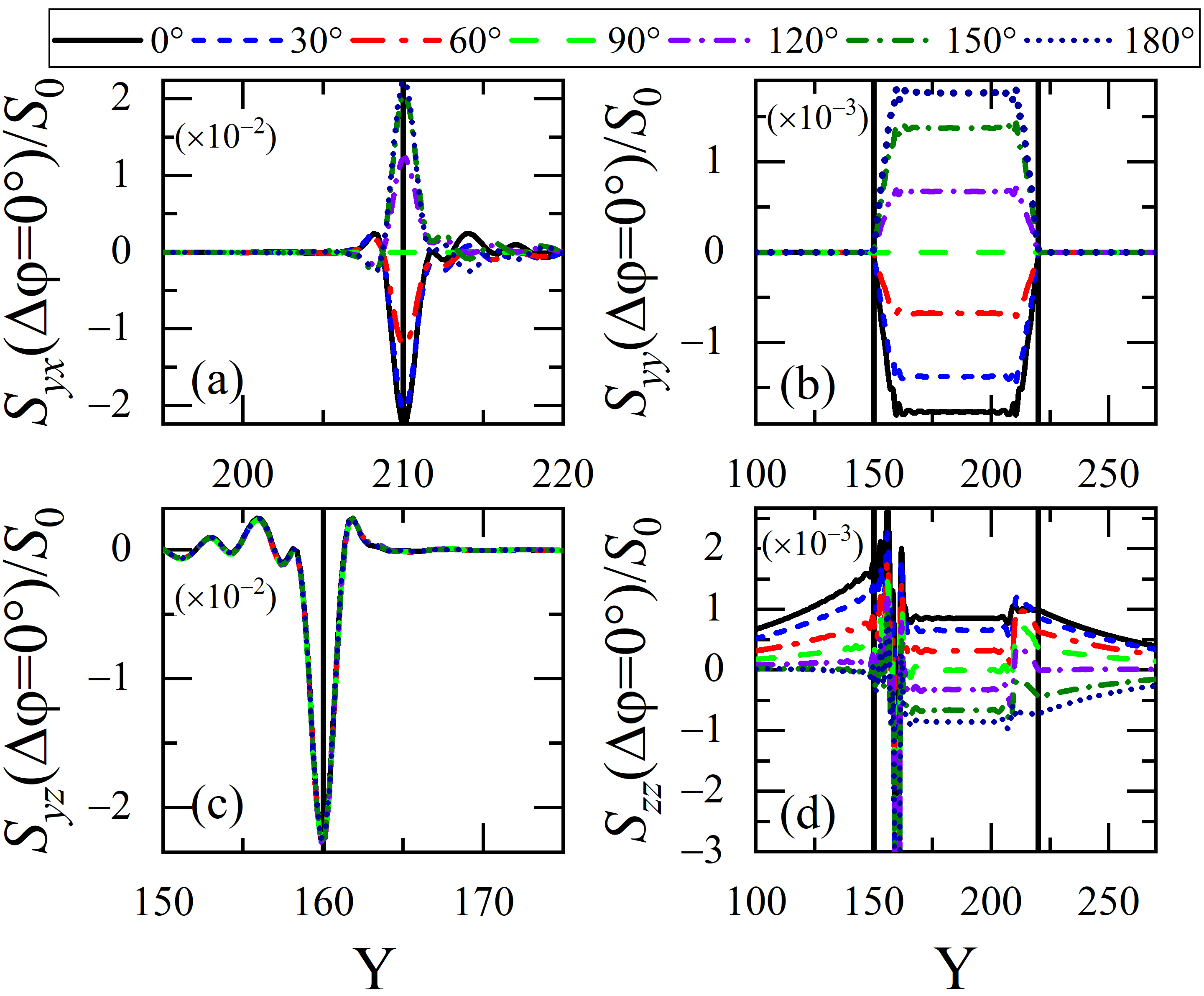}
\caption{
The nonzero components of the spin current tensor
are shown as a function of the normalized coordinate $Y=k_Fy$.
The phase difference between the two superconducting banks is set to $\Delta\varphi=0$.
The central ferromagnet is a half metal ($h_2/E_F=1$), while the outer ferromagnets
of the trilayer are fixed at $h_1/E_F=h_3/E_F=0.1$. The exchange field vector 
${\bm h}_3$ is allowed to vary in $y$-$z$ plane (the legend gives the rotation angle $\phi_3$).
The  vertical lines denote ferromagnet interfaces.
} 
\label{spin_vs_y}
\end{figure}

We now 
explore 
the spin currents
that arise when there is
a zero phase difference between the superconducting electrodes and 
 a finite supercurrent.
The anomalous supercurrent that is generated 
can become spin-polarized when passing through the ferromagnetic trilayer.
In Figs.~\ref{spin_vs_y}(a-c), the spin currents flowing in the $y$ direction $S_{y\sigma}$ are shown for each 
spin $\sigma$. All spin currents are plotted as functions of the dimensionless coordinate $Y$, with $Z=W/2$ fixed.
 In Fig.~\ref{spin_vs_y}(d) the spin current
for the $z$ component of spin, flowing in the transverse $z$ direction $S_{zz}$ is shown.
The legend identifies the different angles of rotation $\phi_3$ of the exchange field vector ${\bm h}_3$.
The Josephson junction has ${\bm h}_1$ and ${\bm h}_2$ in the trilayer
oriented along the $x$ and $y$ directions, respectively. For the type of 
magnetic arrangement studied here, the components of the spin current tensor $S_{zx}$ and
$S_{zy}$ are zero.
As seen in Fig.~\ref{spin_vs_y}(a), 
the $x$ component of spin has its current magnitude largest
at the interface between $\rm F_2$ and $\rm F_3$ (the vertical line at $Y=210$)
for $\phi_3=0^\circ,180^\circ$. This corresponds to the magnetic configurations where 
the exchange field vectors in the magnets are each orthogonal to one another.
For all other $0^\circ <\phi_3 <180^\circ$, the magnitude of the spin current reduces, including
vanishing at $\phi_3=90^\circ$.
Similar trends are observed in Fig.~\ref{spin_vs_y}(b) for the $y$ projection of spin,
with now a reduction in the overall magnitude of the spin currents. 
Additionally, we find that $|S_{yy}|$ increases approximately linearly in the thin outer
magnets, before entering the half-metal, where it modulates slightly while levelling out.
For both Figs.~\ref{spin_vs_y}(a) and \ref{spin_vs_y}(b), the spin currents vanish when $\phi_3=90^\circ$,
which for this orientation, corresponds to when the anomalous current is zero.
For the $z$ component of spin, 
the magnitude of the spin current peaks at the interface between $\rm F_1$ and $\rm F_2$ ($Z=160$).
Unlike what was observed in Fig.~\ref{spin_vs_y}(a), the spin current is insensitive to changes in $\phi_3$.
A transverse spin current also flows, as shown in Fig.~\ref{spin_vs_y}(d), where the $z$ component of the spin current flowing 
in the $z$ direction, $S_{zz}$, is shown. Within the left superconductor ($Y<150$), 
there is a weakly decaying spin current that oscillates upon entering the adjacent ferromagnet
and then $|S_{zz}|$ increases rapidly near the interface (at $Y=160$)
separating the ferromagnet from the half-metal. Within the half-metal and 
ferromagnet $\rm F_3$ ($160<Y<220$), the spin current is relatively constant (aside from a slight modulation at the interface).
Finally within the second superconductor ($Y>220$), 
the spin current again undergoes a slow decay that depends on the
particular angle $\phi_3$.

\begin{figure}[t!]
\centering
\includegraphics[width=0.47\textwidth,scale=0.01]{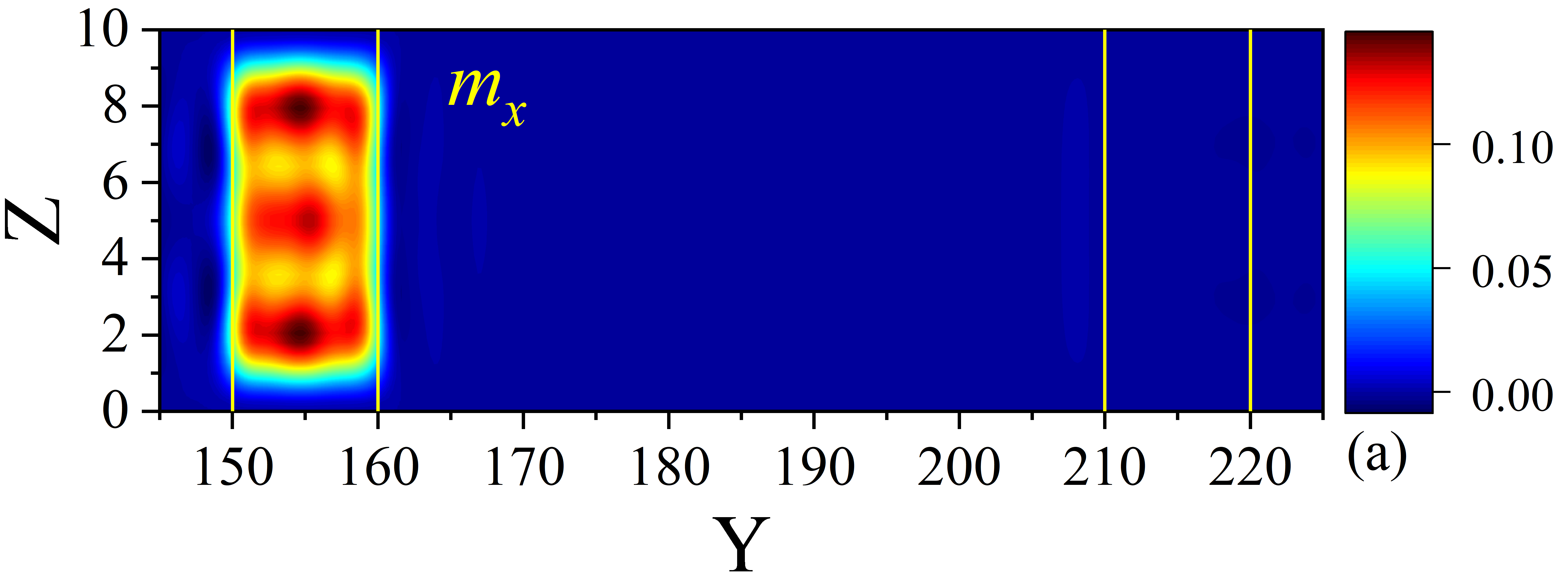}
\includegraphics[width=0.47\textwidth,scale=0.01]{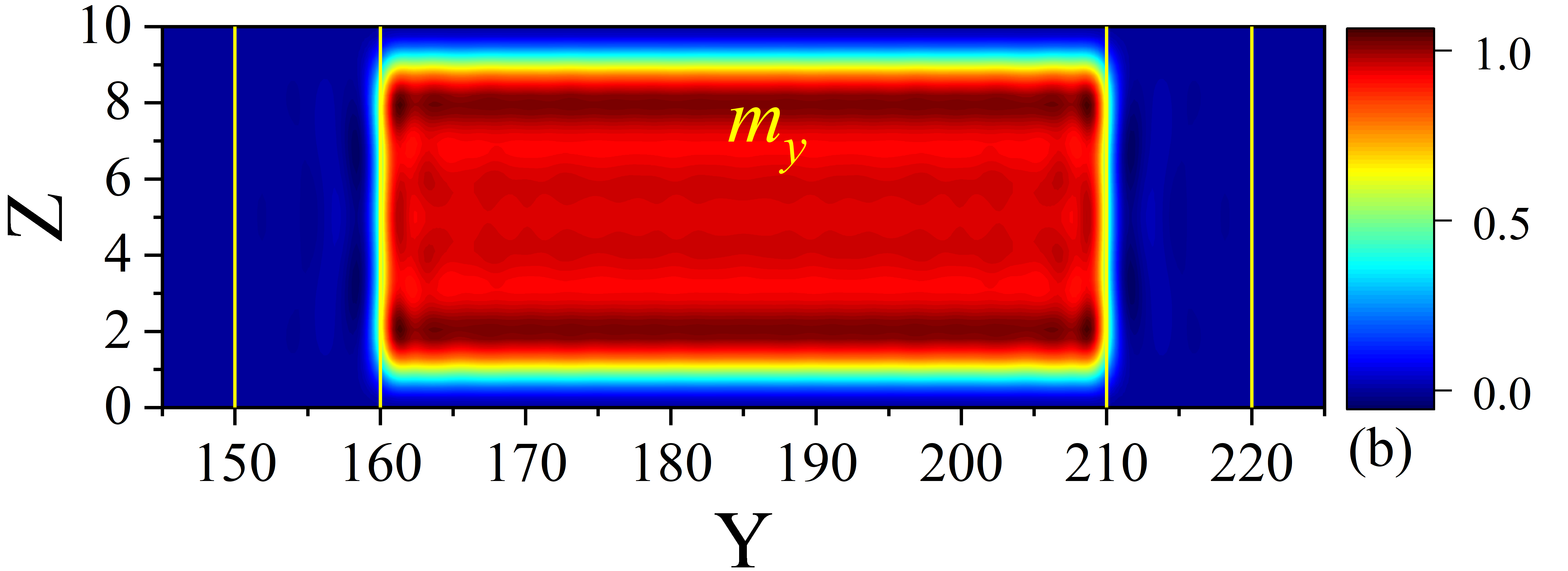}
\includegraphics[width=0.47\textwidth,scale=0.01]{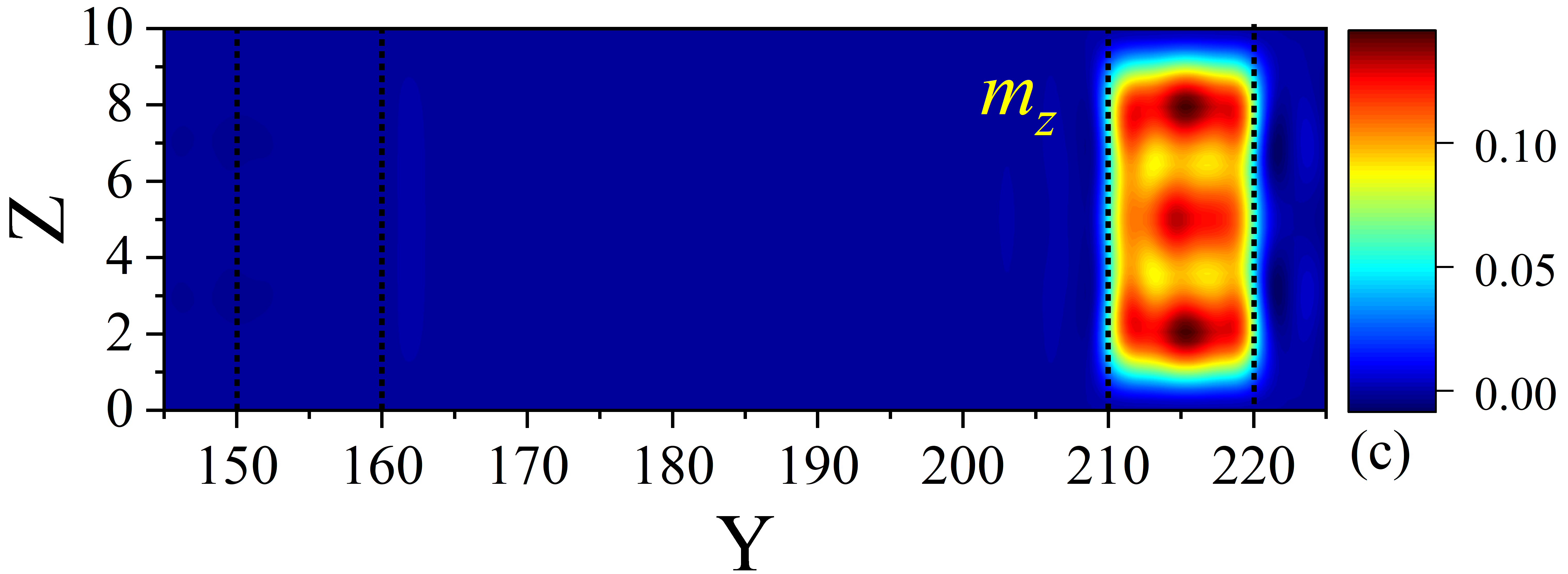}
\includegraphics[width=0.47\textwidth,scale=0.01]{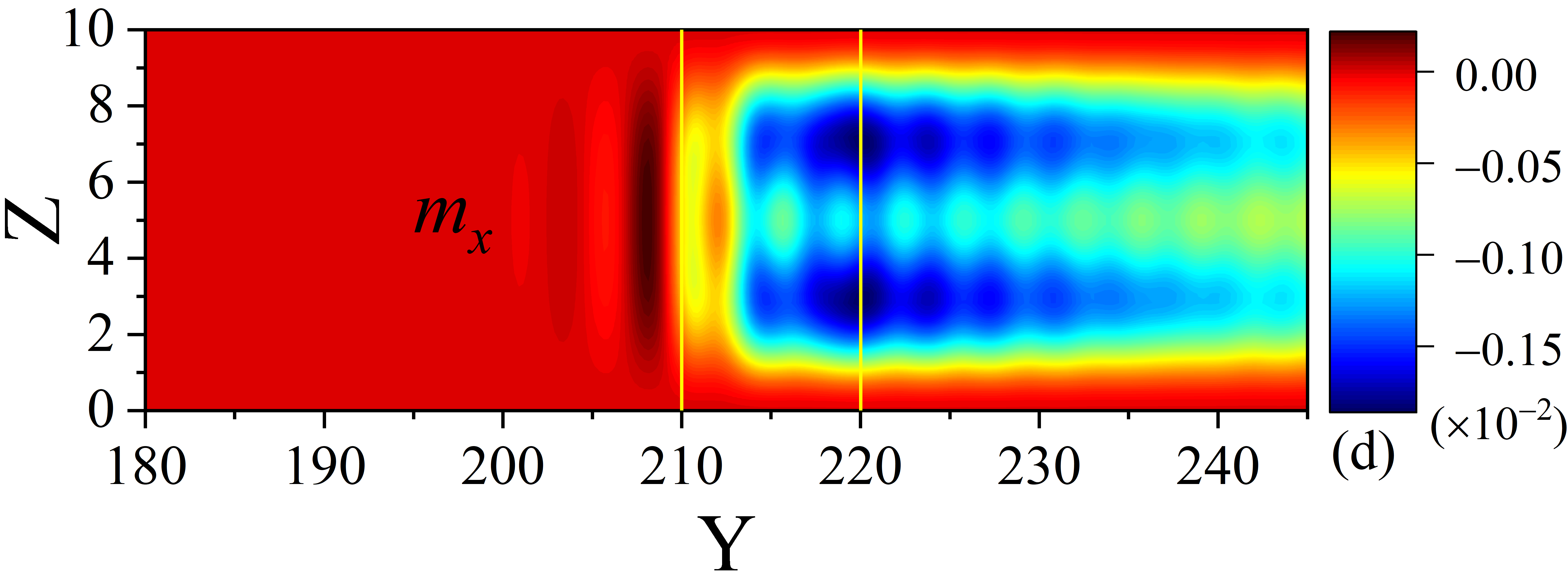}
\caption{The components of magnetic moment are shown (as labeled)
as a function of the dimensionless coordinates $Y$ and $Z$.
Panel (d) shows $m_x$ from (a) over a different scale to illustrate the
spin polarization that extends into $\rm F_3$ and the adjacent superconductor.
The yellow vertical lines enclose the ferromagnet regions.
The phase difference between the S banks is set to $\Delta\varphi=0^\circ$,
and we consider the $xyz$ configuration for the magnets.
} 
\label{2D_mm}
\end{figure}
In discussing spin transport quantities, we expect that 
the presence of an anomalous supercurrent and
proximity effects due to the exchange interactions,
 will lead to a spin torque transfer and a corresponding 
  leakage of magnetism.
The interaction of the spin current with the magnetization
is also relevant for memory technologies,
where the storage of information depends on the 
relative orientation of the magnetizations.
In Fig.~\ref{2D_mm}, a spatial mapping of each component of the
magnetic moment is shown for a half-metallic Josephson junction with the exchange field vectors in the $xyz$
configuration. The  phase difference $\Delta\varphi$ is set to zero, so
that an anomalous supercurrent is generated. The thin outer magnets have $h_{1,3}/E_F=0.1$.
In Fig.~\ref{2D_mm}(a), the $x$ component of magnetic moment $m_x$ (normalized by $\mu_B n_e$)
is shown. As seen, $m_x$ is confined mainly to the $\rm F_1$ region ($150<Y<160$),
vanishing in the vicinity of the hard wall boundaries at $Z=0$ and $Z=10$.
Due to the mutual proximity effects between the surrounding superconductor and half-metal, 
there is also a leakage of magnetization into those regions, as well as an increase in
$m_x$ from its bulk values within $\rm F_1$. 
The same behavior is also observed in Fig.~\ref{2D_mm}(c) for the 
$z$ component of magnetic moment.
The half metal region exhibits a more uniform magnetic moment [Fig.~\ref{2D_mm}(b)], but also exhibits similar 
 proximity effects, including rapid variations near the top and bottom of the structure, as well as an enhancement of its spin polarization near the edges at $Z=2$ and $Z=8$.
 We also present in Fig.~\ref{2D_mm}(d) $m_x$ over a different scale, to
demonstrate the typical long-ranged spin-imbalance that occurs, which in this case originates 
from $\rm F_1$ and is transferred into $\rm F_3$ ($210<Y<220$)
and the adjacent superconductor ($Y>220$).
 
 \begin{figure}[t!]
\centering
\includegraphics[width=0.49\textwidth]{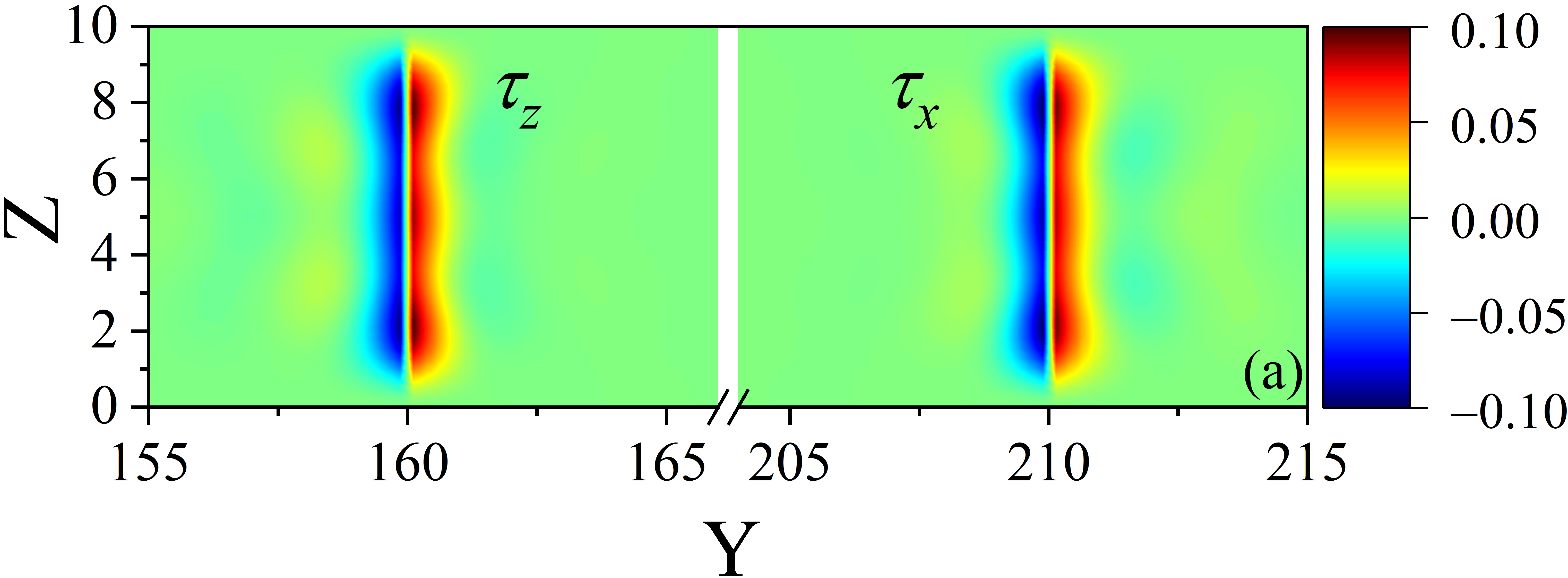}
\includegraphics[width=0.49\textwidth]{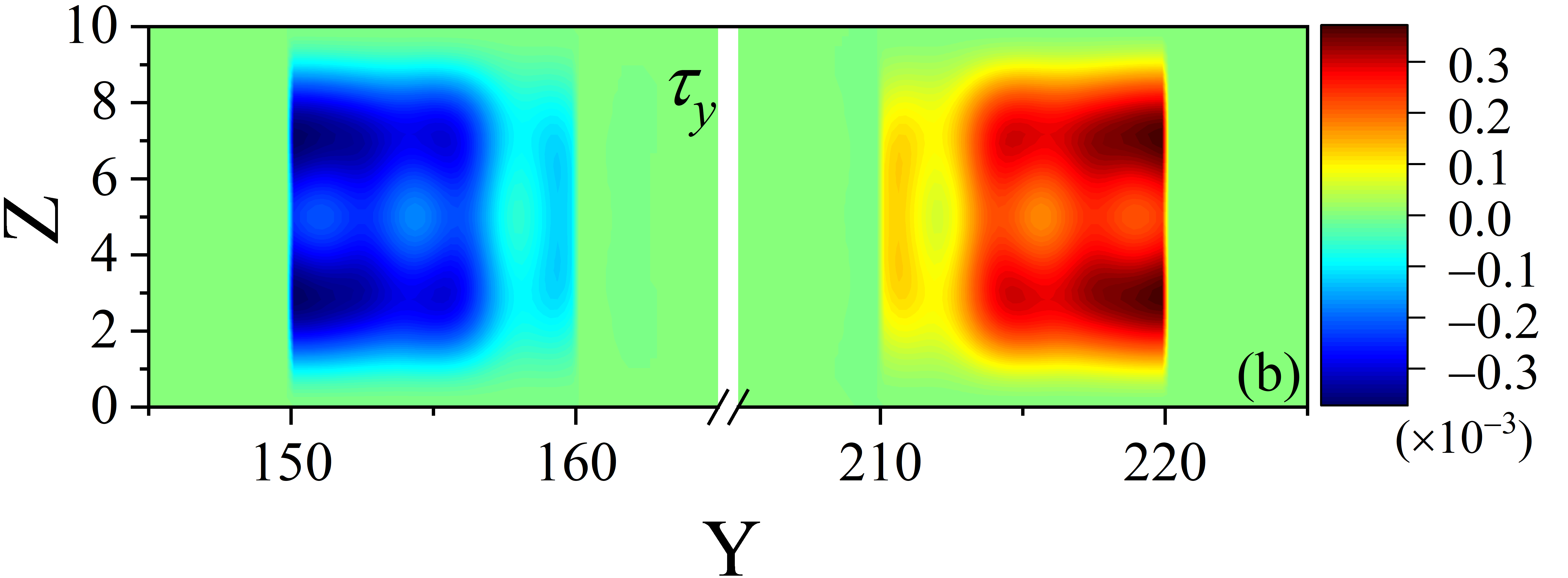}
\includegraphics[width=0.49\textwidth]{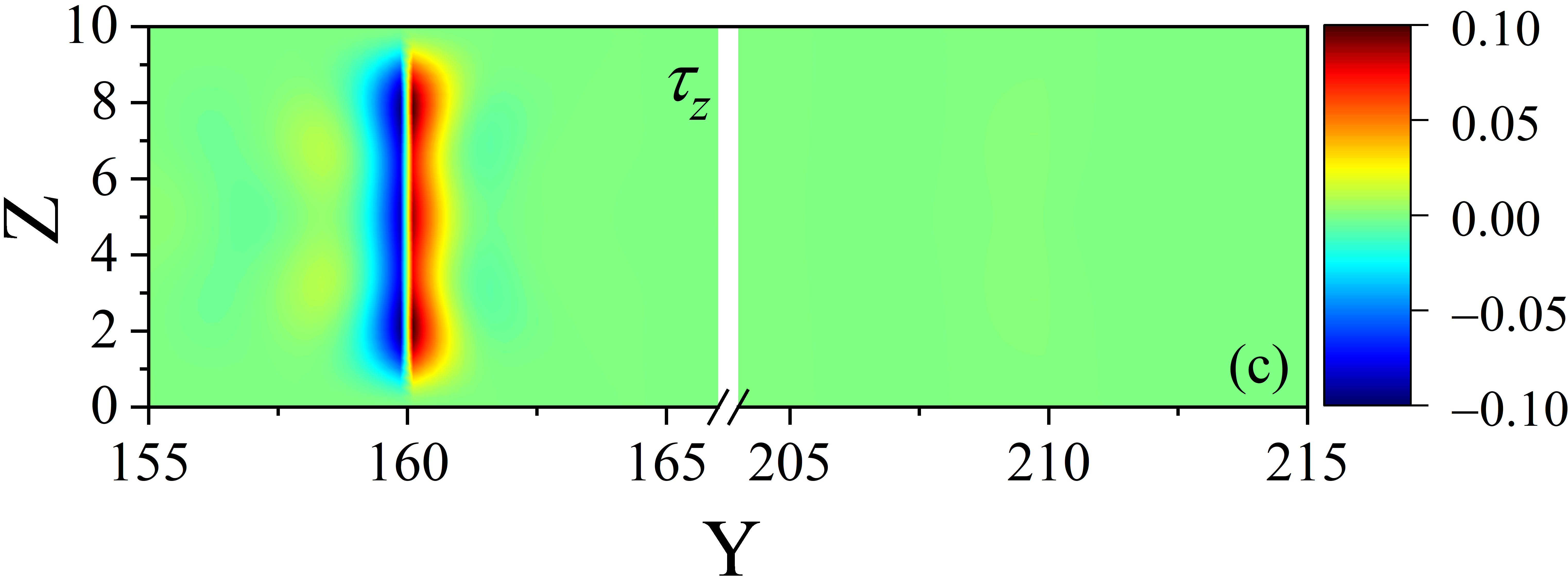}
\caption{
Spatial mapping of the $x$, $y$, and $z$ components of the spin torque.
The phase difference between the two superconducting banks is set to $\Delta\varphi=0$,
which established a zero-phase supercurrent.
The central ferromagnet is a half metal ($h_2/E_F=1$), while the outer ferromagnets
of the trilayer are fixed at $h_1/E_F=h_3/E_F=0.1$.
Here we fix $\theta_3$ so that $\theta_3=90^\circ$, 
 and 
variations in $\phi_3$ occur solely in the $y$-$z$ plane.
In (a) and (b), $\phi_3=0^\circ$, and in (c), $\phi_3=90^\circ$.
For panel (c),   $\tau_z$ 
is the only nonzero component.
} 
\label{tork}
\end{figure}
 The spatial behavior of the spin currents discussed in relation to Fig.~\ref{spin_vs_y} can 
   be modified by the local magnetizations due to the spin-exchange interaction,
    via the mechanism 
of spin torque transfer. 
The use of STT in half-metallic Josephson junctions
can result in faster magnetization switching times in
random access memories. The STT effect has been widely found to
occur in a broad variety of ferromagnetic materials, including
half-metals, making it widely accessible experimentally \cite{ilya}.
 It is thus important
not only to understand the behavior of the spin-polarized currents in ferromagnetic Josephson junctions, 
but also the various ways in which to manipulate them for practical spintronics type of
devices.
We therefore investigate in Fig.~\ref{tork} the equilibrium spin torques throughout the junction regions as
functions of the normalized coordinates $Y$ and $Z$.
We calculate  ${\bm \tau}$
from the magnetization and prescribed exchange fields \cite{Halterman2015:PRB} via: 
\begin{align} \label{tau1}
{\bm \tau} = -\frac{2}{\mu_B} {\bm m} \times {\bm h}.
\end{align}
In Figs.~\ref{tork}(a) and \ref{tork}(b), the outer free magnetization angle is
set to $\phi_3=0^\circ$ (the $xyz$ configuration)
so that an anomalous supercurrent is generated.
In Fig.~\ref{tork}(c), $\phi_3=90^\circ$ (the $xyy$ configuration), and hence no self-biased supercurrent is present.
Beginning with the top row, Fig.~\ref{tork}(a), the two spin torque components $\tau_x$ and $\tau_z$
 are shown superimposed on the same density plot. 
 The half metal boundaries 
 are clearly seen at $Y=160$ and $Y=210$.
It is observed that the STT is most pronounced near these boundaries, peaking antisymmetrically on
each side. This behavior is consistent with the spin currents shown in Fig.~\ref{spin_vs_y} and the expressions in
Eq.~(\ref{torky}), where 
for weak spatial variations in the $z$ direction,
Eq.~(\ref{torky}) states that
$\tau_x=\partial S_{yx}/\partial y$ and $\tau_z=\partial S_{yz}/\partial y$.
Therefore the slope of the curves in Fig.~\ref{spin_vs_y}(a,c)
gives a good approximation for $\tau_x$ and $\tau_z$ directly.
In particular, the normalized $S_{yx}$ in
Fig.~\ref{spin_vs_y}(a) exhibits a slope that undergoes an abrupt sign change when crossing the interface at $Y=210$, where the
slope is zero, in agreement with the sign change of $\tau_x$ about the interface [see in  Fig.~\ref{tork}(a)].
The same analysis shows consistency between Fig.~\ref{tork}(b) and $\tau_z$, with the notable feature 
that changing the angle $\phi_3$ in the free layer has no effect on the $z$ component of the spin currents flowing in the $y$ direction.
From Eq.~(\ref{scon}), we  see that $\bm{\tau}$ is
responsible for the change in local magnetizations due
to the flow of spin-polarized currents.
In Fig.~\ref{tork}(b), the weaker $y$ component of the spin torque is shown.
Unlike the other two components in Fig.~\ref{tork}(a), $\tau_y$ resides solely in the two outer magnets $\rm F_1$ and $\rm F_3$.
This follows from the leakage of magnetism discussed in relation to Fig.~\ref{2D_mm}(d),
where the $x$ component of the magnetization in $\rm F_1$ extends all the way to 
$\rm F_3$ via the spin currents $S_{yy}$ in 
Fig.~\ref{spin_vs_y}(b).
It should be emphasized that for
there to be a $y$ component of spin torque in 
$\rm F_3$, there must also be a finite $m_x$ there since 
$\tau_y$ in $\rm F_3$ is proportional to $h_{z} m_{x}$. 
Finally, we illustrate in Fig.~\ref{tork}(c) the $z$ component of the spin torque
for the rotation angle $\phi_3=90^\circ$. For this magnetic configuration, 
the anomalous current vanishes [See Fig.~\ref{j_vs_h}(c)].
As seen in Fig.~\ref{spin_vs_y}, all spin currents flowing in the $y$ direction also
vanish, except for those with a $z$ spin projection [Fig.~\ref{spin_vs_y}(c)],
which do not depend on $\phi_3$.
Since $\tau_z \approx \partial S_{yz}/\partial y$,
the torque in Fig.~\ref{tork}(c) is centered on the 
$\rm F_1/F_2$ interface at $Y=160$.

\section{Conclusions} \label{conclusions}
Motivated by recent advances and interest in making use of half-metallic materials (supporting one spin species) in superconducting hybrid structures, we have studied the spatial mappings
 and superconducting phase dependency of the
charge supercurrents, spin supercurrents, spin torques, and density of states in ballistic \sfffs~Josephson configurations where the
magnetizations in the F layers can have arbitrary directions and strengths. To this end, we have generalized a wavefunction approach, incorporating 
efficient computational techniques for optimizing the diagonalization of large-size matrices in a
parallel computing environment.
This allows our method to
 properly account for realistic three-dimensional systems,
 including  planar \sfffs~hybrids where the system is confined in two dimensions, and the third dimension is considered to be infinite. Interestingly, the charge supercurrent as a function of magnetization strength in the $\rm F_2$ layer shows that the critical supercurrent reaches its maximum value at high values close to the half-metallic phase so that it becomes only slightly different from its \fnf~counterparts.
 In a situation where the magnetization in the three F layers are mutually orthogonal (the "$xyz$" configuration), we find that a self-biased spontaneous supercurrent exists, which is largest when the $\rm F_2$ layer is a half-metal. We study the behavior of the spontaneous supercurrent for various magnetization orientations in the F layers and determine the
 most favorable configurations. 
We find that for intermediate exchange field strengths of the central ferromagnet layer,
a superconducting diode effect emerges, whereby a unidrectional dissipationless current can flow
by appropriately tuning the phase gradient between the outer superconductor electrodes.
The DOS studies illustrate robust zero energy peaks in the $xyz$ configuration that disappears when there is no three orthogonal magnetization components in the system. The zero energy peaks are spatially located in the $\rm F_1$ and $\rm F_3$ layers and become largest when $\rm F_3$ is a half-metal. Further studies on the spatial distributions of other important physical quantities  demonstrate an extremely 
 long-ranged magnetic moment  and spin currents extending from one 
 superconductor  in the \sfffs~system to the other.
 This long-ranged behavior  and anomalous supercurrent flow
generated spin torques controllable by rotations of the magnetization in the outer ferromagnet.

Apart from the results obtained for the specific \sfffs~system in the half-metallic regime, this generalized approach can be further expanded to systems with spin-orbit coupling, various types of impurities, 
and an external magnetic field. Unlike the limitations that the quasiclassical method demands, our approach can cover systems far beyond the quasiclassical approximations and properly account for band curvatures and arbitrary ratios among the energy scales, such as the exchange field and chemical potential, 
found in many types of systems.                

\acknowledgements
This material is based upon work supported by, 
or in part by, the Department of Defense High Performance Computing Modernization Program
(HPCMP) under User Productivity, Enhanced Technology Transfer, 
and Training (PET) contracts \#47QFSA18K0111, Award PIID 47QFSA19F0058.
K.H. is also supported in part by the NAWCWD In Laboratory Independent Research (ILIR) program.

\appendix

\section{Numerical method} \label{appA}

When expanding the quasiparticle amplitudes $u_{n,\sigma}$, and $v_{n,\sigma}$
in terms of a complete set of basis functions [see  Eq.~(\ref{expanse})], 
for numerical purposes
the sums are cut off at \cite{khold,ct2D}
$N_{l} {\sim} k_F l/\pi \sqrt{1+\max\{h_1,h_2,h_3\}}$, and $N_w {\sim} k_F w/\pi \sqrt{1+\max\{h_1,h_2,h_3\}}$,
thus ensuring that all of the quantum  states are included over the wide range of energy scales present  in the problem.
Inserting Eqs.~(\ref{expanse}) into the matrix eigensystem (Eq.~(\ref{Hk})) and invoking the orthogonality of the  basis functions,
we get the following set of matrix elements:
\begin{widetext}
\begin{align}
[{\cal H}_0]_{pqp'q'}&=\langle pq |H_0| p' q'  \rangle
= \frac{4}{wl} \int_0^l \int_0^w  dz dy  \sin \Big(\frac{p \pi y}{l}\Big) \sin \Big(\frac{q \pi z}{w}\Big) 
\left[-\frac{1}{2m}\Big(\frac{\partial^2}{\partial z^2} +\frac{\partial^2}{\partial y^2}\Big)+\frac{1}{2m}k_x^2-\mu \right]
 \sin \Big(\frac{p' \pi y}{l}\Big) \sin \Big(\frac{q' \pi z}{w}\Big), \nonumber \\
& + \frac{4}{wl} \int_0^l \int_0^w  dz dy  \sin \Big(\frac{p \pi y}{l}\Big) \sin \Big(\frac{q \pi z}{w}\Big) 
U(y,z)
 \sin \Big(\frac{p' \pi y}{l}\Big) \sin \Big(\frac{q' \pi z}{w}\Big), \label{Hoij}
 \end{align}
 \begin{align}
[{\cal D}]_{pqp'q'}& = \langle pq |\Delta| p' q' \rangle = 
\frac{4}{wl} \int_{d_N}^l \int_0^w dz dy \sin \Big(\frac{p \pi y}{l}\Big) \sin \Big(\frac{q \pi z}{w}\Big) \Delta(y,z)
\sin \Big(\frac{p' \pi y}{l}\Big) \sin \Big(\frac{q' \pi z}{w}\Big), 
\end{align}
\begin{align}
[{\cal I}_j]_{pqp'q'}& = \langle pq | h_j |  p' q' \rangle = \label{bigI}
\frac{4}{wl} \int_{d_S}^{d_S+d_{F1}} \int_0^w dz dy \sin \Big(\frac{p \pi y}{l}\Big) \sin \Big(\frac{q \pi z}{w}\Big) h_{j1} 
\sin \Big(\frac{p' \pi y}{l}\Big) \sin \Big(\frac{q' \pi z}{w}\Big) \nonumber \\
&+\frac{4}{wl} \int_{d_S+d_{F1}}^{d_S+d_{F1}+d_{F2}} \int_0^w dz dy \sin \Big(\frac{p \pi y}{l}\Big) \sin \Big(\frac{q \pi z}{w}\Big) h_{j2} 
\sin \Big(\frac{p' \pi y}{l}\Big) \sin \Big(\frac{q' \pi z}{w}\Big)\\
&+\frac{4}{wl} \int_{d_S+d_{F1}+d_{F2}}^{d_S+d_{F1}+d_{F2}+d_{F3}} \int_0^w dz dy 
\sin \Big(\frac{p \pi y}{l}\Big) \sin \Big(\frac{q \pi z}{w}\Big) h_{j3}  
\sin \Big(\frac{p' \pi y}{l}\Big) \sin \Big(\frac{q' \pi z}{w}\Big),\quad j=x,y,z, \nonumber
\end{align}
\end{widetext}
where $q,q',p,p'$ are integers.
In what follows, 
we scale the expressions in Eqs.~(\ref{Hoij}-\ref{bigI}) by the Fermi energy $E_F$
to make them dimensionless.
The integrations for the free-particle term in  Eq.~(\ref{Hoij}) are easily done, 
which after dividing each term by the Fermi energy $E_F = k_F^2/(2m)$,  takes the following dimensionless form:
\begin{align}
[{\cal H}_0]_{pqp'q'}=\left[  \left\{ \left(\frac{p \pi }{k_F l}\right)^2 + \left(\frac{q \pi }{k_F w}\right)^2  \right\} 
+\frac{\varepsilon_\perp}{E_F}-1
\right] \delta_{q q'}\delta_{p p'}.
\end{align}
Here, for simplicity, the scattering potential $U(y,z)$ is assumed to be zero in this work. 
The above dependence on four indices $q,q',p,p'$ can be reduced to a system of two indices $k$ and $k'$
suitable for matrix operations by
changing the index operations so that,  $\{p,q\} \rightarrow k$, and $\{p',q'\} \rightarrow k'$.
This creates a more manageable and compact system of equations

\begin{figure}[t!]
\centering
\includegraphics[width=1\columnwidth]{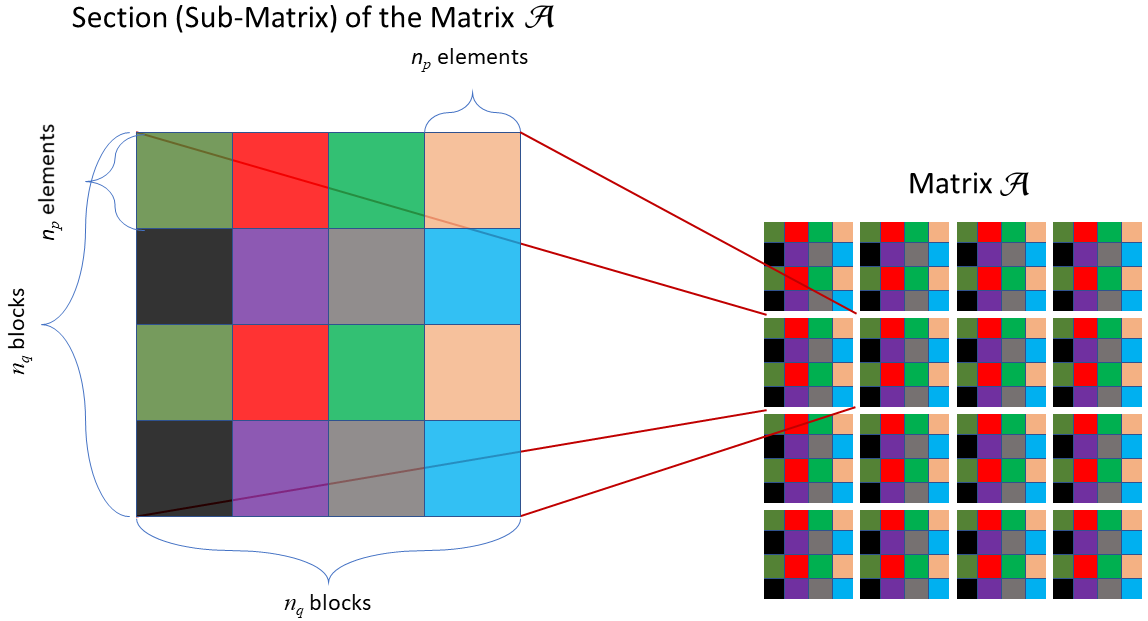}
\caption{A representative case of the distributed memory approach where one matrix is comprised of 16 sections. Each matrix is distributed
within its block cyclic distribution scheme. 
The image shows one of the 16 sections distributed over 2 processor row and 4
processor columns.
For large problems, the number of processor rows and columns for each grid could be significant. 
} 
\label{comp}
\end{figure}

We can now cast
 the matrix eigensystem in Eq.~(\ref{Hk}) into a following form suitable for numerical purposes:
 ${\cal A} \Psi_n = \epsilon_n \Psi_n$, where the matrix ${\cal A}$ is given by,
 \begin{align}      
{\cal A}= 
&\begin{pmatrix}
{\cal H}_0 -{\cal I}_z&-{\cal I}_x+i {\cal I}_y&0&{\cal D} \\[5 pt]
-{\cal I}_x-i {\cal I}_y&{\cal H}_0 +{\cal I}_z&{\cal D}&0 \\[5 pt]
0&{\cal D}^*&-({\cal H}_0 -{\cal I}_z)&-{\cal I}_x-i{\cal I}_y \\[5 pt]
{\cal D}^*&0&-{\cal I}_x+i {\cal I}_y&-({\cal H}_0+{\cal I}_z) 
\end{pmatrix}
\label{Hk2},
\end{align}
Here $\overline{\epsilon}_n = \epsilon_n/E_F$, and
 the expansion coefficients
are written in vector form:
 $\Psi_n=({\bm u}_{n,\uparrow},{\bm u}_{n,\downarrow},{\bm v}_{n,\uparrow},{\bm v}_{n,\downarrow})^T$,
with 
 ${\bm u}_{n,\sigma}=(u_{n,\sigma}^1,\cdots,u_{n,\sigma}^{N})^T$,
and ${\bm v}_{n,\sigma}=(v_{n,\sigma}^1,\cdots,v_{n,\sigma}^{N})^T$.
The rank of the matrix  in Eq.~(\ref{Hk2}) is $N=4 N_l N_w$.
The numerical implementation of the
solver used in solving the $4N \times 4N$ eigensystem
in Eq.~(\ref{Hk2}) was done using 
object-oriented Fortran with the
MPI \cite{mpi}, OpenMP \cite{openmp}, and  ScaLAPACK \cite{scala} libraries. 
 
The  energy $\varepsilon_\perp$ corresponds to the kinetic energy of the quasiparticles that are moving
transverse to the longitudinal $x$ direction. This parameter is effectively a good quantum number
and when performing the sum  over states for a given  physical quantity, a fixed number $n_{\varepsilon_\perp}$ 
of these transverse states must be included.
During each step of the solution process, 
and for the final calculations, the eigenvectors and eigenenergies for $n_{\varepsilon_\perp}$ matrices need to be calculated. 
Each of these $n_{\varepsilon_\perp}$ matrices are distributed in a block cyclic manner onto a subset of the available MPI ranks called a processor grid (PG). Due to the construction of the problem, the matrix ${\cal A}$ can be split into 16 equally sized sections, each containing one or many PGs, depending on the $ n_{\rm row} \times n_{\rm col}$ form of a PG for a given problem. 
A diagram  representing this distributed memory process  is shown in Fig.~\ref{comp}.
Each section is made of $n_q \times n_q$ blocks and each block contains $n_p\times n_p$ matrix elements. Each one of these sections must be perfectly tessellated by the block size, which then allows individual ranks to own the same block in each section of the matrix ${\cal A}$. This allows for reduced computational and communication complexity. 

\begin{figure}[t!]
\centering
\includegraphics[width=0.98\columnwidth]{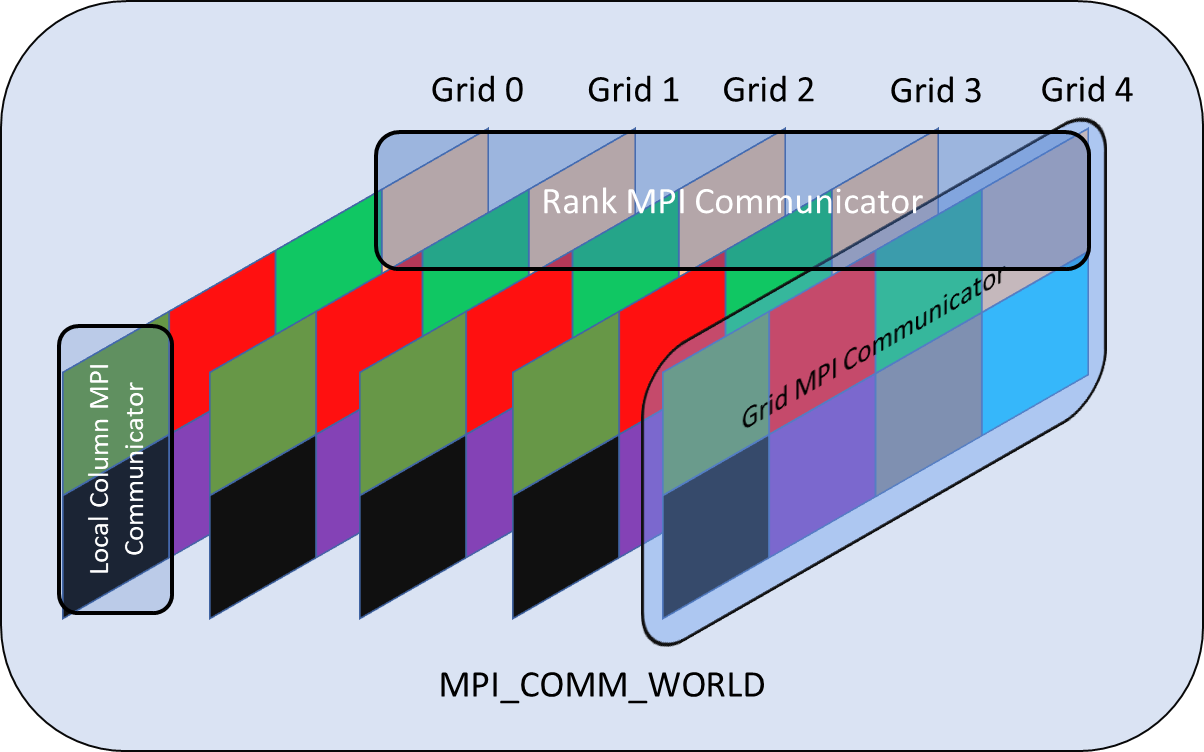}
\caption{MPI\_COMM\_WORLD is partitioned into multiple processor grids. Each rank within the program uses 4 separate communicators at different times. MPI\_COMM\_WORLD, a communicator for the PG (Grid communicator), a communicator for the specific column within the processor grid (local column communicator), and a communicator with the corresponding grid rank in all of the other processor grids (Rank communicator).
} 
\label{MPI}
\end{figure}

For communication, MPI ranks are grouped into four communicators, which are graphically represented in Fig.~\ref{MPI}.
These four are: MPI\_COMM\_WORLD, a communicator for the PG (Grid communicator), a communicator for the specific column within the PG
 (local column communicator), and a communicator with the corresponding grid rank in all of the other processor grids (Rank communicator). At the end of every iteration, all system properties of interest within the solver are summed using MPI collectives across one, some or all of the communicator groups depending on the particular requirements for a given property.

\section{Optimizations and Computational Techniques}
\label{appB}
In addition to parallelization via MPI, thread-level parallelization was implemented using OpenMP. 
A na\"ive
 coding for the calculation of $u_{n,\sigma}(y,z)$, and $v_{n,\sigma}(y,z)$ 
 would require a loop over $q$, within a loop over $p$, 
within a loop over $z$, and finally within a loop over $y$ for 4 levels of looping. 
The quasiparticle expansions in Eq.~(\ref{expanse}) that involve the terms,
$\sum_p u_{n,\sigma}^{p,q} \sin(p\pi y/l) $ and $\sum_p v_{n,\sigma}^{p,q}\sin(p\pi y/l)$,
could be calculated and stored within a vector for every $q$ within each iteration of the loop over $y$, outside of the loop over $z$.
This
reduces  the number of nested loops required.
 In addition, inner loops were written to be of the form: 
\texttt{c = sum(a(:)~$\cdot$~b(:))}, making it possible to 
 fully exploit vector-based compiler optimizations. 

\section{Memory requirements} 
\label{appC}

\begin{figure}[t!]
\centering
\includegraphics[width=0.99\columnwidth]{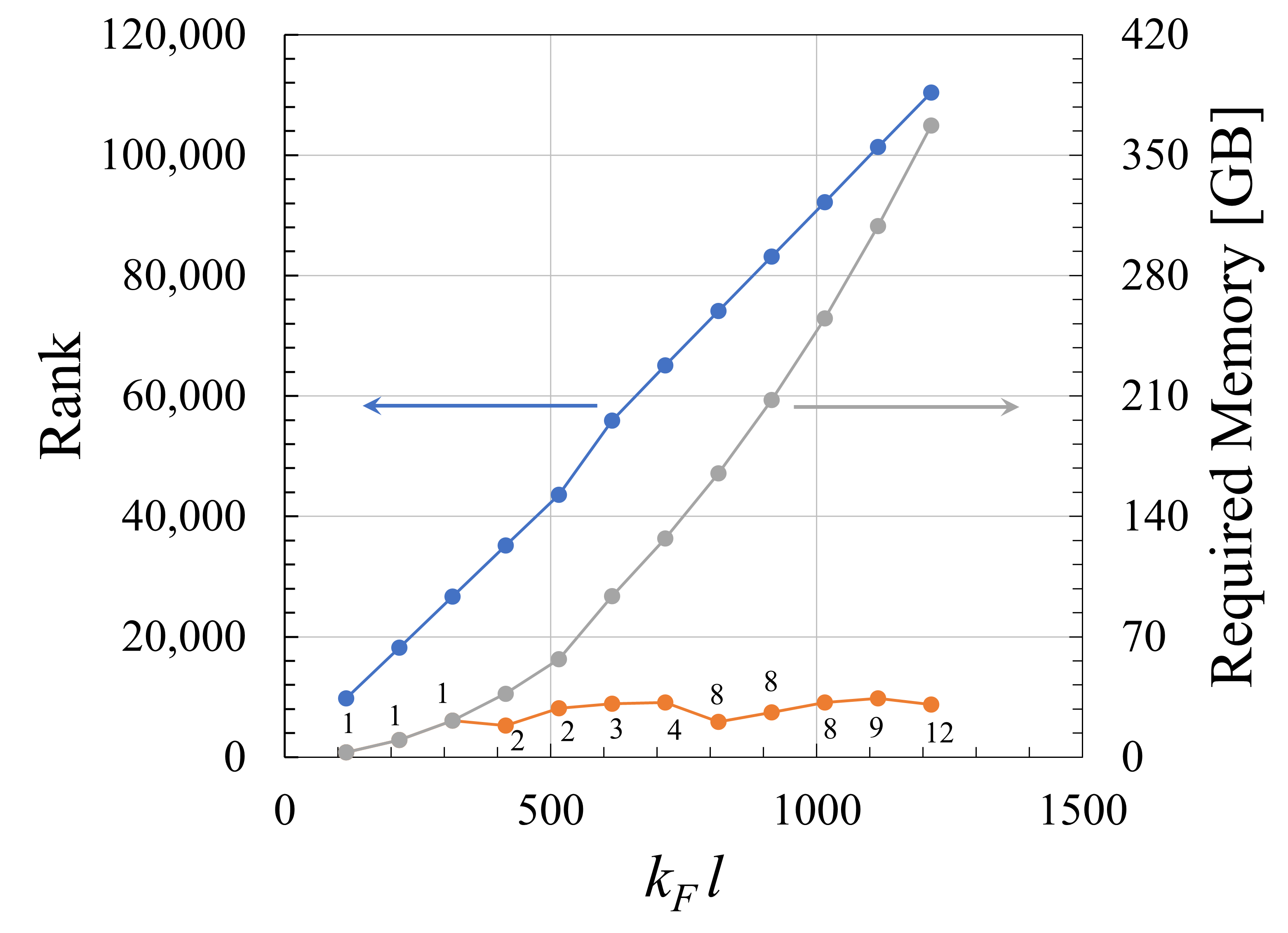} 
\caption{
Matrix Rank and Minimum Total Memory Requirements.
Growth of the rank of matrix ${\cal A}$ and associated required memory to calculate the eigenvectors of a representative matrix. 
Here, $W=100$, blocking is set to $n_p$, and cluster size=1.
The grey curve is the required memory to 
solve the eigenvectors for one matrix.
The orange curve is the associated memory requirement on each process,
with the
grid size listed adjacent to each point.
} 
\label{mem}
\end{figure}
The memory requirements for the problem can be categorized in three major categories: the matrix ${\cal A}$, 
its eigenvectors, and the workspace required by ScaLAPACK to calculate the eigenvectors. The memory required for ${\cal A}$
 and its associated eigenvectors is equivalent and scales as 
 $\alpha (4n_p n_q)^2$. Calculations were performed using 32-bit precision, so the value of $\alpha$ is 8 bytes 
 (4 bytes each for the real and imaginary portions). 
 The amount of workspace required by ScaLAPACK was dependent upon the rank of $\cal{A}$, with
 the number of rows and columns being held by that specific process, and a clustersize parameter (additional memory to orthogonalize eigenvectors). Due to library constraints, the number of 4-byte elements in the ScaLAPACK work array was limited to $2^{31}$, or 8 GB. This was a per-process limit on just the work array, and each rank could use more than 8 GB total. Furthermore, more than one process could be launched on a single compute node to make efficient use of available resources.

The growth of the memory requirements is shown in Fig.~\ref{mem}, 
using nominal values for the system parameters.
The  width of the junction was kept constant,
and its  normalized length was increased from 115 to 1215 in increments of 100. 
 As can be seen, the total required memory for a single matrix increases with the square of the matrix rank. By increasing the number of processes used to evaluate ${\cal A}$, the per-process memory could be kept roughly level. 
 The per-process memory usage and grid size is selected by maximizing the amount of memory managed by a single process.

\end{document}